\documentclass[useAMS,usenatbib]{mnras}

\usepackage{amsmath}
\usepackage{graphicx}
\usepackage[T1]{fontenc}
\usepackage{aecompl}
\usepackage[dvipsnames]{color}
\usepackage[normalem]{ulem}
\usepackage{threeparttable}
\usepackage{multirow}
\usepackage{array}
\usepackage{journals}

        \usepackage[shortlabels]{enumitem}
\setlist[enumerate, 1]{1\textsuperscript{o}}

\newcommand{\SII}{[S~{\sc ii}]\ }
\newcommand{\OIII}{[O~{\sc iii}]\ }
\newcommand{\NII}{[N~{\sc ii}]\ }
\newcommand{\HII}{H~{\sc ii}\ }

\newcommand{\HI}{H~{\sc i}\ }
\newcommand{\Ha}{H$\alpha$\ }
\newcommand{\kms}{\,\mbox{km}\,\mbox{s}^{-1}}

\newcommand{\HST}{\textit{HST}\ }
\newcommand{\SIIHa}{[S~{\sc ii}]/H$\alpha$}
\newcommand{\NIIHa}{[N~{\sc ii}]/H$\alpha$}
\newcommand{\OIIIHb}{[O~{\sc iii}]/H$\beta$}
\newcommand{\be}{\begin{equation}}
\newcommand{\ee}{\end{equation}}

\def \gtsima{$\, \buildrel > \over \sim \,$}
\def \ltsima{$\, \buildrel < \over \sim \,$}
\def \simgt{\lower.5ex\hbox{\gtsima}}
\def \simlt{\lower.5ex\hbox{\ltsima}}

\begin{document}

	\title[Star formation in the SGS of Holmberg~I] {Star-formation complexes in the `galaxy-sized' supergiant shell of the galaxy Holmberg~I}
	
	\author[Egorov et al.]{
		Oleg V.~Egorov$^{1, 2}$\thanks{E-mail: egorov@sai.msu.ru},
		Tatiana A.~Lozinskaya$^{1}$,
		Alexei V.~Moiseev$^{2, 1, 3}$     \newauthor
		and  Grigory V.~Smirnov-Pinchukov$^{1, 4}$ \\
		$^{1}$ Lomonosov Moscow State University, Sternberg Astronomical Institute,	Universitetsky pr. 13, Moscow 119234, Russia 
		\\
		$^{2}$ Special Astrophysical Observatory, Russian Academy of Sciences, Nizhny Arkhyz 369167, Russia
		\\
		$^{3}$ Space Research Institute, Russian Academy of Sciences, Profsoyuznaya ul. 84/32, Moscow 117997, Russia
		\\
		$^{4}$ Max Planck Institute for Astronomy, Konigstuhl 17, D-69117 Heidelberg, Germany
	}
	
	\date{Accepted 2018 Month 00. Received 2018 Month 00; in original
		form 2018 Month 00}
	
	\pagerange{\pageref{firstpage}--\pageref{lastpage}} \pubyear{2018}
	
	\maketitle
	
	\label{firstpage}

\begin{abstract}

We present the results of observations of the galaxy Holmberg~I carried out 
at the Russian 6-m telescope in the  narrow-band imaging, long-slit spectroscopy, and scanning Fabry-Perot interferometer modes.
A detailed analysis of gas kinematics, ionization conditions, and metallicity of star-forming regions in the galaxy is presented.
The aim of the paper is to analyse  the  propagation of star formation in the galaxy and to understand the
role of the ongoing star formation in the evolution of the central `galaxy-sized' supergiant \HI shell (SGS),
 where all regions of star formation are observed.
We show that star formation in the galaxy occurs in large unified complexes rather than in individual giant \HII
regions. Evidence of the triggered star formation is observed both on scales of individual complexes and of the whole galaxy.
We identified two supernova-remnant candidates and one late-type WN star and analysed their spectrum and surrounding-gas kinematics.
We provide arguments indicating that the SGS in Holmberg~I is destructing  by the influence of star formation occurring on its rims.

\end{abstract}


\section{Introduction}

The origin of multiple kpc-scale supergiant shells (SGSs) and holes in the \HI discs of
galaxies is a topic of lively discussions, yet it is now not an intriguing issue. Dwarf irregular (dIrr) galaxies
 served as the main testing ground to study SGSs  with sizes up to 2--3 kpc and a lifetime of hundreds Myr.
Today it is clear that dwarf galaxies sustain a high SF efficiency over several hundred Myr and even up to 1000~Myr
 in some cases \citep{Mcquinn2009, Mcquinn2010a, Mcquinn2010b} with local short starbursts occurring during this period.
 Such a long period of intense star formation provides sufficient energy from stellar winds and supernovae  from multiple generations of stars to drive formation of the observed holes and SGSs \citep[see, e.g.,][]{Weisz2009a,Weisz2009b,Cannon2011a,Cannon2011b}.

Today the study of short local current starburst events usually occurring on the rims of  SGSs
  seems to be most interesting. Such regions of star formation are observed as the extended complexes of ionized gas
 embedded in the walls of \HI SGSs.
It is important to understand what triggers these local starburst events and how the ongoing star formation affects the evolution of an \HI SGS.
A new burst of star formation can either help an SGS grow to larger sizes or push the neutral gas back into the central hole and/or outside the SGS destroying the giant structure.
According to \cite{Warren2011}, the interplay between these effects will ultimately define whether a galaxy
will or will not form a large \HI hole.

In order to shed light on the highlighted topic, we previously studied two well-known SGSs in the irregular galaxies IC~2574  \citep{Egorov2014} and Holmberg~II \citep{Egorov2017}.
Both galaxies contain a large number of \HI SGSs, while bright regions of the ongoing star formation are observed only in the walls of few of them. Such a picture of the \HI distribution and star formation is typical of many irregular and spiral galaxies \cite[see, e.g., ][]{Bagetakos2011}. In \cite{Egorov2017}, we showed  that star formation on the rims of an SGS can often be a result of its collision with other neighbouring \HI SGSs. The existing models of collided superbubbles also support this possibility \citep[e.g.,][]{Chernin1995, Ntormousi2011, Kawata2014, Vasiliev2017}.

In this work we focus on another type of galaxy with SGSs. Some dIrr galaxies contain a galaxy-sized central \HI supershell that becomes the main feature of the ISM encompassing the whole optical disc. The most striking examples of such galaxy-sized SGSs might be observed in
DDO 88 \citep{Simpson2005},   Sextans  A \citep{Skillman1988, vanDyk1998},  M81 dwA \citep{Sargent1983, Westpfahl1994}, SagDIG \citep{Young1997}, Holmberg I \citep{Tully1978, Ott2001}. The latter galaxy is the object of our study in this paper.

\begin{table}
	\centering
	{\scriptsize
		\caption{Published parameters of Holmberg~I.}\label{tab:galaxies}
		\begin{tabular}{lcc}
			\hline
			Parameter & Value & Reference \\
			\hline
			& 	DDO 63 & \\
			Alternative names     &  UGC 5139 & \\
			& KDG~57 &  \\
			RA (2000)     & ${09^h40^m32.3^s}$ & \\
			Dec (2000)   &  $+71^\circ 11\arcmin 11\arcsec$ &\\
			$D$ (Mpc)   &  3.9 & \citet{Dalcanton2009} \\
			$M_B$ (mag)  &  -14.8&  \citet{things} \\
			$R_{25}$ (kpc) & 1.9 &  \citet{things} \\
			$R_\mathrm{HI}$ (kpc) & 4.5 & \citet{Bagetakos2011} \\
			$M_\mathrm{HI}$ ($10^8 M_\odot$)  & 1.4 & \citet{Bagetakos2011} \\
			$V_{sys}$ ($\kms$)  & 140 & \citet{Oh2011} \\
			$\mathrm{\log(SFR_{H\alpha})}$ ($M_\odot\ yr^{-1}$)  & -2.21 & \citet{Karach2013} \\
			$\mathrm{12+\log(O/H)}$ & 7.9 & \citet{Moustakas2010} \\
			\hline
		\end{tabular}
	}
\end{table}

Holmberg~I is a low-mass dIrr galaxy (IAB(s)m according to the \citealt{deVaucouleurs1991} classification) belonging to the M81 group. Its main parameters are listed in Table~\ref{tab:galaxies}.

The first \HI radio observations of Holmberg~I performed at the Westerbork radio telescope were presented
by \cite{Tully1978}.  \cite{Puche1994} and \cite{things} published the data obtained at the VLA.
 According to these observations, the \HI disc in Holmberg~I shows the low-velocity dispersion $\sigma_\mathrm{HI} \sim 9 \kms$ and a slow rotation 
\citep{Oh2011, Stilp2013}. Due to the low inclination ($i = 10-14^\circ$), the estimates of the vertical \HI scaleheight  by different authors are very uncertain: $h=250-550$~pc according to \cite{Ott2001}, 640~pc according to \cite{Bagetakos2011}, and $191\pm41$~pc in \cite{Stilp2013}.

The most detailed analysis of \HI in Holmberg~I was performed by \cite{Ott2001}. The authors showed that the \HI distribution in Holmberg~I is dominated by the central supergiant shell; its morphological center is offset by 0.75 kpc with respect to the dynamical center of the galaxy. From a comparison with isochrones, as well as from dynamical modelling based on the \HI data,
the authors derived that the age of the central supergiant \HI shell is equal to $80\pm20$ Myr.
Later \cite{Bagetakos2011} have detected five other \HI holes of a smaller size in the galaxy.

\cite{Vorobyov2005} carried out numerical simulations of the ISM structure in Holmberg~I. They investigated three scenarios of the central \HI hole formation: 
multiple supernova (SN) explosions, a single gamma-ray burst, and a vertical impact of a high-velocity cloud. Multiple SN explosions were shown to reproduce the \HI morphology  of Holmberg~I more accurately.

\begin{figure}
\includegraphics[width=\linewidth]{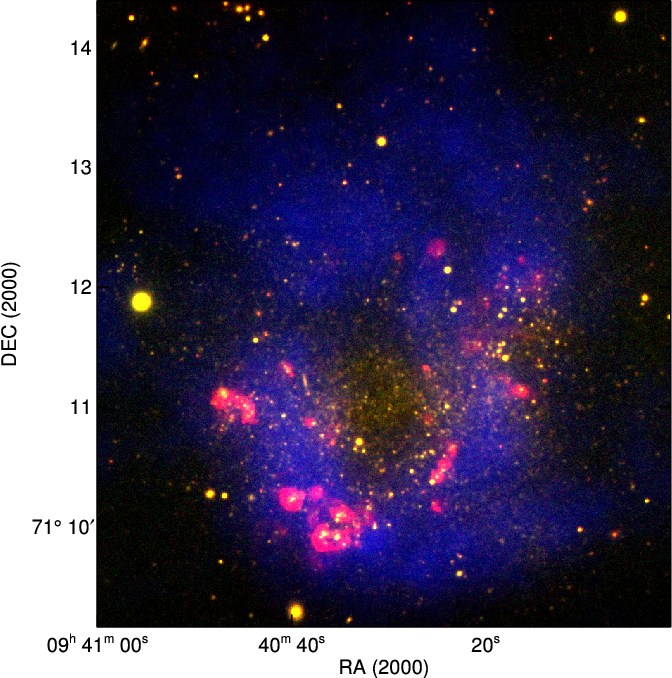}
\caption{False-colour image of Holmberg~I. Blue colour shows the \HI emission in 21 cm taken from the LITTLE THINGS survey \citep{Hunter2012}; red and yellow colours show the optical images presented in this work and trace the emission in the H$\alpha$ line and the stellar population, respectively.}\label{fig:HIHa}
\end{figure}

A comparison of optical images and the \HI distribution shows that the regions of ionized gas are observed on the rims of the central \HI supergiant shell in Holmberg~I (see Fig.~\ref{fig:HIHa}).
The \HII regions in the galaxy were catalogued by  \citet{MH94}. The authors identified 32 \HII regions and measured their sizes and luminosities. Only a few of them were previously spectroscopically observed. The metallicity of a few bright \HII regions in Holmberg~I has been studied earlier by \cite{MH96,Croxall2009, Moustakas2010}. \cite{Berg2012} performed a deep spectroscopy of one \HII region in the galaxy and estimated the oxygen abundance using the direct method. These studies have showed that the galaxy has a low oxygen abundance $12+\log(\mathrm{O/H}) \sim 7.9$  that is typical of its luminosity.

In the present paper, we analyse the gas kinematics and the emission spectrum of star-forming regions in the galaxy Holmberg~I  and discuss  the origin of these regions. Our aim is to recognize how star formation affects  the evolution of \HI SGSs using Holmberg~I as an example. The paper is organized as follows. Section~\ref{sec:obs} presents the details of performed observations and data reduction.  Section~\ref{sec:stars} describes  the young stellar population in the galaxy. In Section~\ref{sec:kinematics} we analyse the general \HI kinematics of the galaxy and perform a search for expanding shells and supershells in \HI and H$\alpha$.  Section~\ref{sec:morphology} is dedicated to the analysis of the ionized gas morphology in Holmberg~I, while in Section~\ref{sec:hiiregs} we describe the spectroscopic properties of the individual \HII regions. In Section~\ref{sec:discuss} we discuss the inferred results and in Section~\ref{sec:summary} summarize the main conclusions.

\section{Observations and data reduction}\label{sec:obs}

 \subsection{Optical FPI-observations}
\label{sec_obs_ifp}

\begin{table*}
\caption{Log of the observations}
\label{tab:obs_data}
\begin{tabular}{llrllllll}
	\hline
	Data set       & Date of obs    & ${T_{exp}}$, s & $FOV$                             & $''/px$             & $\theta$, $''$   & sp. range              & $\lambda_c$, \AA              & $\delta\lambda$ or $FWHM$, \AA        \\ \hline
	FPI \#1  & 2014 Dec 19/20 & $40\times300$  & \multirow{2}*{$6.1'\times6.1'$} & \multirow{2}*{0.71} & 2.1              & \multirow{2}*{8.8~\AA\, around \Ha} &   \multirow{2}*{$-$}  & \multirow{2}*{0.48~($22\kms$) } \\
	FPI \#2  &  2015 Nov 04/05 & $40\times300$  &                                 &                     & 1.5              &                 &                    &  \\
	\hline
	LS PA=142 & 2016 Mar 02/03 & 6600 & \multirow{5}*{$1\arcsec\times6.1\arcmin$} & \multirow{5}*{0.36} & 3.5 &\multirow{5}*{3600--7070} & \multirow{3}*{$-$}  & 5.2 \\
	LS PA=150 & 2017 Dec 09/10 & 3600 &  &  & 1.5 & &  & 4.7 \\
	LS PA=212 & 2017 Dec 08/09 & 8100 &  &  & 2.2 & &  & 4.7 \\
	LS PA=259 & 2016 Feb 10/11 & 4500 &  & & 2.2 &  & & 4.5 \\
	LS PA=259 & 2015 Nov 05/06 & 2700 & & & 1.0 & &  & 4.5 \\
	\hline
	Image FN655 & 2015 Apr 25/26 & 3600 & \multirow{6}*{$6.1\arcmin\times6.1\arcmin$} & \multirow{6}*{0.36} & 1.2 & \Ha+\NII & 6559  & 97 \\
	Image FN674 &  2017 May 28/29 & 2700 & && 1.3 & \SII 6717+6731 & 6733  & 60  \\
	Image FN501 & 2017 May 28/29 & 1800 & & & 1.4 & \OIII 5007 & 5012  & 119  \\
	Image FN608 & 2015 Apr 25/26 & 1200 &  &  & 1.2 & continuum & 6099 & 166 \\
	Image FN712 & 2015 Apr 25/26 & 1200 &  &  & 1.2 & continuum & 7137 & 209 \\
	Image SED525 & 2017 May 28/29  & 900 &  &  & 1.6 & continuum & 5271 & 251  \\
	\hline
\end{tabular}
\end{table*}

The observations were carried out at the prime focus of the 6-m
telescope of the Special Astrophysical Observatory of the Russian Academy of Sciences (SAO RAS) using the scanning Fabry--Perot interferometer (FPI) -- IFP751 -- mounted  inside the
SCORPIO-2  multimode focal reducer \citep{scorpio2}. The operating spectral range  around the \Ha emission line was cut by a bandpass filter with the $\mathrm{FWHM}\approx14$~\AA\ bandwidth.
During the scanning process, we have consecutively obtained 40 interferograms  at different distances
between the FPI plates. The log of these observations and the parameters of other data sets are given in Table~\ref{tab:obs_data}, where ${T_{exp}}$ is the exposure time, $FOV$ -- the field of view,  $''/px$ -- the pixel size on the final images, $\theta$ -- the final angular resolution, $\lambda_c$ -- the central wavelength of the filters used,  $\delta\lambda$ -- the final spectral resolution, and $FWHM$ is a bandwidth of the filters used.

\begin{figure}
	\includegraphics[width=\linewidth]{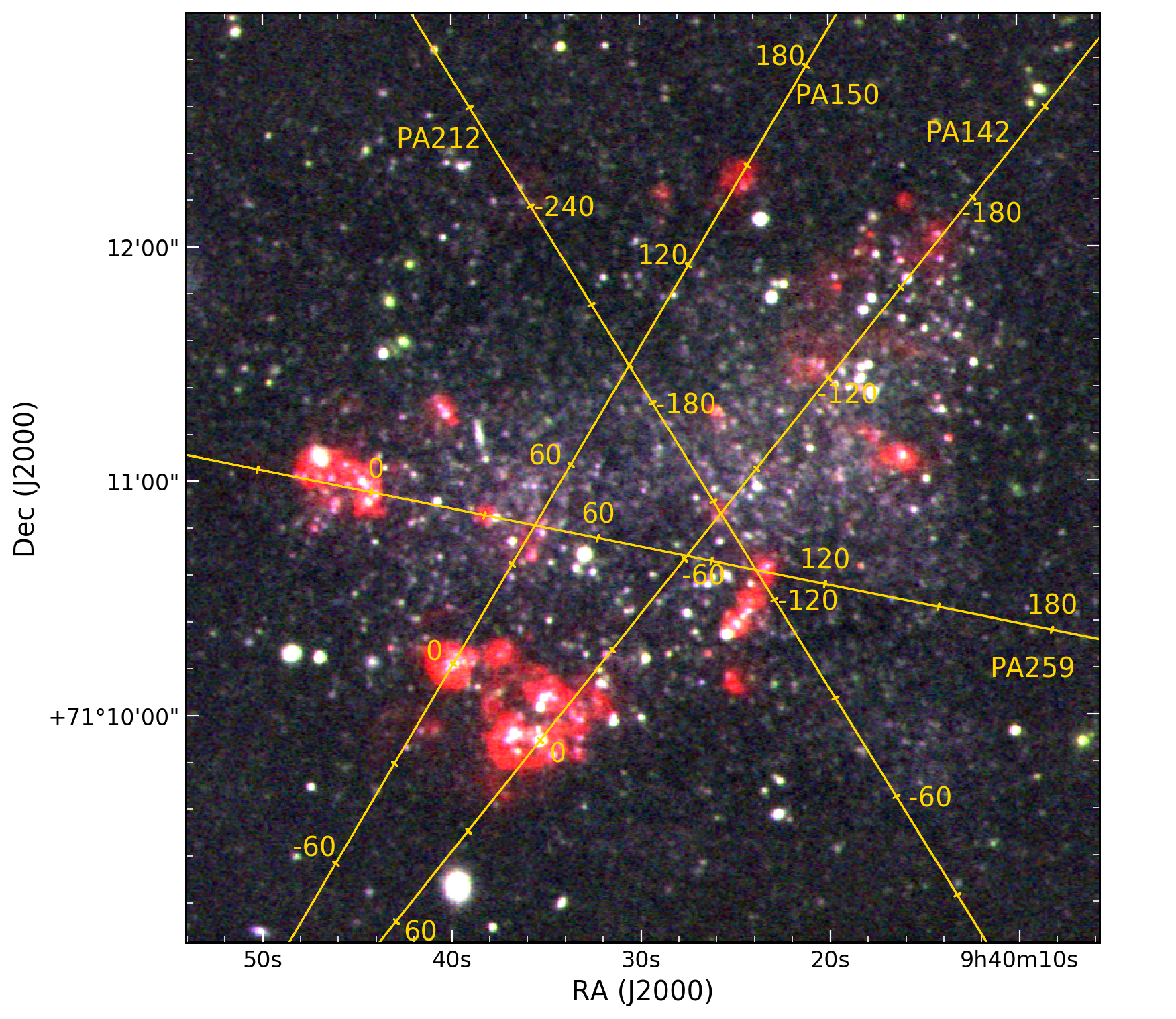}	
	\caption{False-colour image of Holmberg~I. Red colour corresponds to the image obtained with the narrow-band filter centred on the H$\alpha$ emission line, while the blue and green channels show the emission in the continuum (see Section~\ref{sec_obs_img} for details). The slit positions used during spectroscopic observations are overlaid.}\label{fig:slitpos}
\end{figure}

The data reduction was performed using a software package running in the \textsc{idl} environment. For a detailed description of the data reduction algorithms and software, see \citet{Moiseev2002,  Moiseev2008} and \citet{Moiseev2015}.  After the initial reduction, sky line subtraction, photometric and seeing corrections made using the reference stars,
and wavelength calibration, the observed data were combined into data cubes, where each pixel in
the field of view contains a 40-channel spectrum.
We observed the galaxy at two position angles in order to remove the parasitic ghost reflection. These data were reduced separately to get the wavelength cubes of
the object. The analysis of \Ha line profiles
was carried out using the multicomponent Voigt fitting \citep{Moiseev2008} that yields flux, line-of-sight velocity, and velocity dispersion corrected for instrumental broadening for each component as an output.

\subsection{Long-slit spectroscopic observations}
\label{sec_obs_ls}

We observed the galaxy Holmberg~I  in the long-slit mode of the SCORPIO-2 focal reducer at 6-m SAO RAS telescope. Using the VPHG1200@540 grism as a disperser, we have obtained spectra with four slit positions (PA=142, PA=150, PA=212, and PA=259, see Fig.~\ref{fig:slitpos}). The length of the used slit was 6~arcmin, while its width was 1 arcsec. The spectral resolution estimated as the \textit{FWHM} of air-glow emission lines varied from 4.5 to 5.2~\AA.

The data reduction was performed in a standard way using the \textsc{idl}-based pipeline for reducing long-slit
spectroscopic data obtained with SCORPIO-2. The data reduction includes the bias subtraction,
line curvature and flat field corrections, linearisation and air-glow lines subtraction. We linearised the spectra using the reference spectrum of the He–Ne–Ar lamp obtained during observations. In order to calibrate the spectra to the absolute intensity scale, we observed one of spectrophotometric standards (AGK~81d266 and  G191B2B)  at a close zenith distance immediately after or before the object.

The spectrum corresponding to the slit position PA=259 was observed twice with different exposures and at different seeings. In order to increase the signal-to-noise ratio, we have combined the two observations into a single spectrum for a further analysis. Before that, the spectrum with a better seeing was convolved with the Gaussian kernel to achieve the same spatial resolution as for the second one.

It is worth noting that the spectrum with PA=150 was observed in bad weather conditions. The unstable transparency due to clouds has made the flux calibration unreliable for that spectrum, yet the relative flux ratios of nearby emission lines analysed in this paper are correct.

The main purpose of the long-slit observations was to analyse the age, ionization conditions, and gas metallicity of \HII regions in the galaxy. Hence, we were interested in the emission-line fluxes and equivalent width. To measure the fluxes of emission lines we used our own software working in the \textsc{idl} and based on the \textsc{mpfit} \citep{mpfit} routine. The Gaussian fitting was applied to measure the integrated line fluxes of each studied region. To subtract a spectrum underlying stellar population, we performed its modelling using the \textsc{ULySS}\footnote{\url{http://ulyss.univ-lyon1.fr}} package \citep{Koleva2009}.
However, its influence on the measured emission-line fluxes of \HII regions appears to be almost negligible in our case, so we decided to skip this procedure in order to not add systematic errors, which can be produced in case of an unreliable model. 
For estimating the final uncertainties of the measured line fluxes, we quadratically added the errors propagated through all data-reduction steps to the uncertainties returned by the \textsc{mpfit}.

A reddening correction was applied to each spectrum before estimating the line-flux ratios listed in the paper. For that we derived the $E(B-V)$ colour excess from the observed Balmer decrement and then used the \cite{Cardelli1989} curve parametrized by \cite{Fitzpatrick1999} to perform a reddening correction.
In this paper, we use the following abbreviations of the emission-line flux ratios: \SIIHa\, is F([S~{\sc ii}] 6717,6731\AA)/F(H$\alpha$); \NIIHa\, is F([N~{\sc ii}] 6584\AA)/F(H$\alpha$); \OIIIHb\, is F([O~{\sc iii}] 5007\AA)/F(H$\beta$).

\subsection{Narrow-band imaging}
\label{sec_obs_img}

Deep optical images of Holmberg~I  in the H$\alpha$, [S~\textsc{ii}], and \OIII  emission lines were taken at the prime focus of the 6-m SAO RAS telescope  with the SCORPIO \citep{scorpio} and SCORPIO-2 multimode focal reducers using the narrow-band filters FN655, FN674, and FN501, respectively. The transmission curves of each used filter can be found on the SCORPIO-2 website\footnote{\url{https://www.sao.ru/hq/lsfvo/devices/scorpio-2/filters_eng.html}}.

We used the broader-band FN608, FN712, and SED525 filters centred on the continuum to subtract  stellar contamination from the images obtained in the same night. 

In order to calibrate the images to the absolute intensity scale, we observed the spectrophotometric standard AGK~81d266 with each line-centred filter immediately after the object at a close zenith distance. The obtained images were additionally corrected for the interstellar reddening using the \cite{Cardelli1989} curve parametrized by \cite{Fitzpatrick1999} and the colour excess $E(B-V) = 0.05$ derived from our spectra.

Because the FWHM of the FN655 filter is broader than the distance between the \Ha and [N~{\sc ii}]\, emission lines, the image in this filter is contaminated by the [N~{\sc ii}] 6548, 6584~\AA\, emission. According to our spectroscopic observations, the \NIIHa\, flux ratio is $0.05 - 0.07$ for Holmberg~I (see Section~\ref{sec:hiiregs}). Since the transmission of the FN655 filter is lower in the region of \NII emission lines than that of H$\alpha$, we  conclude that the \NII contamination of \Ha images does not exceed 5 per cent.

A similar situation also takes place for the FN501 filter -- the \OIII 5007~\AA\, image is contaminated by the \OIII 4959~\AA\, line that is 3 times fainter. Taking it into account, as well as the differences of the FN501 filter transmission at 5007~\AA\, and 4959~\AA, we multiplied the fluxes of the final \OIII image to 0.8 to reveal only \OIII 5007~\AA\, emission-line contribution.

Our final narrow-band images show surface-brightness value at the $3\sigma$ level corresponding to $(3.3, 2.4, 3.6) \times 10^{-17} \mathrm{erg\ s^{-1} cm^{-2} arcsec^{-2}}$ in the H$\alpha$, [S~\textsc{ii}], and \OIII emission lines, respectively.

In order to get the \OIIIHb\, flux ratio from our images, we took into account a theoretical ratio of $\mathrm{H\alpha/H\beta} = 2.86$ for $T_e=10000$~K and used the reddening corrected \Ha image divided to this value to estimate the H$\beta$ flux. All the \OIIIHb\, maps and values derived from the images and used further in the paper were obtained in that way. Because of the low metallicity and hence the low dust content in Holmberg~I, as well as of the small variations of $E(B-V)$ (according to our spectra, see Section~\ref{sec:hiiregs}), this procedure yields realistic values of the \OIIIHb\, flux ratio in our case. The performed Monte Carlo simulations showed that typical relative uncertainties of that ratio for our data are 10 per cent for the signal-to-noise ratio S/N=30 and 30 per cent for S/N=5. At the same time these uncertainties should be 5 and 27 per cent, respectively, if direct measurements of the H$\beta$ flux were available.

 \subsection{Multiwavelength archival data used}
\label{sec_obs_other}

We used the archival VLA data in the \HI 21-cm line from the LITTLE THINGS survey \citep{Hunter2012} for studying the \HI gas distribution and kinematics. In this work we used the
natural-weighted  data cube with the angular resolution $beam=14.7\times12.7$~arcsec and velocity scale $2.6 \kms$ per channel.

In order to analyse the stellar population, we used the images of Holmberg~I obtained with the ACS/WFC camera at the \textit{Hubble Space Telescope} (\textit{HST})  with the F555W and F814W wide-band filters. These images were obtained as part of a larger \HST program aimed at studying M81 Group dwarf galaxies (GO-10605; PI: Skillman; \citealt{Weisz2008}) and were reprocessed within the ACS Nearby Galaxy Survey Treasury (ANGST) project in \cite{Dalcanton2009}. In this paper we used their published results of stellar photometry.

The archival far-ultraviolet (FUV) images from the \textit{GALEX} observatory were used for analysing the recent star-formation activity in Holmberg~I. The calibrated data published by \cite{Hunter2010} were downloaded from the LITTLE THINGS archive.

\section{Young stellar population}\label{sec:stars}

\begin{figure}
	\includegraphics[width=\linewidth]{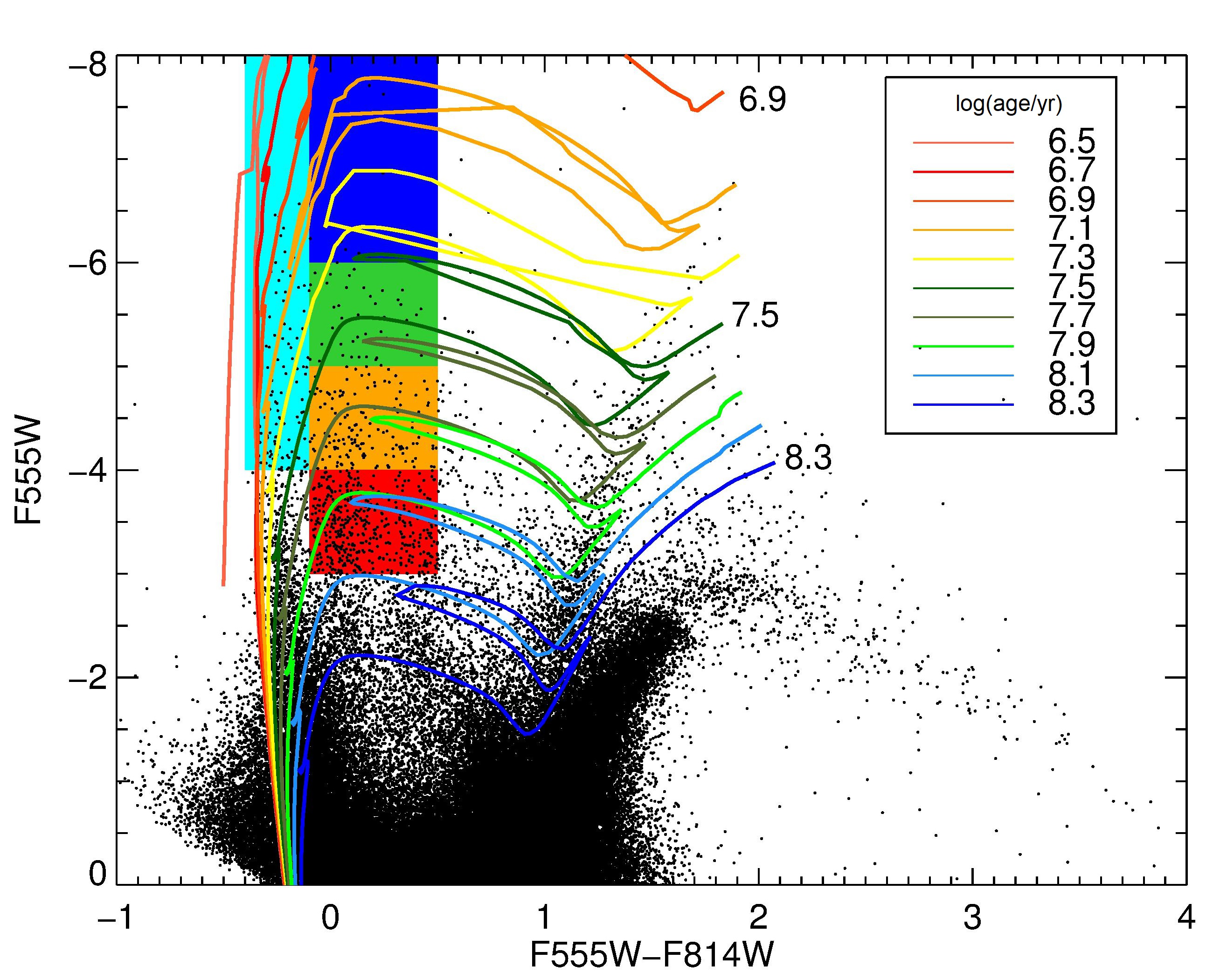}
	
	\includegraphics[width=\linewidth]{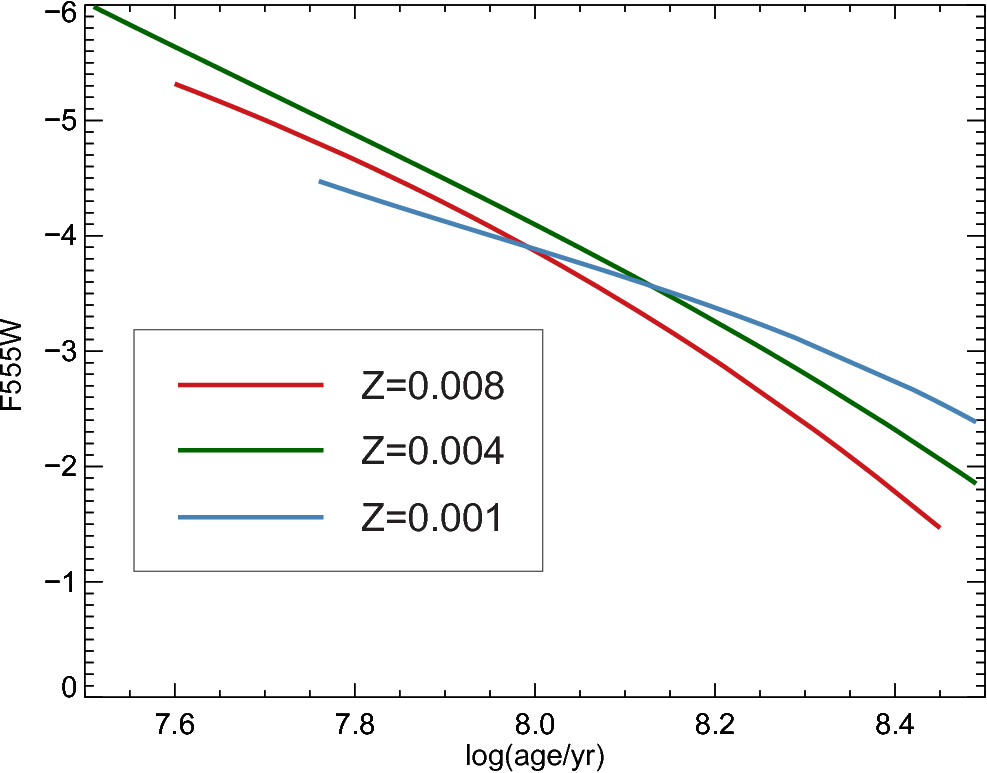}
	
	\caption{Top panel: observed CMD of the galaxy Holmberg~I. The coloured lines correspond to the PARSEC isochrones \citep{bressan12} computed with the CMD~2.9 code for different ages and for the metallicity $Z=0.004$. The coloured rectangles show the areas filled by stars of different ages. Cyan -- OB stars younger than 20 Myr; blue -- BHeB (Blue Helium Burners) stars younger than 30 Myr; green -- BHeB of ages from 30 to 55 Myr;
		orange -- BHeB of ages from 55 -- 115 Myr;
		red -- BHeB of ages from 115 -- 190 Myr.
		Bottom panel: dependence of the F555W magnitude on the age of  BHeB stars for different metallicities. } \label{fig:cmd}
\end{figure}

\begin{figure*}
	\includegraphics[width=0.95\linewidth]{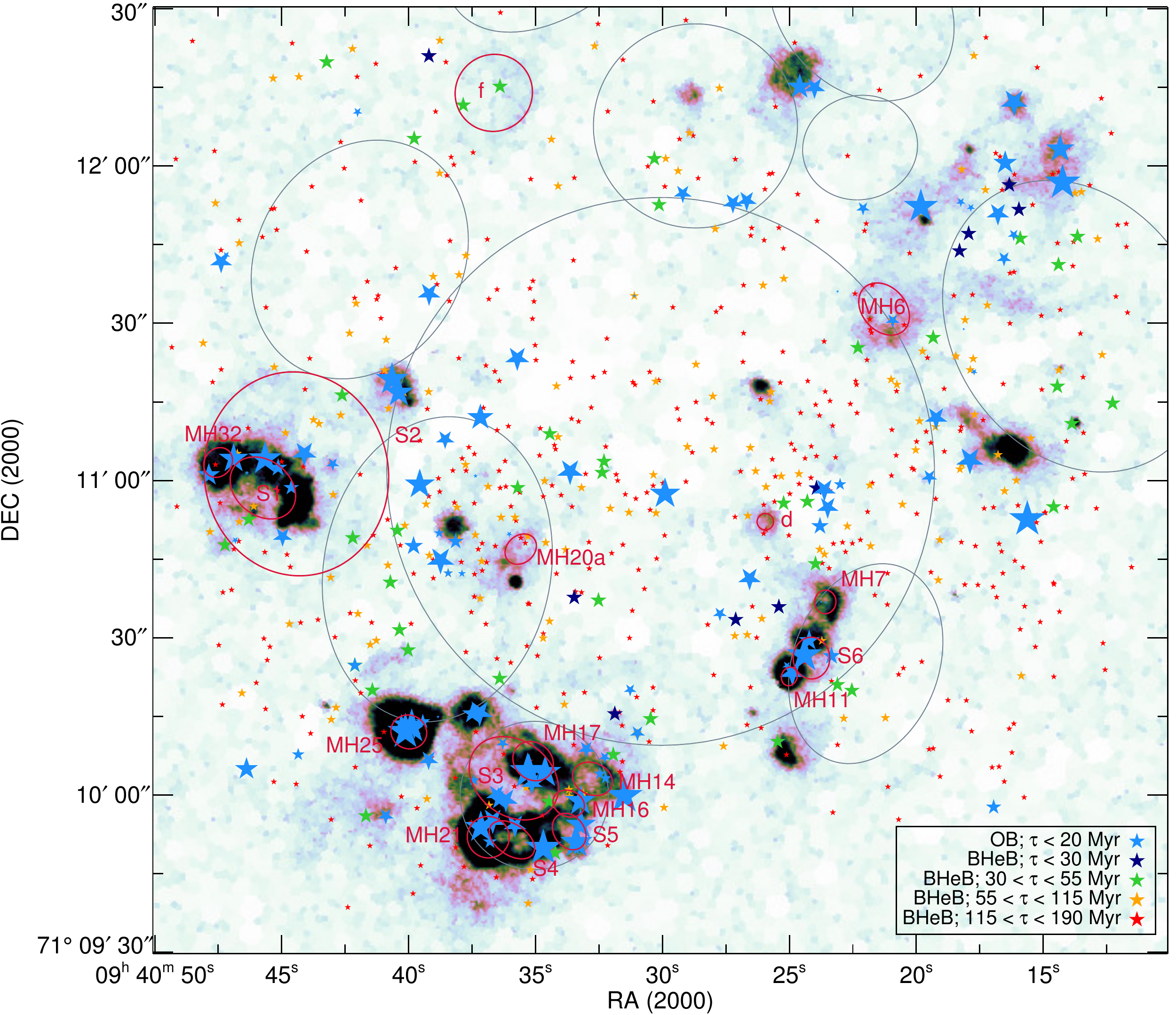}
	
	\caption{Distribution of young stars of different ages shown on the \Ha emission line image of 
Holmberg~I. As well as in Fig.~\ref{fig:cmd}, different colours correspond to different ages: light blue -- OB stars younger than 20 Myr; dark blue -- BHeB stars younger than 30 Myr; green -- BHeB of ages from 30 to 55 Myr;
		orange -- BHeB of ages from 55 -- 115 Myr;
		red -- BHeB of ages from 115 -- 190 Myr. Red and grey ellipses show the location of the identified
  \Ha and \HI shells, respectively (see Section~\ref{sec:shells}).} \label{fig:HoI_stars}
\end{figure*}

The first studies of the young stellar population in Holmberg~I were performed by \cite{Hoessel1984} with
an early CCD imaging. They found that the galaxy has a small number of stars with the masses $\mathrm{M} > 20\mathrm{M}\odot$, and that even the majority of the $16\mathrm{M}\odot$ stars appear to have evolved significantly to the red.

The abilities of \HST images allowed one to study the resolved stellar population and star-formation history in the galaxy. 
 \cite{Weisz2008} presented the spatial density distribution of blue and red stars in Holmberg~I  (see their fig.~9). They found that older red stars are highly
 concentrated in the interior of the central \HI SGS, while the blue stars location correlates with the \HII regions. The authors showed that Holmberg~I had two episodes of the intense star formation -- the ancient one that  occurred several Gyr ago and the current one that started in the last 100 Myr.

We used the available results of the \HST photometry of individual stars in Holmberg~I to identify main-sequence OB stars which could be responsible for ionization of  the observed \HII complexes. The selection of these stars was performed following the procedure described in \cite{Bastian2011}. Analysing the colour--magnitude diagram (CMD), shown in the top panel of Fig.~\ref{fig:cmd}, we identified 121 objects that should be young OB stars, 40 stars among them have $M_{F555W} < −5$ and give the major contribution to  ionization.
Their location is shown in Fig.~\ref{fig:HoI_stars} in light blue; the size of star symbols correlates with their number of ionizing quanta on the logarithmic scale. It is worth noting, however, that since the main-sequence branch is not very clearly resolved in the CMD, some of the detected OB stars might not be OB stars actually.

Many previous studies \cite[see, e.g.,][]{Dohm-Palmer1997, Weisz2008, Bastian2011} used the position of Blue Helium Burner stars (BHeB) in the CMD as a good indicator of their age. To select  BHeB stars of different ages in Holmberg~I, we used the PARSEC isochrones \citep{bressan12,chen14,chen15,tang14} computed with the CMD~2.9 code\footnote{http://stev.oapd.inaf.it/cmd}. As is shown in the bottom panel of Fig.~\ref{fig:cmd}, the F555W luminosity of such stars  almost linearly correlates with their age. We plotted the isochrones for the metallicity $Z=0.004$ and age $\log\mathrm{\tau}=6.7 \div 9.0$ over the CMD shown in the top panel of Fig.~\ref{fig:cmd}. The coloured regions in the CMD represent different stages of stellar evolution. The cyan region is for young main-sequence OB stars (mostly O stars) with absolute F555W magnitude less than $-4$; their age is smaller than 20 Myr. The colours from blue to red show the location of BHeBs of different ages. The blue colour corresponds to the stars younger than 30 Myr, green -- to those between 30 and 55 Myr, orange -- to the stars with ages between 55 and 115 Myr, and the red area contains the stars with ages from 115 to 190 Myr. Note that stars at other evolution stages should also appear in the mentioned areas, but the BHeB stage is noticeably longer, and we can expect most of the detected stars to be at that stage.
The locations of the detected OB and BHeB stars are shown in Fig.~\ref{fig:HoI_stars}.

 As follows from Fig.~\ref{fig:HoI_stars}, the main-sequence O stars that  trace the areas of the ongoing star
formation are indeed located  mainly in the \HII regions. The distribution of BHeBs traces the location of regions
 of the recent star formation and as one can see from Fig.~\ref{fig:HoI_stars}, most BHeBs aged between 30 and 55
 Myr are located inside the central SGS and northwestern supershell. The older BHeBs are distributed more or less
uniformly which agrees with the results of \cite{Bastian2011}  for other galaxies.

\section{Ionized- and neutral-gas kinematics}\label{sec:kinematics}

\subsection{General neutral-gas kinematics}

The previous studies \citep{Ott2001, Oh2011} of the neutral-gas kinematics of Holmberg~I have shown that this galaxy rotates slowly and demonstrates significant radial motions.
The shapes of  rotation curves reported by different authors look similar, yet the maximum velocity and the corresponding radius differ (see fig.~9 in \citealt{Ott2001} and fig. A.10 in \citealt{Oh2011}).
These discrepancies might be explained by a low inclination of Holmberg~I 
 and significant radial motions. All these factors make the analysis difficult and a bit controversial.

For analysing the gas kinematics, we would like to subtract the component corresponding to the regular galaxy rotation
from any data cube we obtained. Such a procedure allows us to highlight the local gas motions caused by a stellar feedback
 or gas inflow/outflow processes. This `derotation' method and its advantages are described in \cite{Egorov2014}.

We tried to model the observed \HI velocity field by tilted rings. Significant discrepancies were found towards the
 northwestern and southern outer parts of the galaxy, where the influence of the ram pressure or tidal interaction are
 probable. Because of that, we found that a simple tilted-ring model is not applicable in the case of Holmberg~I, and the
 inclusion of warp and radial flows is necessary. However, since we are interested mostly in the inner part of the galaxy, the construction of such a detailed model is out of the scope of our paper.

In this paper  for reconstruction
 of the velocity map corresponding to the galaxy rotation, we used the results of a more detailed modelling performed by \cite{Oh2011}. The inclination $i=14^\circ$ and the position angle
$PA=45^\circ$ were adopted. Using this model, we have found that the residual motions in the \HI do not exceed 8~$\kms$ in
 the area surrounded by the SGS. We subtracted the obtained  model of the Holmberg~I circular rotation from all \HI and \Ha
data cubes used further.

\subsection{Searching for \HI and \Ha superbubbles}\label{sec:shells}

\begin{figure*}
	\centering
		\includegraphics[width=0.8\linewidth]{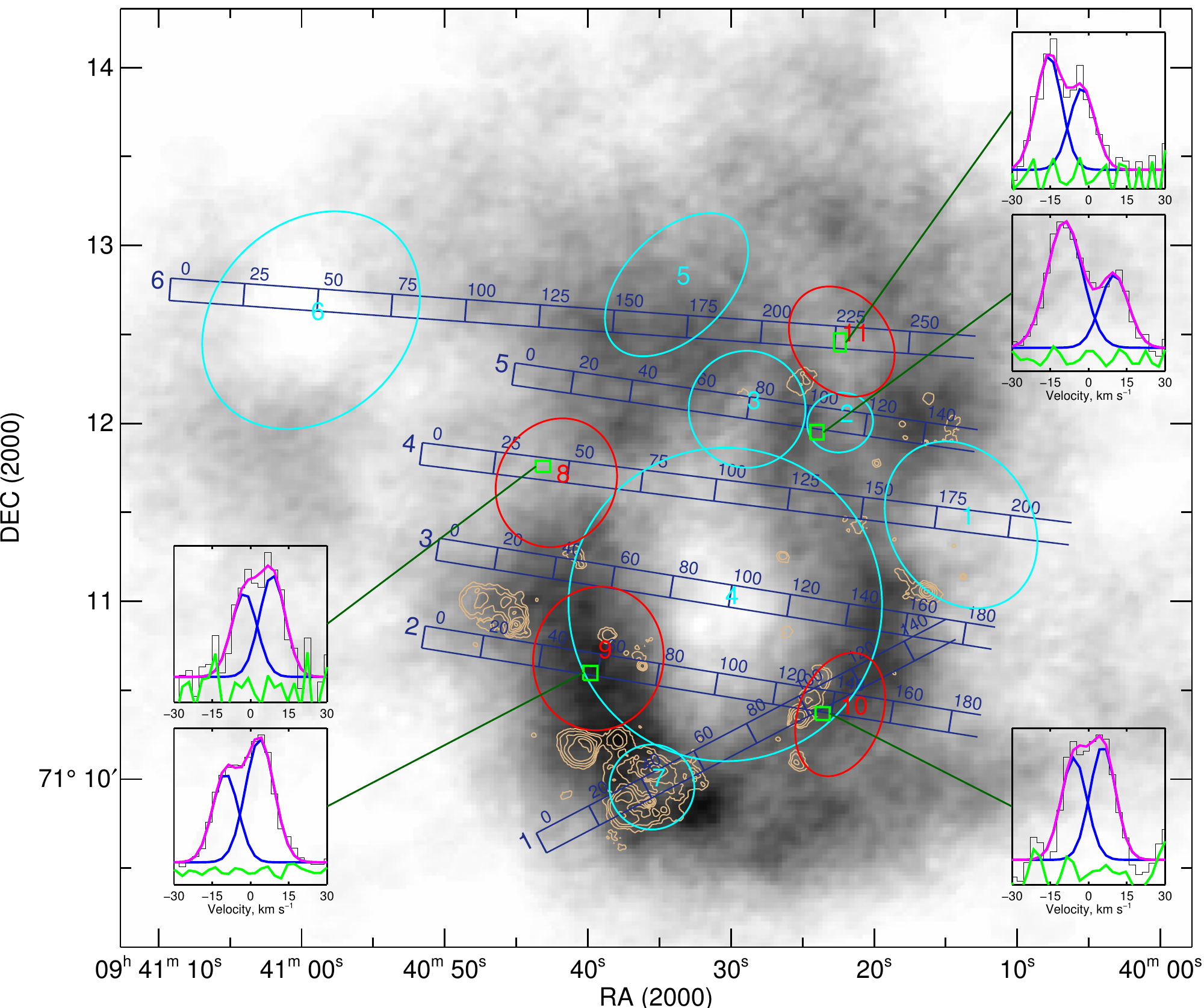}

	\includegraphics[width=\linewidth]{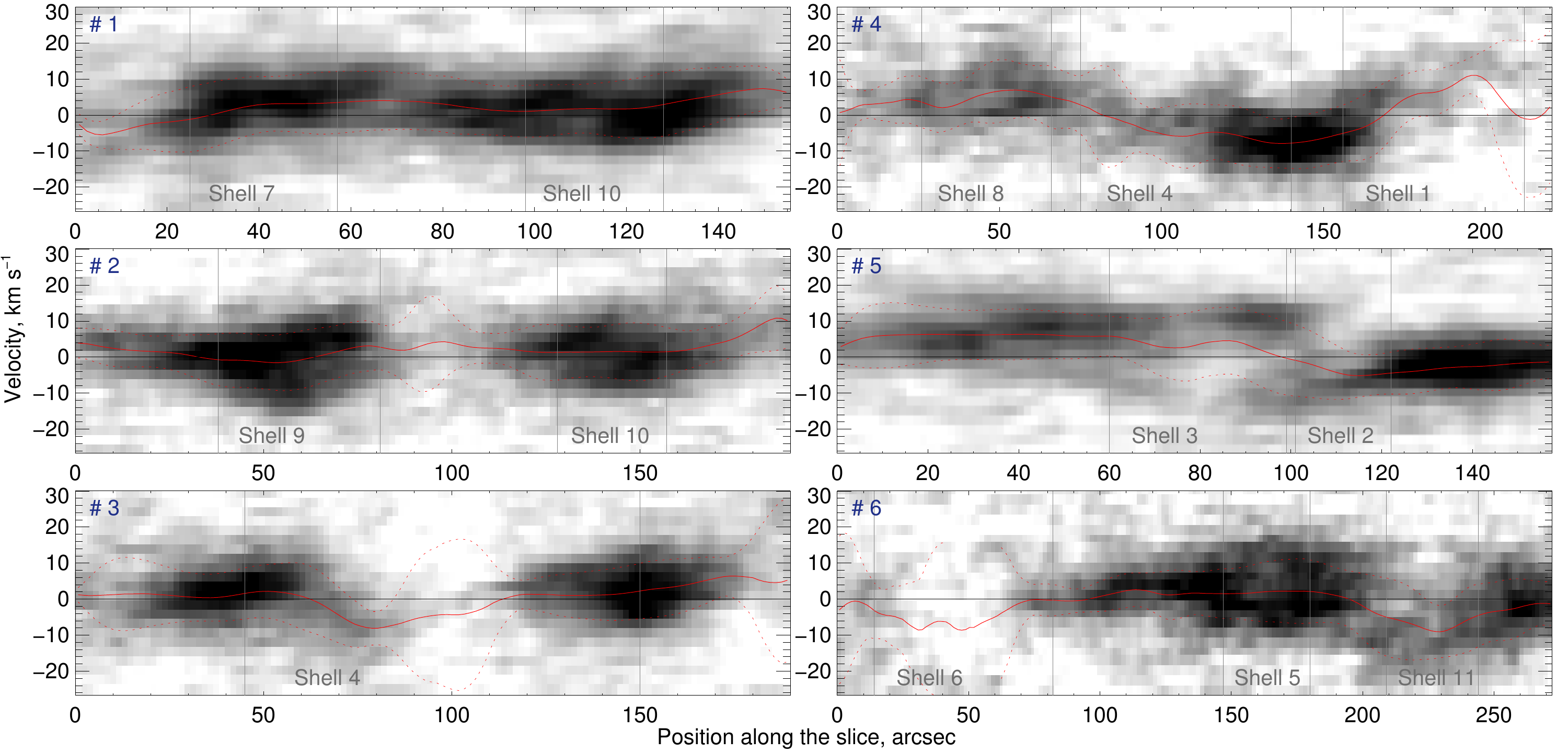}
\caption{Location of the slices for the PV diagrams in the \HI image of Holmberg~I (top), which are shown as examples
 in the bottom panels. The contours show the \Ha flux distribution. Cyan ellipses denote the location of the previously
 known \HI supershells; the red colour shows the borders of the expanding \HI supershells discovered in this work.
 The examples of the \HI line profiles integrated over the area inside green squares and the results of their multicomponent
 fitting are shown: black line corresponds to the observed line profile, red -- to the model, blue -- to individual
 components, and green -- to residuals. Vertical grey lines in the PV diagrams show the edges of supershells. Red solid and dotted lines there trace the intensity-weighted mean velocity and its standard deviation, respectively.}\label{fig:HoI_pv}
\end{figure*}

Up to now, seven \HI supershells in Holmberg~I were known. The central galaxy-sized SGS was studied in detail by
 \cite{Ott2001}; these authors also mentioned the existence of a smaller \HI supershell (\#7 in Fig.~\ref{fig:HoI_pv})
 in the southeastern part of the SGS which is clearly seen on the \HI map. \cite{Bagetakos2011} found six \HI holes and
 expanding supershells (including the central SGS)  by a visual inspection of channel maps and
 further confirmation by analysing the position-velocity (PV) diagrams. \cite{Warren2011} derived a slightly larger size
 of the SGS than other authors using the reprocessed \HI data.
The location of all  the mentioned \HI shells is shown in
Figs.~\ref{fig:HoI_stars},~\ref{fig:HoI_pv}, and their properties are listed in
Table~\ref{tab:shells}. We adopt a diameter of shell \#7 given by \cite{Ott2001} (recalculated for a distance of 3.9 Mpc)  and that for shell \#4 (SGS) given by \cite{Warren2011}. For all the other \HI shells, we used the parameters listed in table 7 of \cite{Bagetakos2011}\footnote{We corrected  a misprint in their table -- radii are listed instead of diameters for Holmberg~I which follows from a comparison with other works and the available \HI data.}.

In this paper we are interested in the distribution and kinematics of the \HI supershells in the context of their relation to
 the young stellar population and to the regions of the current star formation. We analysed the \HI line profile in the vicinity of the star-forming regions in the SGS and constructed a large number of PV diagrams crossing different parts of the SGS. Fig.~\ref{fig:HoI_pv} shows several examples of them. One can see that almost all the \HI supershells mentioned above exhibit approaching and/or receding sides in these PV diagrams (so-called `velocity ellipse'). Based on the analysis of the PV diagrams and two-component \HI line fitting, we confirm previously published values of the expansion velocities for all of these supershells except for \#2. For the latter, we found $V_{exp}$ to be significantly lower than that published by \cite{Bagetakos2011}.
Moreover, since the largest separation between two components of the \HI line profile inside  shell \#2 is observed
 at the border with shell \#3, we propose that shell \#2 might be just a part of larger shell \#3.

\begin{table}
	\caption{Parameters of the identified \HI and \HII shells. The expansion velocities of \HII regions were measured from shifted details in the wings of the \Ha line profiles. The values of $V_\mathrm{exp}$ and the age in brackets were obtained using Eq.~\ref{eq:expansion}. }\label{tab:shells}
\centering
	\begin{tabular}{lccc}
		\hline
		Name &  $D$ (pc) & $V_\mathrm{exp}\ (\kms)$ & Age (Myr) \\
		\hline
		\multicolumn{4}{c}{\HI shells} \\
		\hline
		1  & 1008 & 6  & 50 \\
		2  & 400 & 10 &  12 \\
		3  & 745 & 12 & 19 \\
		4  & 2000 & 6 & 50 \\
		5  & 858 & 12 & 21 \\
		6  & 1371 & 6 & 69 \\
		7  & 528 & 7 & 23\\
		8   & 796 & 5 & 48 \\
		9 & 874 & 7 & 37 \\
		10  & 660 & 6  & 33 \\
		11 & 679 & 6 & 34  \\	
		\hline
		\multicolumn{4}{c}{\HII shells} \\
		\hline
		S1 &  229 & 38  &  1.8 \\		
		S2 &  708 & <22 & >9.7  \\		
		S3 &  310  &  (34)  & (2.7)  \\		
		S4 &  144 & (32) &  (1.3) \\		
		S5 &  126  & (35) & (1.1)  \\
		S6 &   143   & 	27		& 	1.6	\\	
		MH6  & 184  & 20 & 2.8 \\
		MH7  & 82 &  85& 0.3\\
		MH11  & 63 & (20) & (1.0)\\
		MH14  &132 & (29) & (1.3)\\
		MH16   & 106 & (39) & (0.8)\\
		MH17   & 145 & (24)& (1.8)\\
		MH20a   & 111 & 25(17) & 1.3 (2.0) \\
		MH21   & 152 & (29) & (1.6)\\
		MH25  &127 & (20) & (2.0)	 \\
		MH32 &106 & 31 (27) & 1.0 (1.2)\\
		d & 60 & 74& 0.2\\
		f   & 282 & <22 & >3.8 \\
		\hline
	\end{tabular}
\end{table}

As  follows from the demonstrated PV diagrams, clear signs of `velocity ellipses' are  also seen outside the previously known \HI supershells.
Based on that, we detect four new expanding \HI supershells denoted as \#8 -- \#11. Their borders were derived from a visual inspection of the PV diagrams crossing the shell in different directions. Borders of  shells \#8 and \#11 are seen on the \HI integral map, while shells \#9 and \#10 do not appear on it. The last two are implicitly seen on the channel maps. The expansion velocities of these new detected \HI shells were derived with the \HI line fitting (see examples in Fig.~\ref{fig:HoI_pv}); they correspond to $V_{exp} = 5-7 \kms$.

We estimated the kinematic ages of all  the \HI supershells in terms of the \citet{Weaver1977}
model by the  relation
$
t=0.3 D/V_{\rm exp},
$
where $D$ is the diameter in pc,  $V_{\rm exp}$ -- the expansion velocity  in $\kms$, and $t$ --
the age in Myr. All the values  are listed in Table~\ref{tab:shells}.

Thus, the central SGS exhibits a number of local expanding \HI supershells of a smaller size on its rim; all \HI supershells
are plotted in Fig.~\ref{fig:HoI_stars}. 

 By analysing the \Ha image we identified 18 shell-like
 emission structures. They  mostly correspond to \HII regions, while 6 of them are faint shells between several \HII regions named  S1 -- S6.
We show all the identified ionized shells in Fig.~\ref{fig:HoI_stars} by the red ellipses.

\begin{figure}
	\includegraphics[width=\linewidth]{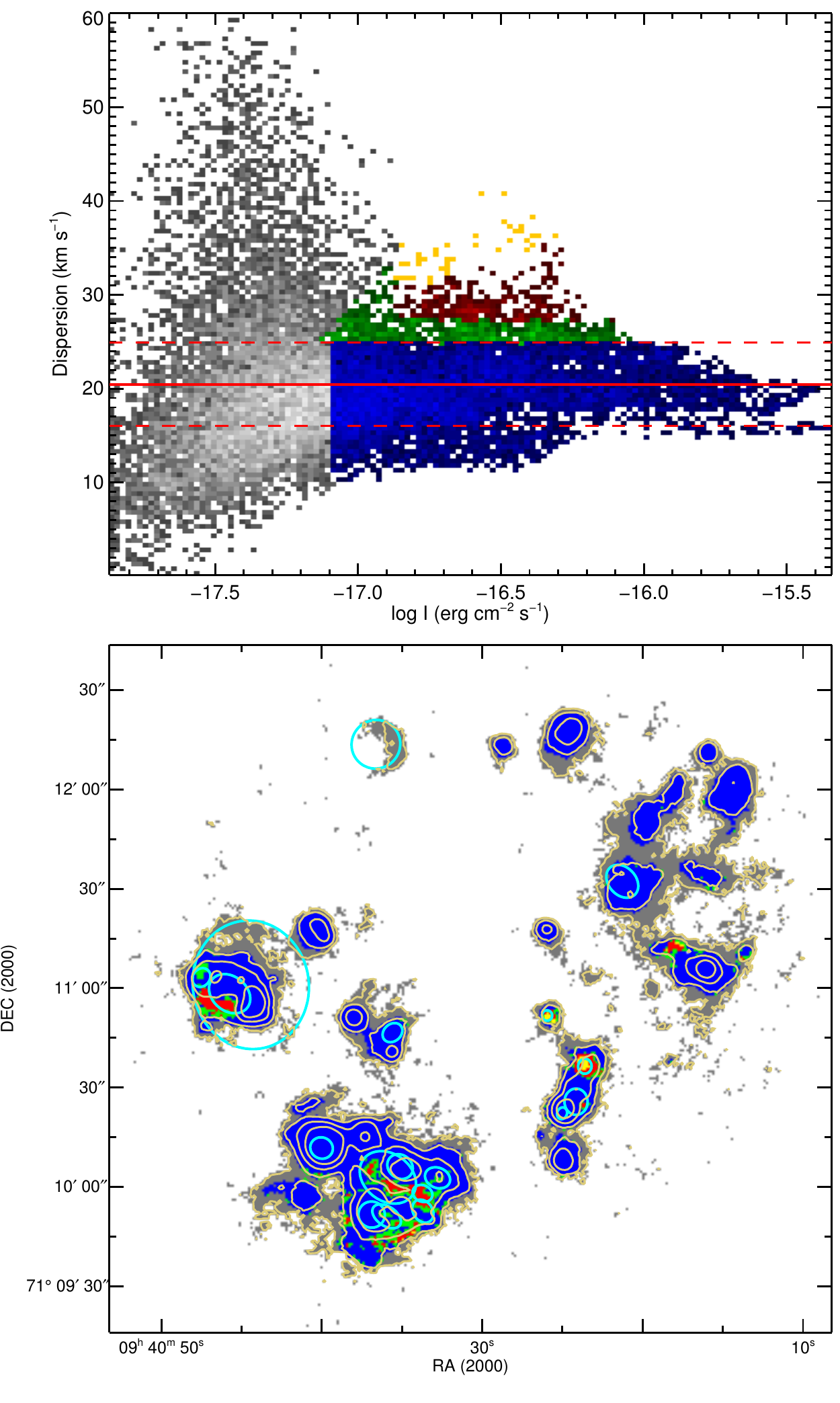}
	\caption{\Ha intensity -- velocity dispersion diagram for Holmberg I (top). The red solid horizontal line marks the intensity-weighted mean velocity dispersion of ionized gas in the galaxy and the dashed lines -- its standard deviation.
		The bottom plot shows the location of regions marked by different colours in the $I-\sigma$ diagram. The contours correspond to the lines of equal \Ha intensity. Cyan ellipses show the borders of shell-like structures appearing in the \Ha image (see Fig.~\ref{fig:HoI_stars}).}\label{fig:isigma}
\end{figure}

It has been shown in many studies \cite[][see also references therein]{MunozTunon1996, Moiseev2012, Vasiliev2015a} that the intensity – velocity dispersion $(I - \sigma)$ diagrams for ionized  gas can be successfully used to identify areas where the increased velocity dispersion is caused by expanding shells. We fitted a single-component
Voigt profile to the H$\alpha$ line profile in each pixel obtained from the FPI data cube and constructed the $I - \sigma$ diagram shown in Fig.~\ref{fig:isigma}. The red line in the
diagram marks the intensity-weighted average velocity dispersion of ionized gas in the galaxy -- $\sigma_m = 20.5 \pm 4.5 \kms$. The identical colours in the diagram (top panel) and map (bottom panel) are used to highlight the characteristic areas. Thus, the horizontal strip with a relatively low velocity dispersion and high surface brightness marked in blue corresponds to bright \HII regions and includes 50 per cent of galaxy's H$\alpha$ flux. The areas of the increased dispersion, which are most likely associated with spectroscopically unresolved expanding shells, are shown in green, whereas the red colour corresponds to shell-like structures with the expected clear separation of the line profile components. As was shown in \cite{Moiseev2012}, the diagonal yellow lane could correspond to the presence of compact objects, which highly influence the ISM (like WR, LBV stars, supernova remnants). The remaining areas of low surface brightness corresponding to the diffuse gas and to the edges of \HII regions are shown in grey.

As it follows from a comparison of $I-\sigma$ with the location of the identified ionized shells, a half of them are clearly
 expanding. Additionally, two areas not showing the corresponding shell-like nebula appear in the $I-\sigma$ diagram as
expanding superbubbles. One of them is located at the southern edge of the complex of ionized shells and might
 represent faint expanding superbubbles without resolved morphology, while another (in the western complex of ionized
 nebulae in the galaxy) coincides with the relatively bright \HII region -- MH4.

We tried to estimate the expansion velocities of the identified ionized superbubbles applying a multicomponent Voigt fitting to the observed \Ha line profile towards the inner part of each of them (see the examples of the line profiles in Figs.~\ref{fig:profs_all} -- \ref{fig:profs_nw}). Unfortunately, with the current spectral and spatial resolution only two regions -- MH7 and d -- show clear separated components in the line profiles. These objects are marked by the yellow colour in the $I - \sigma$ diagram and identified as probably unique energy sources. In Section~\ref{sec:hiiregs} we will show
that they are good supernova remnant candidates.

Most of the remaining shells reveal the enhanced velocity dispersion inside them, and the blue- or redshifted faint components are observed in the wings of the \Ha line profile in some of them. For those objects, we 
used a separation between the components to estimate their expansion velocity.

A small angular size of some of the observed ionized shells results in smoothing the line profiles (even for relatively fast-expanding regions) and makes impossible the direct detection of signs of their expansion from the line profile asymmetry at the given spatial resolution.
At the same time, for several \HII regions we observe the maximum line-of-sight velocity dispersion $\sigma$ in its centre, while it slowly decreases towards the edge.
As was shown in \cite{Guerrero1998}, in such a case, the expansion velocity correlates well with the gradient of  the line-of-sight velocity dispersion, and 
as we will show in Smirnov-Pinchukov \& Egorov (in prep.), it can be estimated as
\begin{equation}\label{eq:expansion}
V_{exp}=\alpha\times\sqrt{\sigma_{HII}^2-\sigma_{0}^2},
\end{equation}
where $\sigma_{HII}$ is the velocity dispersion inside the nebula, $\sigma_{0}$ -- in its unperturbed vicinity. The coefficient $\alpha$  depends on the spectral resolution and on $\sigma_{0}$; in the case of our data  $\alpha=1.59\pm 0.04$. The results are listed in Table~\ref{tab:shells}.

Using the estimated expansion velocities, we measured a kinematic age of each ionized superbubble using the same approach as for \HI supershells. The obtained values do not exceed 3 Myr for all the regions except S2 and f. These older faint supershells will be discussed in Sections~\ref{sec:morphology} and \ref{sec:hiiregs}, respectively.

 It is worth noting that the estimated values of the kinematic age  correspond to the expansion velocities measured by faint details in the wings of the \Ha line profiles for almost all regions. Since these faint components can be of origin not related to an expanding superbubble (e.g., a local gas inflow/outflow), the calculated values of $V_{exp}$ can be significantly overestimated which leads to possible underestimation of the age.

\section{Complexes of ionized gas and ionization balance}\label{sec:morphology}

\begin{figure*}
	\centering
	\includegraphics[width=0.97\linewidth]{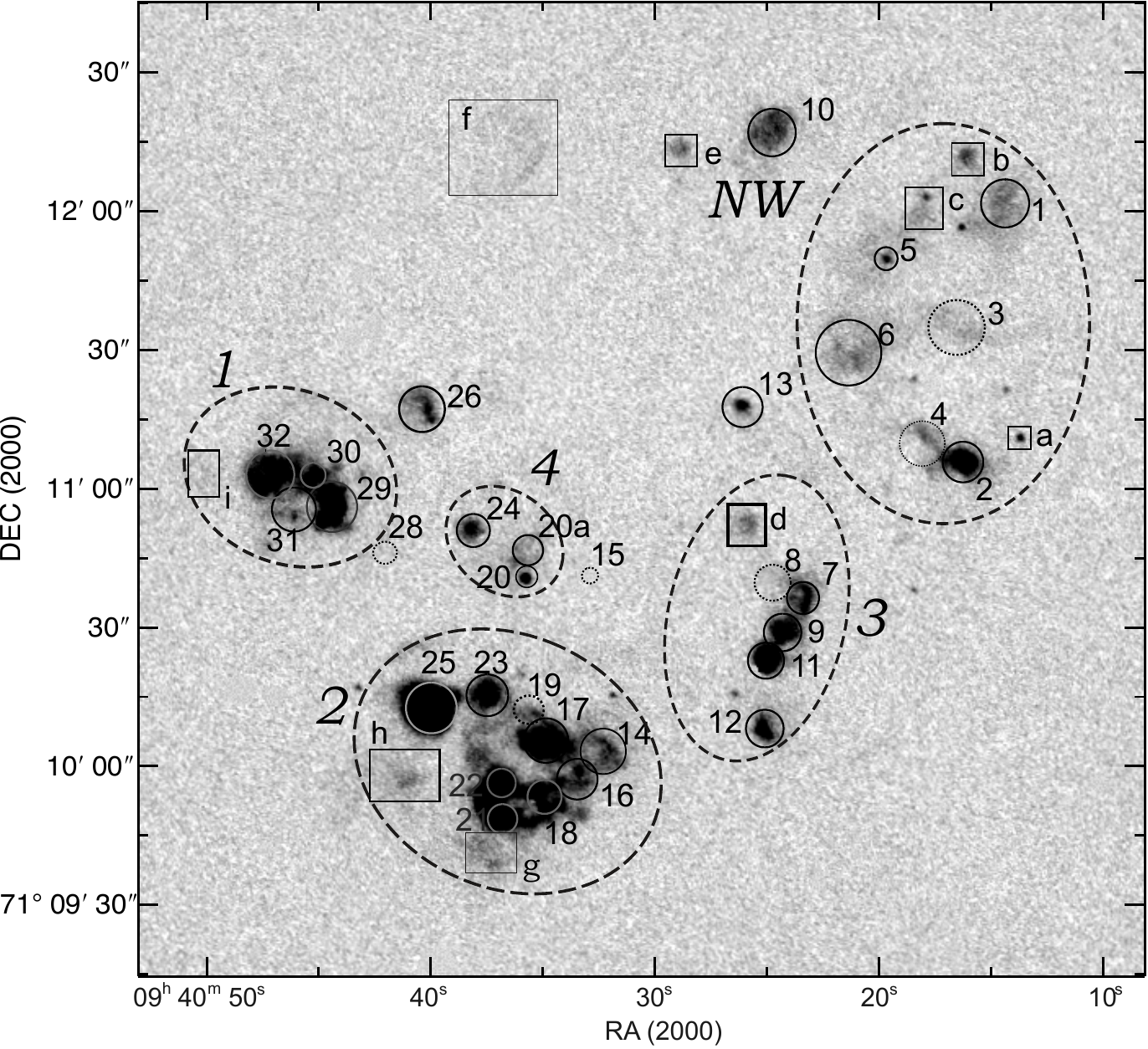}
	\caption{\Ha image of Holmberg~I. The identification of the \HII regions according to the list from \citet{MH94} is shown along with their names.
		Dotted circles show the objects from their list which appear not to be \HII regions. The new faint \HII regions identified in this paper are shown by squares and
		named by letters. Dashed ellipses denote the borders of the \HII complexes discussed in Section~\ref{sec:morphology}.}
	\label{fig:hoi_loc}
\end{figure*}

\begin{table*}
	\centering
	\caption{Parameters of the \HII regions of Holmberg~I}\label{tab:HIIregs}
		\scriptsize	
		\begin{tabular}{lcccccccccr}
		\hline

		Complex  &  LyC$\star$ & LyC \HII  &  \HII reg. &   Size  &  $v_{\mbox{los}}$  &  $\sigma_v$  & $\mathrm{F(H\alpha)}$ & $\mathrm{EW_{H\alpha}}$ &  Age    		 & morphology \\
		&  $10^{50}$  & $10^{50}$ &                       & (pc)  	  &   $(\kms)$  &     $(\kms)$ 		      	 & $10^{-15}$ erg s$^{-1}$ cm$^{-2}$  	&  \AA                                        &  (Myr)  		 &            \\
		\hline
		total    &     16.0    &    4.5    &     &    								  &  			&        			 &  	    &         		 &  		&	  \\
		\hline
		&        	   &           &  29 &  	$165$     &       $ -12.2 \pm 2.5 $  &  $17.3 \pm 1.7$  	 &  25.7  & $ 629$ & 4.6 & multiple   \\
	  &  	   &       &  30 &     	90	 		&    	$ -8.7 \pm 0.9 $ 	&    $21.7\pm1.2$  &   4.2   & $ 463$ & 5.0 &	compact \\
1		&     1.52          &    0.86       &  31 &     			90				   &       $-18.3 \pm 3.0$ &   $ 29.1 \pm 0.8$   	 &   1.2    & $119$ & 7.2 &   diffuse \\
		&             &           &  32 &     		220					  &       $-6.4 \pm 4.8$  &  	$ 24.4 \pm 3.6$   	 &   16.7    & $ 231^*$ & $6.1^*$ &     shell \\
		&             &           &  i  &     		$\sim 170$					  &       $-$  &  	$ - $   	 &   0.6    & $190$ & $6.4$ &     diffuse \\
		\hline
		&        	   &           &  14 &  	$230$     &       $ 0.0 \pm 3.2 $  &  $24.4 \pm 2.2$  	 &  10.0  & $ 290$ & 5.5 & shell   \\
		&        	   &           &  16 &  	$160$     &       $ 4.1 \pm 3.4 $  &  $28.9 \pm 2.0$  	 & 6.7  & $ 289$ & 5.5 & multiple   \\
		&        	   &           &  17 &  	$160$     &       $ -2.2 \pm 1.4 $  &  $22.0 \pm 1.7$  	 &  14.7  & $ 325$ & 5.4 & comp.+shell   \\
		&             &           &  18 &     	$60$	 &    $2.3 \pm 3.3$ &   $ 22.3 \pm 0.6$    	 &
		6.8   & $ 241$ &    6.0 & compact\\
       &           &         &  19 &     $85$	  &   $ -5.2 \pm 1.7$    &    $21.3 \pm 0.9$    &
		1.3 & 648 & 4.6 & diffuse \\
2		&      7.06       &   2.49        &  21 &   	$220$	  &      $-1.4 \pm 3.4$&   $ 25.1\pm 1.2$      &    20.8   & $ 769$ &  4.4 & shell \\
	&             &           &  22 &    $90$	  &   $-4.1 \pm 2.7$  &  	$ 23.1 \pm 1.1$  	 &
	8.6    & $ 331$ &   5.3 &   compact \\
		&             &           &  23 &    $110$	  &       $0.7 \pm 4.1$  &  	$ 20.5 \pm 1.3$   	 &    5.6    & $826$ &  4.3 &  compact \\
		&             &           &  25 &  	$200$	  &   $-0.6 \pm 1.8$  &   $ 19.9 \pm 0.8$   	 &   50.0    & $ 980$ & 4.1 &     shell \\
		&             &           &  g &  	$150$	  &   $-4.8 \pm 2.2$  &   $ 22.3 \pm 1.6$   	 &   2.2    & $ 816$ & 4.3 &     diffuse \\
		&             &           &  h &  	$115$	  &   $-3.0 \pm 2.6$  &   $ 19.7 \pm 1.4$   	 &   1.3    & $ 291$ & 5.5 &     diffuse \\
		\hline
		&             &           &  7  &      	$150$	  &    $-0.1 \pm 2.8$ &  	$ 28.8 \pm 4.1$   	 &    4.7    & $ 299$ &  5.5 &  shell \\
		&             &           &  8  & 130    &       $-2.3 \pm 2.4$  &   $ 20.8 \pm 3.4$  	 &   0.8    & $ 78$ & 7.8     & diffuse \\
		3        &     1.2  &   0.46    &  9  &  $140$	  &  $-7.6 \pm 4.0$  &   $ 20.9 \pm 1.1$   	 &   6.0    & $341$ &  5.3 &   compact \\
		&             &           &  11 &   $120$  &  $-15.5 \pm 3.0$  &   $ 20.7  \pm 1.2$   	 &   10.2   & $ 422$ & 5.0 &   shell \\
		&             &           &  12 &  $140$	  &   $-5.8 \pm 4.0$ &   $ 20.6 \pm 0.6$    	 &   4.2    & $1081$ & 3.8 &     compact \\
		&             &           &  d &  $130$	  &   $-8.6 \pm 4.2$ &   $ 22.8 \pm 5.5$    	 &   1.4    & $93$ & 7.5 &   shell \\
		\hline
		&             &           &  20 &   $60$	  &    $ -14.0 \pm 3.6$ &   $ 18.2 \pm 0.8 $   	 &   	   1.3    & $246$ &  5.9 &     compact \\
		4		 &    0.33        &    0.12       &  20a &   $185$	  &      $-8.0 \pm 1.9$  &   $ 18.5 \pm 1.9 $   	 &   	   2.6    & $70$ &   8.1 &   shell \\
		&             &           &  24 &    130	  &    $-16.1 \pm 3.0$ &   $ 14.6 \pm 0.9 $    &    2.8    & $ 127$ & 7.1 &   compact \\
		\hline
		&             &           &  1  &      	$245$  &       $-11.0 \pm 3.9$  &   $ 18.0 \pm 2.9$    	 &    4.6     &  237  & 6.0  		 & diffuse \\
		&             &           &  2  &    $95$	  &       $-3.5 \pm 3.4$  &   $ 21.0 \pm 0.5$   	 &    9.6    & $657$ & 4.6 &   compact \\
		       &      &   &  3  &     $110$	  &  $-13.5 \pm 1.8$  &   $ 18.7 \pm 1.7$    	 &    0.6     &   86 & 7.6   		 & diffuse \\
		&             &           &  4  &      	110	  &   $-0.8 \pm 1.7$  & 	$ 26.2 \pm 2.3$ 	 &   1.2    & $86$ & 7.6 &    diffuse \\
		NW       &     1.86  & 0.6  &  5  &      60  &    $-15.3 \pm 1.4$  & 	$ 16.6 \pm 0.4$  	 &    0.6    & $526$ &  4.8 &  compact \\
		 &             &           &  6  &   290	  &  $-15.4 \pm 2.5$ &  	$ 17.7 \pm 1.9$  	 &   5.7    & $94$ & 7.6&     shell \\
		&             &           &  a  &      	40	  &   $-1.2 \pm 2.7$  & 	$ 23.9 \pm 0.7$ 	 &   0.4    & $196$ & 6.4 &    compact \\
		&             &           &  b  &      	115	  &   $-9.7 \pm 4.6$  & 	$ 17.2 \pm 1.5$ 	 &   1.5    & $1690$ & 2.9 &    compact \\
		&             &           &  c  &      	205	  &   $-18.8 \pm 4.9$  & 	$ 16.3 \pm 3.0$ 	 &   2.0    & $99$ & 7.4 &    multiple \\
		\hline
		&    0.30         &     0.11      &  10 &     	235 	  &  $-21.3 \pm 1.6$  &  	$ 14.5 \pm 1.3$   	 &   7.4    & $ 1610 $ & 2.9  &   diffuse \\
			 &   0          &   0.03        &  13 &     	75  	  &       $-10.1 \pm 2.3$  &   $ 17.8 \pm 1.3$   	 &   1.2    & $165$ & 6.6 &  compact \\
		separate &     0.83        &   0.06        &  26 &     	200	  &    $-6.3 \pm 2.1$ &   $14.1  \pm 1.2$     &   4.6    & $ 147$ & 6.1  &  multiple \\
		&      0.036       &   0.016        &  e &  		125	  	  &  $4.8 \pm 2.6$    &   $13.2 \pm 2.0$                  &    1.1      &   263       & 5.7   &	compact	 \\
		&     0        &    0.03       &  f &   		310		  &   $-0.6 \pm 4.5$          &      $11.3 \pm 5.6$               &  1.7        &   225             &  6.1	& shell	 \\
		\hline
\multicolumn{11}{l}{$^*$ EW(\Ha) the value is highly contaminated by the presence of a background galaxy. Hence, the age can be overestimated.}
	\end{tabular}
\end{table*}

\begin{figure*}
	\includegraphics[width=0.95\linewidth]{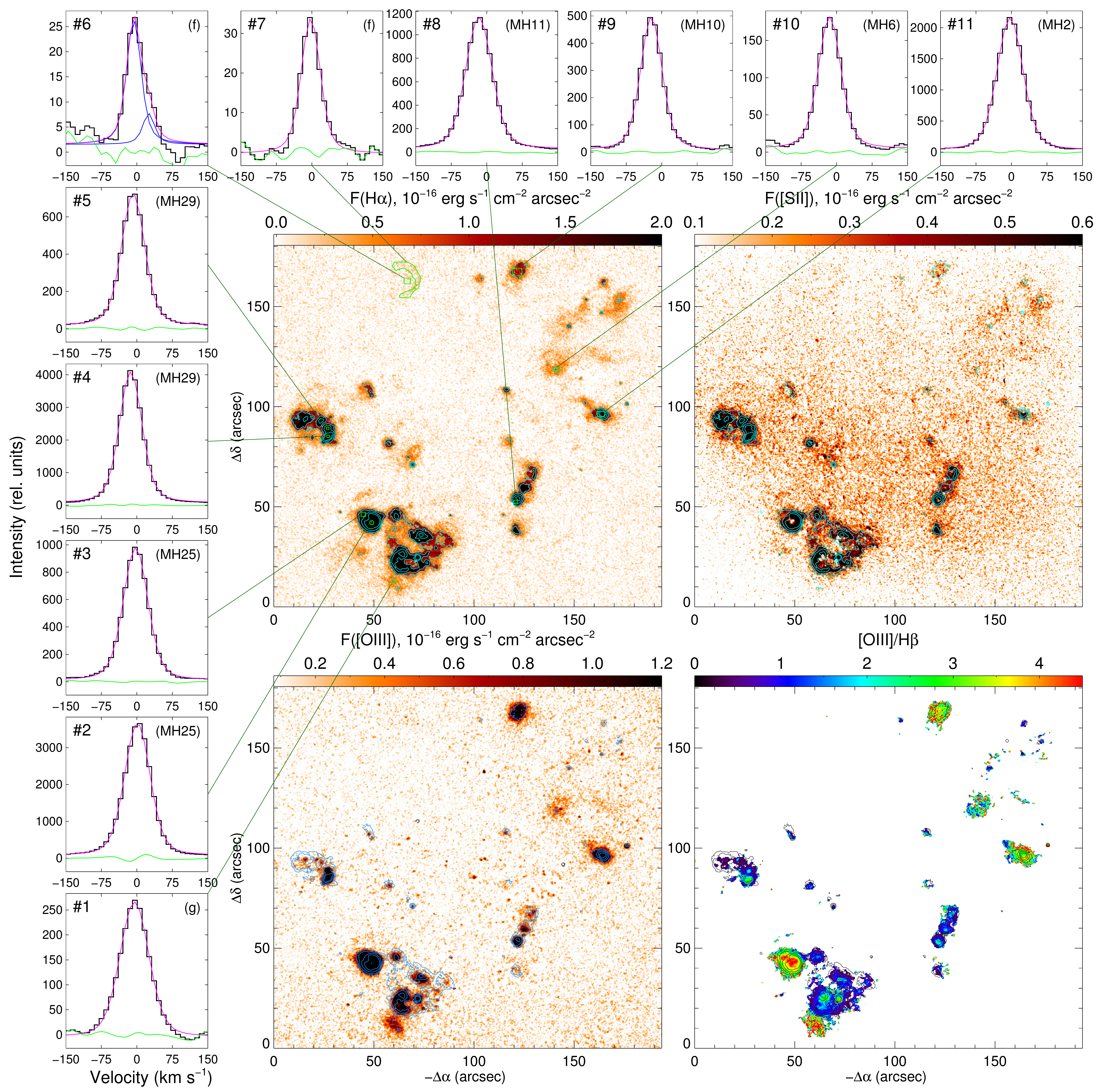}
	\caption{Images of Holmberg~I in the H$\alpha$, [S~\textsc{ii}], \OIII lines and a map of the \OIIIHb\, line ratio (constructed as \OIII/\Ha\, multiplied by 2.86). The contours on each panel correspond to the \Ha intensities of $(0.75, 1.5, 3, 4.5)\times10^{-16}\ \mathrm{erg\ s^{-1}\ cm^{-2}\ arcsec^{-2}}$.   Examples of the \Ha line profile averaged over the areas denoted by the green borders (in the \Ha image) are shown. The observed line profiles are shown by the black line, the results of its decomposition onto the individual Voigt components -- by the blue line, the sum of components -- by the magenta line, and the residuals -- by the green line. The names of the corresponding \HII region are given in parentheses. }\label{fig:profs_all}
\end{figure*}

The ionized gas in Holmberg~I is distributed along the walls of the central \HI SGS revealing the brightest \HII regions in its southeastern part and relatively faint regions in the area between this SGS and another one in the west. Our estimate of the galaxy's \Ha flux corrected for reddening $F(H\alpha) = (3.60\pm0.04)\times10^{-13}$~erg~s$^{-1}$~cm$^{-2}$ is in good agreement with the previous measurements made by \cite{Karach2007}.

Our deep \Ha images, as well as FPI data, allowed us to identify several new faint \HII regions and diffuse structures
 that were never observed before in Holmberg~I. It is clearly seen from our data that the individual \HII regions in the
 galaxy are bound together by faint filament-like structures with a surface brightness of $0.2-1.0\times10^{-16} \mathrm{erg\ s^{-1} cm^{-2} arcsec^{-2}}$ in the \Ha line.

Based on these newly discovered faint structures joining bright nebulae, we select five unified extended complexes  
in Holmberg~I. Further, we will show that they indeed represent physically connected complexes of the ongoing star
 formation. The location of these \HII complexes (\#1, 2, 3, 4, and NW), as well as the names of  bright nebulae
(according to the list of \citealt{MH94}) inside them, are shown in Fig.~\ref{fig:hoi_loc}.  The faint newly
identified \HII regions are named by letters.

Almost all the \HII regions, except for MH10, 13,  26, e, and f are part of these selected complexes. Note, however,
 that a diffuse emission is observed between complex \#1 and the \HII region MH26. Similarly, a faint bridge in \Ha is also seen between the complex NW and the \HII regions MH10, 13, and e.
Despite that, we do not include them in the mentioned complexes because of a large relative distance and
 different environment in \HI (e.g., the region MH10 is located in the place where 3 \HI supershells collide, and is
 separated from the complex NW by \HI shell \#2).

\begin{figure*}
	\includegraphics[width=0.95\linewidth]{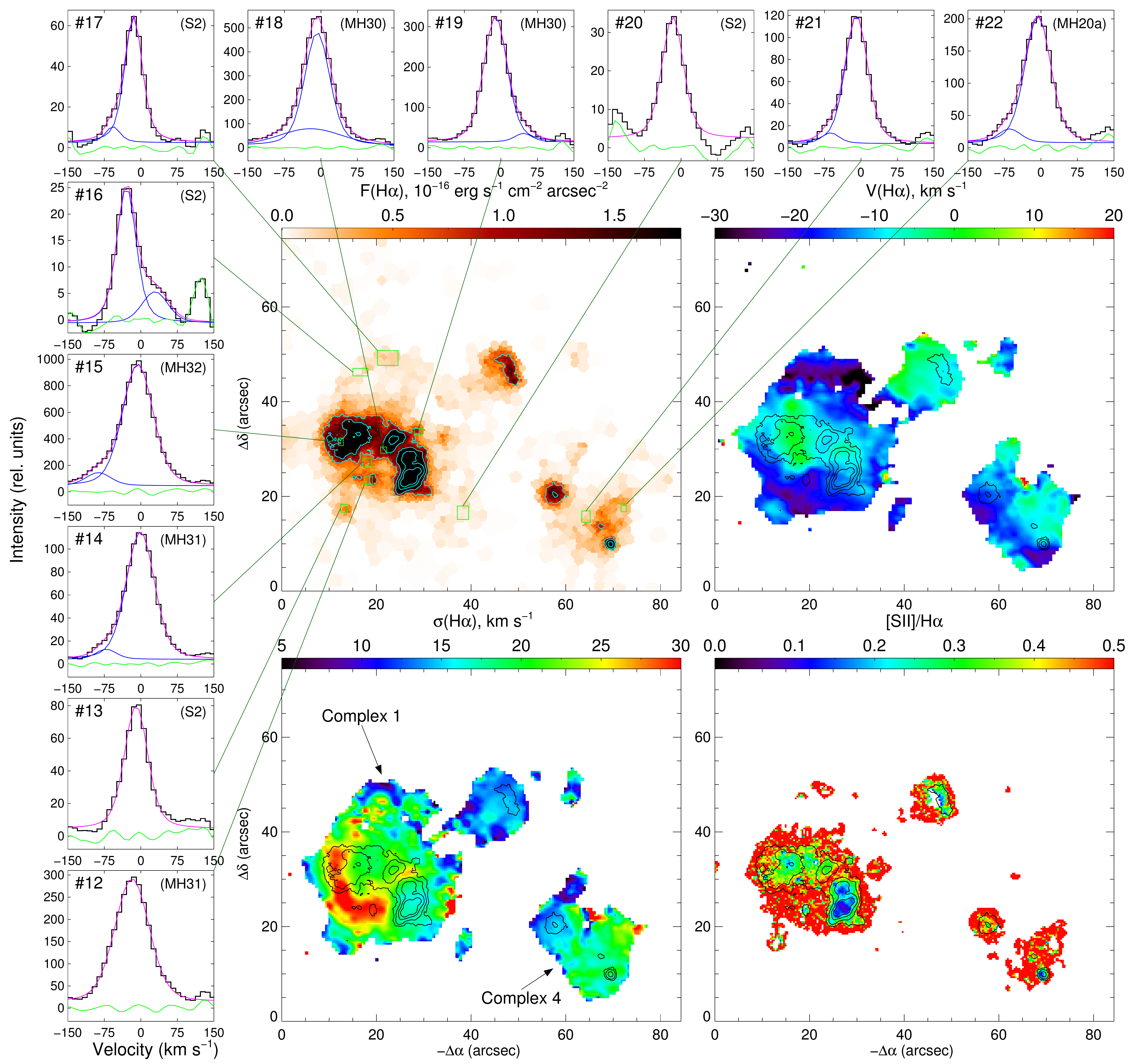}
	\caption{Complexes \#1 and \#4: their image in the \Ha line (top-left), line-of-sight velocity (top-right) and velocity dispersion  (bottom-left) maps, and the distribution of the \SIIHa\ ratio (bottom-right). The contours on each panel correspond to \Ha intensities of $(0.75, 1.5, 3, 4.5)\times10^{-16}\ \mathrm{erg\ s^{-1}\ cm^{-2}\ arcsec^{-2}}$.   The examples of the \Ha line profile averaged over the areas denoted by the green borders (in the \Ha image) are shown. The observed line profiles are shown with the black line, the results of its decomposition onto the individual Voigt components -- with the blue line, the sum of components -- with the magenta line, and the residuals -- with the green line. The names of the corresponding \HII region or shell (if any) are given in parentheses. }\label{fig:profs_c1_4}
\end{figure*}

\begin{figure*}
	\includegraphics[width=0.95\linewidth]{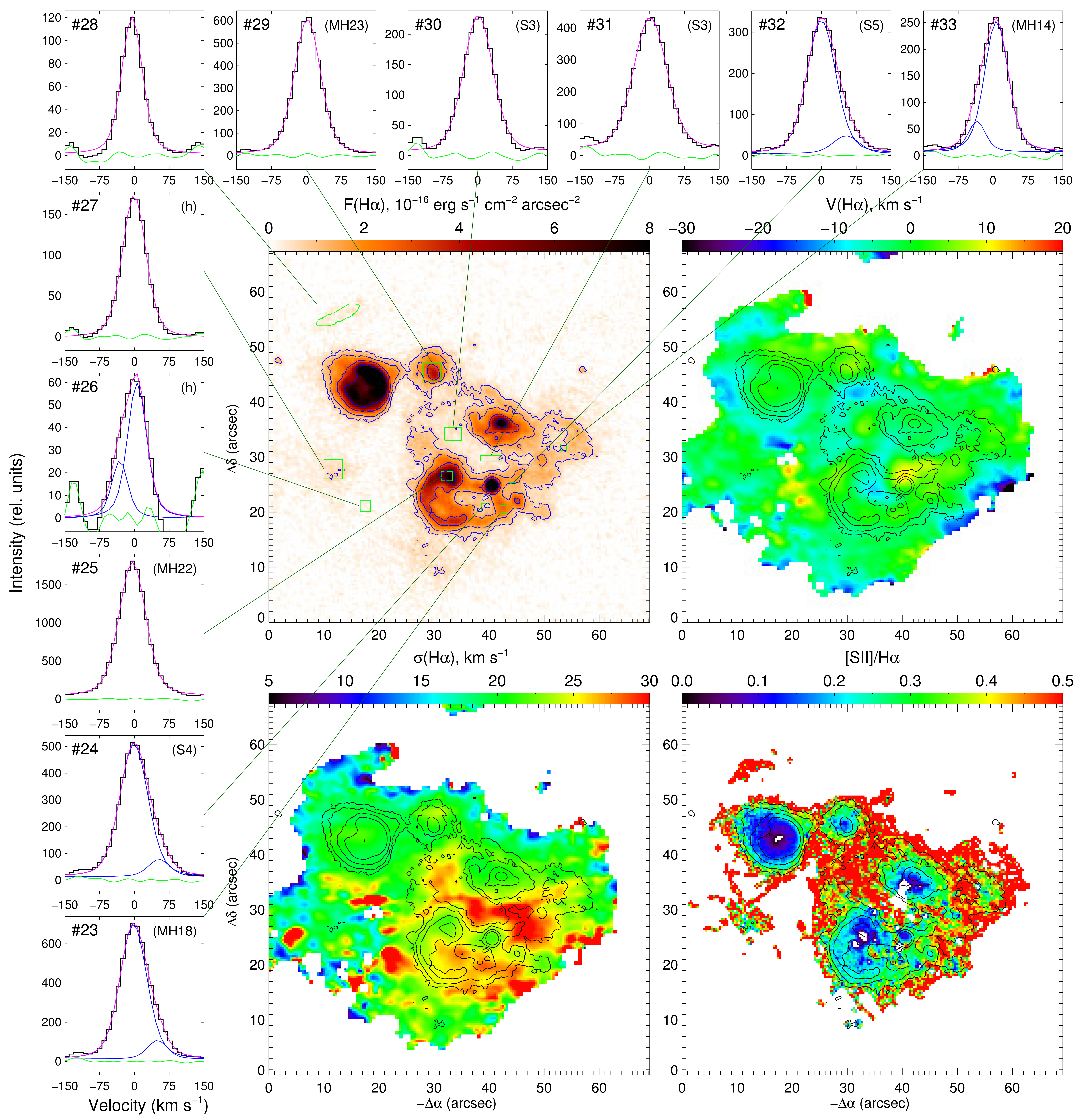}
	\caption{Same as in Fig.~\ref{fig:profs_c1_4} but for complex~\#2.}\label{fig:profs_c2}
\end{figure*}

\begin{figure*}
	\includegraphics[width=0.95\linewidth]{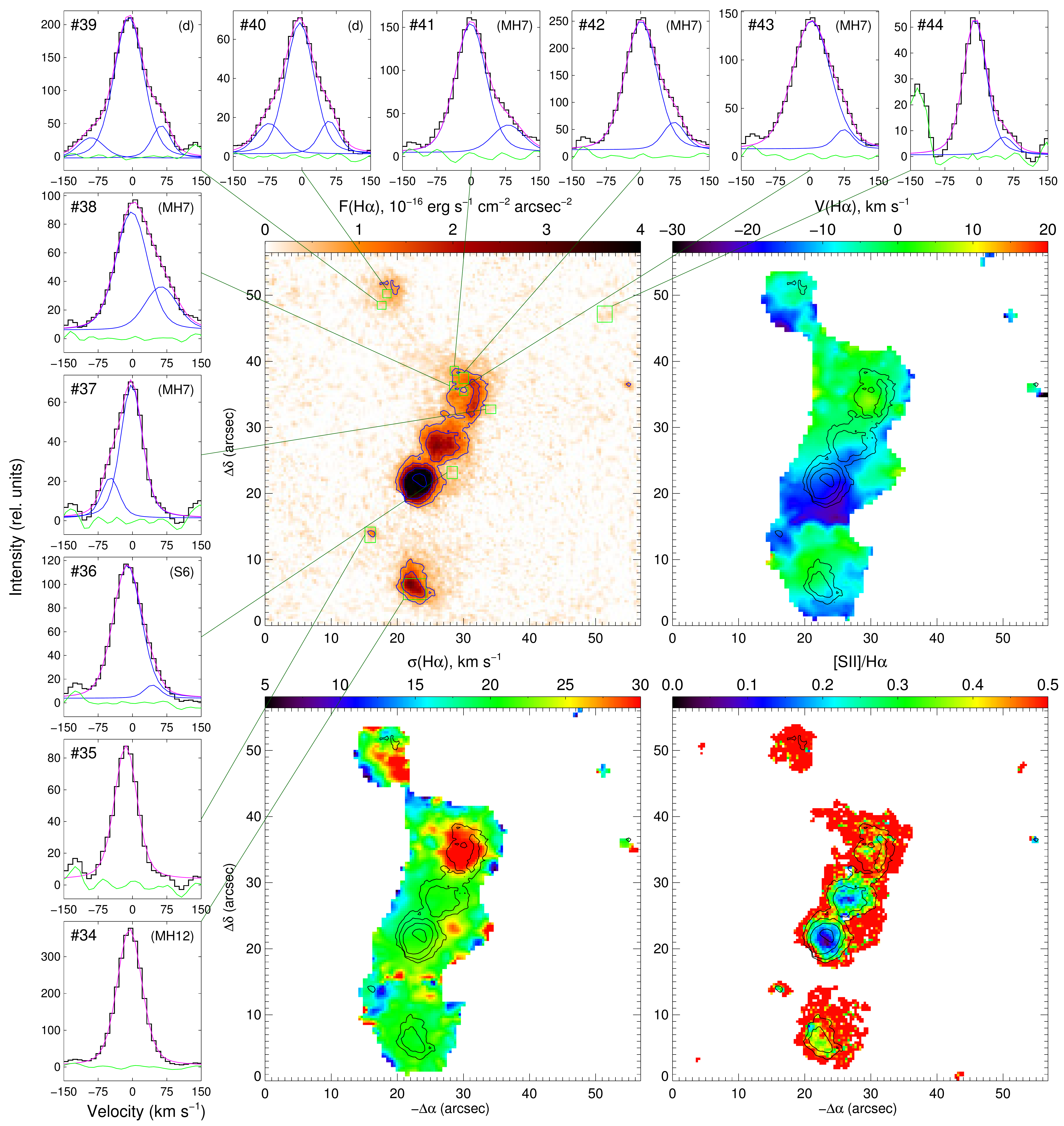}
	\caption{Same as in Fig.~\ref{fig:profs_c1_4} but for complex~\#3.}\label{fig:profs_c3}
\end{figure*}

\begin{figure*}
	\includegraphics[width=0.95\linewidth]{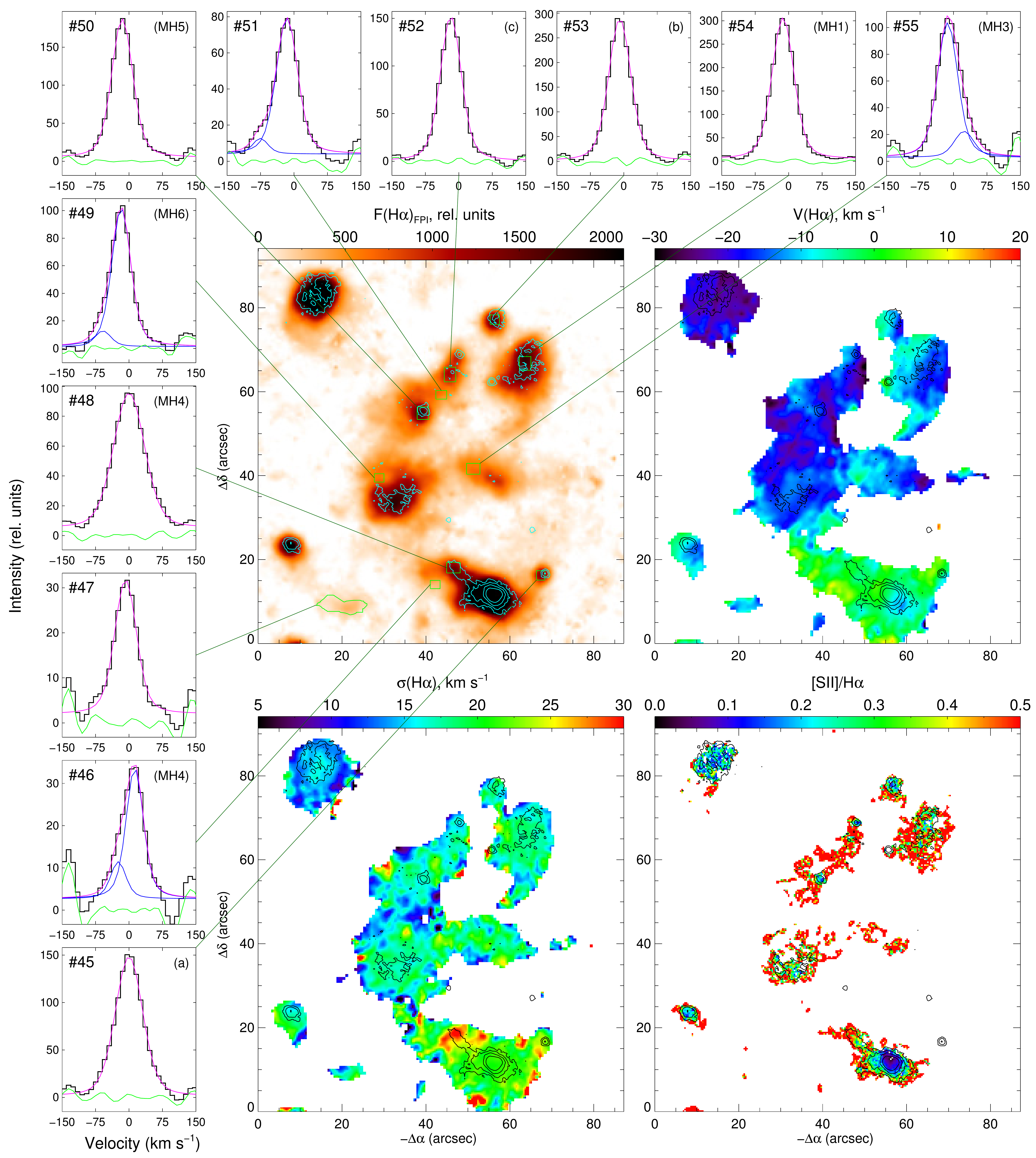}
	\caption{Same as in Fig.~\ref{fig:profs_c1_4} but for the complex~NW. The \Ha flux distribution from the FPI data is used here instead of the \Ha narrow-band image because of a better sensitivity to faint emission features. The contours on each panel correspond to \Ha intensities of $(0.5, 1, 2, 4)\times10^{-16}\ \mathrm{erg\ s^{-1}\ cm^{-2}\ arcsec^{-2}}$. }\label{fig:profs_nw}
\end{figure*}

 In general, the morphology of ionized gas in Holmberg~I in the \SII and \OIII lines (see Fig.~\ref{fig:profs_all}) is consistent with the \Ha distribution, yet the images in \SII and \OIII are much more noisy. Note that several \HII regions appear to be brighter in \OIII than in H$\alpha$, that is especially clear in the bottom right-hand panel of Fig.~\ref{fig:profs_all}. We will discuss these regions  in Section~\ref{sec:hiiregs}. The maps of the \SIIHa\, ratio shown in the bottom right-hand panels of Figs.~\ref{fig:profs_all}--\ref{fig:profs_nw} for each complex reveal a picture typical of other Irr galaxies: most \HII regions have a low value of \SIIHa\, in their centres, while it grows  towards their edges.
In the majority of regions, the ratio \SIIHa$\  > 0.4$ is observed only in the areas of faint filaments and diffuse emission outside the bright \HII regions. Most probably,  we observe an emission of the diffuse ionized gas (DIG) there which is detected in many other galaxies.
Its emission and enhanced line ratios in star-forming galaxies are usually explained by the leakage of hard ionizing quanta from \HII regions with an additional influence of shock waves propagating through the low-density medium and turbulent mixing layers \citep[see, e.g.,][and references therein]{Hoopes2003, Seon2009, Zhang2017}.

If the \HII complexes we selected are indeed unified structures, their \Ha emission should be consistent with the location of the young
 stars inside a complex, and the age of  \HII regions in a complex should be more or less identical.
We estimated the age of each \HII region from the equivalent width ($EW$) of its \Ha emission line derived from the images in this line and in the nearby continuum ($EW_\mathrm{H\beta}$ derived from the spectra is also used in Section~\ref{sec:hiiregs}). 

The possibility of using the $EW$ of Balmer lines as an age indicator was illustrated by
a number of authors (e.g., \citealt{Copetti1986, Schaerer1998, Leitherer1999}). We use the model published by
\cite{Levesque2013} for the sub-solar metallicity ($Z=0.004$). The results are listed in Table~\ref{tab:HIIregs}. Taking into account that $EW_\mathrm{H\alpha}$  becomes much less sensitive to ages older than $\sim5-6$~Myr, we may state that the ages of \HII regions are almost identical in each \HII complex, yet a few younger or older regions than median value are observed.

As follows from a comparison between Table~\ref{tab:HIIregs} and Table~\ref{tab:shells}, the age of the \HII regions derived from $EW_\mathrm{H\alpha}$ is systematically higher than kinematic-age estimates for the same regions. Such a problem of inconsistency between these two methods of age estimation was discussed previously in \cite{Wiebe2014}. 
As we mentioned in Section~\ref{sec:kinematics}, because of the assumption used for estimating the expansion velocity, the obtained values of kinematic ages can be underestimated. On the other hand, the relation between $EW_\mathrm{H\alpha}$ and the age of \HII region was obtained assuming a single stellar population and is highly dependent on the used models and initial mass function. The presence of an additional older stellar population in the region (that is very probable in Holmberg~I since stars with ages of about 100 Myr are observed in the same place as OB stars, see Fig.~\ref{fig:HoI_stars}) will increase the observed continuum level and decrease the age derived with this method.

We analysed the ionization balance for each complex based on the selection of O stars made in Section~\ref{sec:stars}.
Using the models from \cite{Martins2006}, we evaluated the bolometric luminosity and the number of ionizing photons LyC$\star$ from each O star. Similar values can be inferred from the observed \Ha flux. We estimated the number of ionizing photons $Q_0$ needed for gas ionization in  each \HII complex with the procedure described in \cite{Osterbrock2006}.
For the case of optically thick nebulae, its \Ha luminosity    $L_\mathrm{H\alpha}$ depends on $Q_0$ as

\begin{equation}
\frac{L(\mathrm{H}\alpha)}{h\nu_{\mathrm{H}\alpha}} \simeq \frac{\alpha^{eff}_{\mathrm{H}\alpha}}{\alpha_B}Q_0 \simeq 0.45Q_0,
\label{eq:ionization}
\end{equation}

where the \Ha effective recombination coefficient $\alpha^{eff}_{\mathrm{H}\alpha} \simeq 1.17\times10^{-13}\ \mathrm{cm^{3}\ s^{-1}}$ and the total recombination coefficient of hydrogen $\alpha_B \simeq 2.59\times10^{-13}\ \mathrm{cm^{3}\ s^{-1}}$ for $T = 10\,000$~K. Hereafter we refer to the estimated amount of needed ionizing photons $Q_0$ from (\ref{eq:ionization}) as LyC$_\mathrm{HII}$ to note that it has been converted from the \Ha luminosity.

The estimated values of LyC$_\mathrm{HII}$ and LyC$\star$ for each \HII complex are shown in Table~\ref{tab:HIIregs}. As follows from this table, the identified young massive stars provide even more than enough ionizing 
quanta to produce the observed \Ha luminosity of each complex.

The excess of ionizing quanta points to their possible leakage from the \HII complexes because of the porosity of the ISM. About 60 per cent of the available ionizing
 radiation should escape from each  \HII complex to provide the observed \Ha flux.  A recent modelling of the feedback of a star cluster onto the ISM performed by \cite{Rahner2017}
has shown that the escape fraction of ionizing radiation from low-metallicity (the same as for Holmberg~I)
 massive molecular clouds might indeed be large -- up to 60 per cent and even higher in some short periods of its evolution.

The large escape fraction of ionizing radiation from the \HII complex should provide a large amount of the diffuse \Ha emission in the galaxy.
Thus, \cite{Oey2007} showed that about 60 per cent of the total \Ha emission of a galaxy (independently on its type) is represented by its diffuse component. Our values of the escape fraction are in agreement with these estimates. Our images actually reveal a number of elongated diffuse structures towards the inner part of the SGS and on the rims of smaller \HI supershells. Earlier in \cite{Egorov2014} and \cite{Egorov2017}, we have identified faint ionized giant (1.5--2 kpc sized) supershells on the inner rims of \HI SGS of the galaxies IC~2574 and Holmberg~II. Probably, such a large structure could
 also exist in Holmberg~I, but its brightness is below the detection limit of our observations. Indeed, the ionized supershell in Holmberg~II was observed at a surface brightness level of  $3-6\times10^{-18} \mathrm{erg\ s^{-1} cm^{-2} arcsec^{-2}}$ that is nearly the detection limit during our observations of Holmberg~II but below that for  Holmberg~I.

Note also that some stars identified as being of the OB type could be not OB actually (see Section~\ref{sec:stars}). Some young stars classified as OB in Fig.~\ref{fig:HoI_stars} lie significantly outside the \HII regions and probably are misclassified. This fact will reduce the estimated fraction of
ionizing quanta especially calculated for the whole galaxy ($f_{esc} \sim 70$ per cent).

Below we describe each of the identified \HII complexes and analyse their ionized gas kinematics (see Figs.~\ref{fig:profs_all}--\ref{fig:profs_nw}).

\subsubsection*{\HII complex \#1}

This complex contains an arc of several bright \HII regions and the faint diffuse shell S1, and is limited by a faint ring with a diameter of about 700 pc (the shell S2). The brightest \HII regions in the complex are located at the outer edge of the SGS.

Faint blueshifted (and even broad underlying) components in the \Ha line profiles are observed in the central part of the complex near the bright nebulae (see profiles \#14, 18 in Fig.~\ref{fig:profs_c1_4}), while the faint redshifted emission is detected in the line profiles outside the bright nebulae towards the outer part of the complex (profile \#19).

The most intriguing structure observed in this complex is a surrounding ring -- shell S2. While it is only noticeable in other \Ha images as a faint filament
 (see, e.g., Fig.~\ref{fig:HoI_stars}), it is much better seen in the top left-hand panel of Fig.~\ref{fig:profs_c1_4}, where adaptive binning was used to reach a signal-to-noise ratio better than 10 in each pixel. An analysis of the FPI data cube shows that this structure is clearly seen at blueshifted velocities and almost disappear at red channels. The top right-hand panel of Fig.~\ref{fig:profs_c1_4} also shows that the shell S2 is blueshifted relative to other part of the complex and of the whole galaxy as well.

Several faint clumps in the southern part of the ring S2 demonstrate the unperturbed single component \Ha line profiles (see profiles \#13 and \#20 in Fig.~\ref{fig:profs_c1_4}), while its northern part  reveals the slightly asymmetrical \Ha line profiles (\#16 and \#17) with blue- or redshifted components in their wings. Note, however, that the signal-to-noise ratio for these components is low.

Only with  this information at hand, we can speculate on a possible nature of the shell S2. It might be created by the ionizing radiation of a few O stars
 outside bright \HII regions as well as the UV photons leakage from those regions influencing surrounding neutral gas. As follows from Figs.~\ref{fig:HIHa}, \ref{fig:HoI_pv},  complex \#1 is located in the low-density medium (in comparison with the other part of the SGS) but at the edge of the relatively bright \HI cloud. We propose that the complex is located at the front edge of this dense \HI cloud. In that case we should observe the faint ionizing front moving towards us through the low-density \HI medium even at a great distance from the complex, while its bright receding side should be much slower and will be interpreted as a non-shifted component of the line profile. Such a picture is able to explain the observed blue shift of the ring on the line-of-sight velocity map and is consistent with the observed \Ha line profiles (the presence of a faint blueshifted component) from different regions of complex \#1.

\subsubsection*{\HII complex \#2}

The brightest complex of star formation is located in the southeastern part of the SGS in the area of its maximum \HI density that explains a large number of O stars and \HII regions there. Most \HII regions are encircled by \HI supershell \#7,
and the northeastern \HII regions are located at the edge of  \HI supershell \#9 (see Figs.~ \ref{fig:HoI_stars}, \ref{fig:HoI_pv}). Many \HII regions
 there have the shell-like morphology, and their  \Ha line profiles (e.g., \#24, 32, 33 in Fig.~\ref{fig:profs_c2}) show signs of  expansion with velocities corresponding to kinematic ages of $1-2$~Myr. The second largest ionized superbubble in the galaxy -- S3 -- is observed in the central part of the complex and joins several bright \HII regions. Its kinematic age ($\sim 3$~Myr) is slightly greater than that of other shells in the complex, that can point to its relation to the previous generation of stars formed there.
Most  O stars found in the complex are located inside the \HII regions, and none of them was found
 towards the centre of the S3 shell.

An extended area of the diffuse ionized gas (including two faint \HII regions g and h) is observed at the southeastern edge of the complex in the low-density environment showing the multicomponent \Ha line profile at some places (e.g., profile \#26 in Fig.~\ref{fig:profs_c2}). The faint filaments extending towards the centre of the SGS and of \HI supershell \#9 are also observed. All these features are probably the result of the ionizing quanta leakage from \HII regions due to  inhomogeneities of the ISM.

In general, all the \HII regions and ionized shells are of more or less the same age (see Tables~\ref{tab:shells} and \ref{tab:HIIregs}), that together with the detection of a large amount of diffuse filaments and the presence of surrounding \HI supershell \#7 indicates that  complex \#2 is indeed a unified structure.
Probably, at a higher spatial and spectral resolution it would be worth  exploring this complex as a separate \HI supershell with star formation in its walls.

Note that one of the brightest \HII regions in the galaxy -- MH25 -- may not be related to  \HII complex \#2. We consider it as a part of  complex \#2 because of a very close location to it, but neither obvious emission filaments nor diffuse gas connecting them are observed. In addition, it is located at the edge of \HI supershell \#9 and far away from supershell \#7, while the latter joins all other \HII regions in  complex \#2.

\subsubsection*{\HII complex \#3}

We joined a chain of the \HII regions in the southwestern part of the SGS to  complex \#3. As follows from Figs.~\ref{fig:HoI_stars}, \ref{fig:HoI_pv}, most part of this complex is located at the edge of \HI supershell \#10.

In contrast with the previous two, this complex does not show any extended diffuse emission around the \HII regions, yet a filament connecting all of them can be clearly seen. The velocity field (the top right-hand panel in Fig.~\ref{fig:profs_c3}) shows a significant difference of the line-of-sight velocity in this complex. The area surrounding the \HII region MH11 is blueshifted at $\simeq 20 \kms$ in comparison with the other part of the complex.
Taking it into account, the physical relation of the \HII regions in this complex with each other is less evident
than that in other complexes.

\subsubsection*{\HII complex \#4}

This small complex consists of two bright nebulae MH20 and MH24, the shell MH20a, and the extended diffuse emission filling the area. The enhanced \SIIHa\, ratio in our long-slit spectra between these nebulae (see also Fig.~\ref{fig:profs_c1_4}) together with the presence of a blueshifted component in the wing of the \Ha line profile (e.g., \#21 in Fig.~\ref{fig:profs_c1_4}) indicates a major role of shocks in excitation of the diffuse emission. The complex is located towards the inner wall of the SGS and the centre of  \HI supershell \#9. Hence, it resides in the   environment of a lower density in comparison with complexes \#1 and 2 which can explain a small number of \HII regions and low star-formation rate there.

In contrast with all other \HII complexes in the galaxy, this complex is seen close to the centre of a large \HI supershell.
We suppose that it is actually located at the approaching side of  \HI supershell \#9. This assumption is consistent with the \HI and \Ha kinematics in the region.
 Indeed, as follows from the PV diagrams and \HI line profiles shown in Fig.~\ref{fig:HoI_pv},  \HI supershell \#9 is clearly expanding and its approaching side
 has the velocity $V \simeq 10 \kms$ at the position of  complex \#4. This estimate  agrees well  with the line-of-sight velocity of ionized gas there: the whole complex \#4 is blueshifted at $\simeq 10-18 \kms$
 (see the top right-hand panel in Fig.~\ref{fig:profs_c1_4}).

As  follows from fig.~9 in \cite{Weisz2008}, the spatial density distribution of both red and blue stars shows a peak
 there. Note, however, that most of the O stars detected there lie close to, although, outside the mentioned \HII regions
and the whole complex as well. At the same time many of BHeB stars of  ages of $50 - 200$ Myr  are observed towards
 this complex. Together with the age estimates of the \HII regions (see Table~\ref{tab:HIIregs}), it allows one
 to propose that this complex is older than those previously discussed.

\subsubsection*{\HII complex NW}

Especially interesting is a group of nebulae located at the northwestern side of the galaxy in the dense layer of
 \HI between the SGS   and supershell \#1.

We classified this `arc' of faint \HII regions as a unified complex based on the presence of the diffuse emission
connecting them with each other that coincides with the distribution of H~\textsc{i}. The additional evidence is that
 almost all the \HII regions show more or less similar ages (see Table~\ref{tab:HIIregs}). Taking into account the localisation of neutral and
 ionized gas, we conclude that the ongoing star formation in this complex was triggered by the collision of supershell \#1
  with the SGS (see the discussion in Section~\ref{sec:discuss}).

\begin{figure*}
	\includegraphics[width=0.92\linewidth]{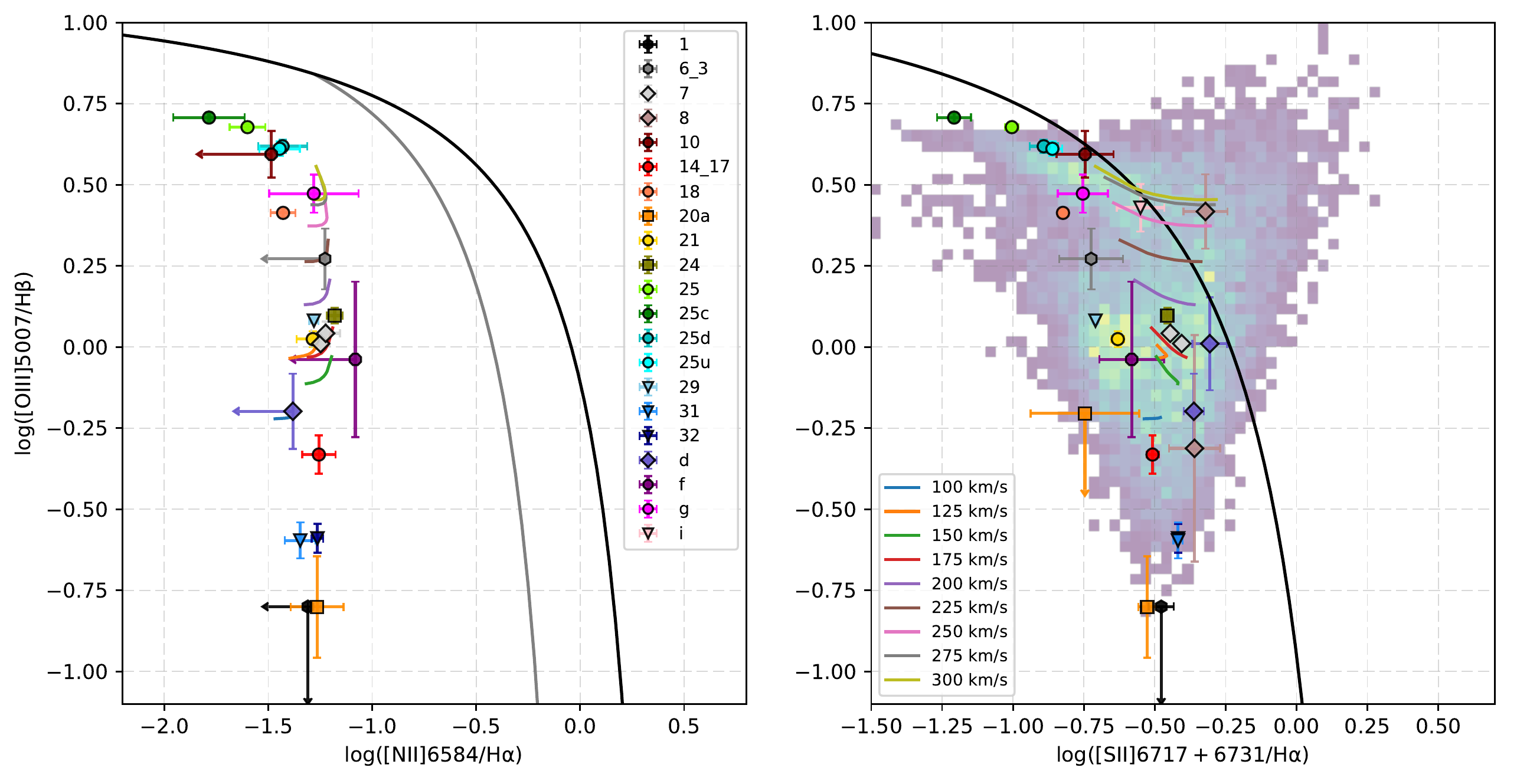}
	
	\caption{Diagnostic diagrams \OIIIHb\, vs \NIIHa\, (left) and \SIIHa\, (right) constructed for the individual \HII regions in Holmberg~I. The models of the shock+precursor emission for the metallicity $z=0.004$ (the SMC set of models) from \citet{Allen2008} are overplotted in both diagrams.  The 2D-histogram in the right-hand panel shows the distribution of \OIIIHb\, vs \SIIHa\, derived from direct images.
		The black line in both panels is for the `maximum starburst line' from \citet{Kewley2001}, and the grey line from \citet{Kauffmann2003} in  the right-hand panel separates pure star-forming regions from those with a composite mechanism of excitation.}\label{fig:bpt}
\end{figure*}

The ionized gas in the complex NW almost does not reveal any high velocity motions except for few diffuse areas,
where the blue- or redshifted components in the wings of the \Ha line profiles are observed (profiles \#46, 49, 51, 55 in
 Fig.~\ref{fig:profs_nw}), and the region MH4 revealing a broad \Ha line profile (\#48 in Fig.~\ref{fig:profs_nw}).
 Note, however, that the northeastern part of the complex shows a significant blue shift of the line-of-sight velocity
(up to $-25 \kms$). Moreover, the kinematics of ionized and neutral gas are different there:  \HI is observed even at the redshifted velocities at some places (see, e.g., the PV diagram \#4 in Fig.~\ref{fig:HoI_pv}).
The collision of  two \HI supershells  and inhomogeneities in the \HI density distribution can be a reason of such
a complex gas kinematics.

\section{Individual \HII regions and analysis of their spectra}\label{sec:hiiregs}

In this Section we analyse the emission spectra and chemical abundance of the \HII regions in the galaxy Holmberg~I. Several most interesting \HII regions are discussed in details.

The most complete catalogue of the \HII regions in Holmberg~I by \cite{MH94} contains 32 \HII regions.
In the current paper, we reveal several new \HII regions (shown with squares and denoted
by letters in Fig.~\ref{fig:hoi_loc}) together with the faint extended structures of ionized gas mentioned in the previous Section. On the other hand, some areas of ionized gas identified by \cite{MH94} as \HII regions seem to be extended filaments of ionized gas created by the stellar wind or supernovae from a nearby OB association: MH3, 4, 8, 19, and 28. The region MH15 does not reveal any emission in the \Ha line and seems to be a bright star. These non-\HII regions from \cite{MH94} are shown in Fig.~\ref{fig:hoi_loc} by the dotted circles.

Table~\ref{tab:HIIregs} lists the properties of each \HII region in the galaxy:  its size,
 median line-of-sight velocity $V_{LOS}$ and velocity dispersion $\sigma$, integrated \Ha flux F(H$\alpha$), equivalent
 width of the \Ha line EW(H$\alpha$), age derived from EW(H$\alpha$), and the type of its morphology according to our classification made by eye. Note that several  \HII regions
are classified as multiple -- they represent several compact (or compact and diffuse) nebulae instead of single as was defined earlier.

Only four brightest \HII regions have been previously spectroscopically studied: MH18, 22, and 29 \citep{Croxall2009}, and MH25 \citep{MH96}. We obtained the spectra for the galaxy with slits crossing a large part of its \HII regions (see Fig.~\ref{fig:slitpos}). For every observed region, we obtained the integrated emission spectrum for a further analysis. The reddening-corrected values of most  measured emission-line fluxes (normalized to the fluxes of the H$\alpha$ or H$\beta$ lines) are listed in Table~\ref{tab:spec}. The adopted values of the interstellar extinction coefficient c(H$\beta$) calculated from the  observed Balmer decrement are also shown in the Table. The values of \SIIHa\, and \OIIIHb\, obtained from direct images agrees well with those measured from the long-slit spectra.

 In Fig.~\ref{fig:bpt} we plot the \HII regions, for which we obtained spectroscopic data in the diagnostic BPT \citep*{BPT} diagrams. The right-hand panel shows also a 2D histogram corresponding to the distribution of line ratios for each pixel of our images. The emission from the regions lying above the black `maximum starburst line' \citep{Kewley2001} cannot be explained by photoionization from massive OB stars. Composite mechanism of excitation should work for regions lying between this line and the grey curve from \cite{Kauffmann2003} (for the solar abundance) in the left-hand panel. As  follows from the right-hand diagram, some pixels in our narrow-band images fall into the area above the maximum starburst line.
These pixels correspond to a faint emission outside \HII regions only, and as we noted in the previous Section, they most probably represent the DIG.

All the \HII regions observed with the long-slit spectrograph (except for the faint diffuse region MH8) show the emission line ratios corresponding to their positions in the BPT diagrams below the separation lines. Despite that, the shock waves should produce lower line ratios due to the low metallicity of  Holmberg~I. In fact, the separation lines should be shifted towards the lower \NIIHa\, and \SIIHa\, line ratios. We add a grid of shock waves ionization models in the BPT diagrams from \cite{Allen2008} which correspond to the shock velocity from 100 $\kms$ and to the metallicity $Z=0.004$ (SMC-like).  As  follows from the comparison of the \SIIHa\,
ratios with these models, a shock excitation can play a significant role in ionization of the regions MH7, 8, 24, d, and i. Two of them -- MH7 and d -- show a high value of \SIIHa > 0.4 estimated from direct images and the multicomponent \Ha line profile with high-velocity components (see Fig.~\ref{fig:profs_c3}). Because of this, we consider these regions as probable supernova remnants. The region MH24 lies close to MH7 in
 the BPT diagram but does not show any peculiarities of the ionized gas kinematics -- that is why, we do not consider it as an SNR
 candidate. The two regions -- MH8 and i -- are faint diffuse areas of ionized gas.

Table~\ref{tab:spec}  provides the estimates of EW$_\mathrm{H\beta}$ and corresponding age for all the observed \HII regions obtained in the way we described in the previous Section. The ages of \HII regions are consistent with those obtained using  EW$_\mathrm{H\alpha}$ derived from direct images (see Table~\ref{tab:HIIregs}).

\begin{table*}
	\caption{Results of the emission spectra analysis}\label{tab:spec}
	{\scriptsize
		\begin{tabular}{lccccccc}
			\hline
			Region & $\mathrm{1}$ & $\mathrm{6\_3^*}$ & $\mathrm{7}$ & $\mathrm{7}$ & $\mathrm{8}$ & $\mathrm{8}$ & $\mathrm{10}$ \\
			\hline
			Slit & PA142 & PA142 & PA259 & PA212 & PA259 & PA212 & PA150 \\
			Pos.(arcsec) & -167 -- -158 & -138 -- -116 & 101 -- 107 & -132 -- -122 & 95 -- 99 & -138 -- -133 & 141 -- 151 \\
			c(H$\beta$) & $0.00 \pm 0.17$ & $0.00 \pm 0.26$ & $0.20 \pm 0.04$ & $0.25 \pm 0.09$ & $0.03 \pm 0.33$ & $0.23 \pm 0.59$ & $0.10 \pm 0.21$ \\
			{[O \textsc{ii}] 3727,3729}/H$\beta$ & $2.79 \pm 0.36$ & $-$ & $-$ & $-$ & $-$ & $-$ & $-$ \\
			He \textsc{ii} 4686/H$\beta$ & $-$ & $-$ & $-$ & $-$ & $-$ & $-$ & $-$ \\
			{[O \textsc{iii}] 5007}/H$\beta$ & $< 0.16$ & $1.87 \pm 0.40$ & $1.02 \pm 0.04$ & $1.10 \pm 0.09$ & $2.62 \pm 0.69$ & $0.49 \pm 0.39$ & $3.93 \pm 0.65$ \\
			{[N \textsc{ii}] 6583}/H$\alpha$ & $< 0.05$ & $< 0.06$ & $0.06 \pm 0.01$ & $0.06 \pm 0.01$ & $-$ & $-$ & $< 0.03$ \\
			{[S \textsc{ii}] 6717,6731}/H$\alpha$ & $0.33 \pm 0.03$ & $0.19 \pm 0.05$ & $0.39 \pm 0.01$ & $0.36 \pm 0.01$ & $0.48 \pm 0.09$ & $0.44 \pm 0.09$ & $0.18 \pm 0.04$ \\
			{[Ar \textsc{iii}] 7136}/H$\beta$ & $-$ & $-$ & $-$ & $-$ & $-$ & $-$ & $-$ \\
			EW(H$\beta$) & $41 \pm 7$ & $18 \pm 4$ & $31 \pm 1$ & $45 \pm 3$ & $9 \pm 2$ & $13 \pm 6$ & $120 \pm 34$ \\
			Age (Myr) & $5.4 \pm 0.2$ & $6.6 \pm 0.4$ & $5.7 \pm 0.1$ & $5.3 \pm 0.1$ & $7.6 \pm 0.3$ & $7.2 \pm 0.6$ & $4.2 \pm 0.3$ \\
			$\mathrm{12+\log(O/H)}$ & $-$ & $< 8.02 $ & $7.71 \pm 0.03$ & $7.74 \pm 0.05$ & $-$ & $-$ & $<7.78$ \\
			\hline
			\hline
			Region & $\mathrm{14\_17^*}$ & $\mathrm{18}$ & $\mathrm{20a}$ & $\mathrm{20a}$ & $\mathrm{21^*}$ & $\mathrm{24}$ & $\mathrm{25}$ \\
			\hline
			Slit & PA142 & PA142 & PA259 & PA150 & PA142 & PA259 & PA150 \\
			Pos.(arcsec) & -20 -- -8 & -3 -- 2 & 41 -- 47 & 34 -- 44 & 4 -- 10 & 28 -- 33 & -6 -- 4 \\
			c(H$\beta$) & $0.11 \pm 0.12$ & $0.14 \pm 0.04$ & $0.06 \pm 0.10$ & $0.36 \pm 0.40$ & $0.00 \pm 0.06$ & $0.00 \pm 0.04$ & $0.08 \pm 0.03$ \\
			{[O \textsc{ii}] 3727,3729}/H$\beta$ & $-$ & $-$ & $-$ & $-$ & $-$ & $-$ & $1.42 \pm 0.68$ \\
			He \textsc{ii} 4686/H$\beta$ & $-$ & $-$ & $-$ & $-$ & $-$ & $-$ & $0.07 \pm 0.02$ \\
			{[O \textsc{iii}] 5007}/H$\beta$ & $0.47 \pm 0.06$ & $2.59 \pm 0.07$ & $0.16 \pm 0.06$ & $< 0.62$ & $1.06 \pm 0.06$ & $1.25 \pm 0.07$ & $4.76 \pm 0.10$ \\
			{[N \textsc{ii}] 6583}/H$\alpha$ & $0.06 \pm 0.01$ & $0.04 \pm 0.01$ & $0.05 \pm 0.02$ & $< 0.04$ & $0.05 \pm 0.01$ & $0.07 \pm 0.01$ & $0.03 \pm 0.00$ \\
			{[S \textsc{ii}] 6717,6731}/H$\alpha$ & $0.31 \pm 0.02$ & $0.15 \pm 0.01$ & $0.30 \pm 0.02$ & $0.18 \pm 0.08$ & $0.23 \pm 0.01$ & $0.35 \pm 0.01$ & $0.10 \pm 0.01$ \\
			{[Ar \textsc{iii}] 7136}/H$\beta$ & $-$ & $0.05 \pm 0.01$ & $-$ & $-$ & $-$ & $-$ & $0.06 \pm 0.01$ \\
			EW(H$\beta$) & $64 \pm 8$ & $72 \pm 3$ & $11 \pm 1$ & $11 \pm 3$ & $253 \pm 46$ & $43 \pm 1$ & $99 \pm 3$ \\
			Age (Myr) & $4.9 \pm 0.1$ & $4.8 \pm 0.1$ & $7.4 \pm 0.1$ & $7.3 \pm 0.4$ & $2.9 \pm 0.4$ & $5.3 \pm 0.1$ & $4.4 \pm 0.1$ \\
			$\mathrm{12+\log(O/H)}$ & $7.40 \pm 0.07$ & $7.84 \pm 0.04$ & $-$ & $-$ & $7.63 \pm 0.05$ & $7.81 \pm 0.03$ & $7.94 \pm 0.06$ \\
			\hline
			\hline
			Region & $\mathrm{25c}$ & $\mathrm{25se}$ & $\mathrm{25nw}$ & $\mathrm{29}$ & $\mathrm{31^*}$ & $\mathrm{32^*}$ & $\mathrm{d}$ \\
			\hline
			Slit & PA150 & PA150 & PA150 & PA259 & PA259 & PA259 & PA142 \\
			Pos.(arcsec) & -2 -- 0 & -6 -- -4 & 2 -- 5 & -3 -- 2 & -11 -- -6 & -19 -- -14 & -78 -- -72 \\
			c(H$\beta$) & $0.02 \pm 0.04$ & $0.00 \pm 0.06$ & $0.32 \pm 0.07$ & $0.07 \pm 0.02$ & $0.00 \pm 0.06$ & $0.07 \pm 0.04$ & $0.09 \pm 0.32$ \\
			{[O \textsc{ii}] 3727,3729}/H$\beta$ & $1.85 \pm 1.24$ & $-$ & $3.07 \pm 2.07$ & $2.88 \pm 0.63$ & $-$ & $1.74 \pm 1.30$ & $-$ \\
			He \textsc{ii} 4686/H$\beta$ & $0.16 \pm 0.05$ & $-$ & $-$ & $-$ & $-$ & $-$ & $-$ \\
			{[O \textsc{iii}] 5007}/H$\beta$ & $5.09 \pm 0.18$ & $4.15 \pm 0.20$ & $4.08 \pm 0.21$ & $1.21 \pm 0.02$ & $0.25 \pm 0.03$ & $0.26 \pm 0.03$ & $1.02 \pm 0.34$ \\
			{[N \textsc{ii}] 6583}/H$\alpha$ & $0.02 \pm 0.01$ & $0.04 \pm 0.01$ & $0.04 \pm 0.01$ & $0.05 \pm 0.00$ & $0.05 \pm 0.01$ & $0.05 \pm 0.00$ & $-$ \\
			{[S \textsc{ii}] 6717,6731}/H$\alpha$ & $0.06 \pm 0.01$ & $0.13 \pm 0.01$ & $0.14 \pm 0.01$ & $0.20 \pm 0.00$ & $0.38 \pm 0.01$ & $0.38 \pm 0.01$ & $0.49 \pm 0.07$ \\
			{[Ar \textsc{iii}] 7136}/H$\beta$ & $0.06 \pm 0.02$ & $0.10 \pm 0.04$ & $0.05 \pm 0.02$ & $0.05 \pm 0.00$ & $-$ & $-$ & $-$ \\
			EW(H$\beta$) & $65 \pm 3$ & $197 \pm 21$ & $118 \pm 9$ & $86 \pm 1$ & $26 \pm 1$ & $51 \pm 2$ & $30 \pm 8$ \\
			Age (Myr) & $4.9 \pm 0.1$ & $3.5 \pm 0.6$ & $4.3 \pm 0.1$ & $4.6 \pm 0.1$ & $6.1 \pm 0.1$ & $5.1 \pm 0.1$ & $5.8 \pm 0.6$ \\
			$\mathrm{12+\log(O/H)}$ & $-$ & $7.99 \pm 0.08$ & $-$ & $7.66 \pm 0.02$ & $7.15 \pm 0.07$ & $-$ & $-$ \\
			\hline
			\hline
			Region & $\mathrm{d}$ & $\mathrm{f}$ & $\mathrm{g}$ & $\mathrm{i}$ &  &  &  \\
			\hline
			Slit & PA212 & PA212 & PA142 & PA259 &  &  &  \\
			Pos.(arcsec) & -150 -- -145 & -252 -- -237 & 13 -- 19 & -37 -- -26 &  &  &  \\
			c(H$\beta$) & $0.08 \pm 0.18$ & $0.29 \pm 0.57$ & $0.00 \pm 0.17$ & $0.00 \pm 0.20$ &  &  &  \\
			{[O \textsc{ii}] 3727,3729}/H$\beta$ & $-$ & $-$ & $-$ & $-$ &  &  &  \\
			He \textsc{ii} 4686/H$\beta$ & $-$ & $-$ & $-$ & $-$ &  &  &  \\
			{[O \textsc{iii}] 5007}/H$\beta$ & $0.63 \pm 0.17$ & $0.92 \pm 0.50$ & $2.97 \pm 0.40$ & $2.69 \pm 0.45$ &  &  &  \\
			{[N \textsc{ii}] 6583}/H$\alpha$ & $< 0.04$ & $< 0.08$ & $0.05 \pm 0.03$ & $-$ &  &  &  \\
			{[S \textsc{ii}] 6717,6731}/H$\alpha$ & $0.43 \pm 0.04$ & $0.26 \pm 0.07$ & $0.18 \pm 0.04$ & $0.28 \pm 0.05$ &  &  &  \\
			{[Ar \textsc{iii}] 7136}/H$\beta$ & $-$ & $-$ & $-$ & $-$ &  &  &  \\
			EW(H$\beta$) & $31 \pm 5$ & $9 \pm 4$ & $162 \pm 55$ & $-$ &  &  &  \\
			Age (Myr) & $5.7 \pm 0.4$ & $7.5 \pm 0.5$ & $3.8 \pm 0.9$ & $-$ &  &  &  \\
			$\mathrm{12+\log(O/H)}$ & $<7.66$ & $<8.21$ & $7.96 \pm 0.15$ & $-$ &  &  &  \\
			\hline
			\multicolumn{8}{l}{$^*$The slit crossed only the edge of the region. EW, age, and oxygen abundance estimates can be unreliable.}
		\end{tabular}
	}
\end{table*}

We estimated the oxygen abundance $\mathrm{12+\log(O/H)}$, which is the indicator of gas metallicity for the most observed \HII regions, yet for several of them we provide only the upper-limit value. The results are reported in Table~\ref{tab:spec}.
Since the weak [O~\textsc{iii}] 4363 \AA\ emission line, which is sensitive to electron temperature, does not appear in our spectra, we were unable to use the `direct' $T_e$ method for this. We applied the empirical S-method by \cite{Pilyugin2016} which is based on the flux ratios of the strong emission lines [S~\textsc{ii}], [N~\textsc{ii}], and \OIII to H$\beta$. This method was selected among many other developing strong-line empirical calibrations, because it is applicable for low-metallicity galaxies, is consistent with the results of the $T_e$ method and does not demand  measurements of [O~\textsc{ii}] 3727, 3729~\AA\ which are very noisy in our data due to the poor sensitivity of the spectrograph in the blue region of the spectrum.

The oxygen abundance of only four brightest regions was measured before. We have obtained spectroscopic data for three of them -- MH18, 29, and 25. \cite{Croxall2009} estimated $\mathrm{12+\log(O/H)} = 8.04$ for the first two of these regions using the `direct' $T_e$ method, while their measurements with the empirical method \citep{McGaugh1991} gave lower values (7.71 and 7.82, respectively). \cite{Moustakas2010} used another empirical method by \cite{Pilyugin2005} and obtained the values $\mathrm{12+\log(O/H)} = 7.55$ and 7.71, and $7.54 \pm 0.36$ for the region MH25. Because of the  well-known problem of discrepancy between different methods of oxygen abundance estimation \cite[see, e.g.,][]{Kewley2008, Lopez-Sanchez2012}, we recalculated the metallicity of these regions with the S method \citep{Pilyugin2005} using the line ratios reported by \cite{Croxall2009} to directly compare with our results. Thus, we obtained $\mathrm{12+\log(O/H)} = 7.83 \pm 0.03$ and $7.87 \pm 0.03$ for MH18 and MH29, respectively.
While our  metallicity estimate  for MH18 is in a very good agreement with the  previous data, its value for MH29 is lower and is in a better agreement with a value reported by \cite{Moustakas2010}. In fact, the latter region consists of four compact and one diffuse \HII regions. Our oxygen-abundance estimate for this region corresponds to its northern part, while \cite{Croxall2009} performed the observations of its brightest central part which can be a reason for the discrepancy. In addition, \cite{Croxall2009} used a much larger value of the extinction coefficient, that can also influence the  reported oxygen abundance. Due to the absence of reported fluxes of the \SII lines in \cite{MH96} and \cite{Moustakas2010}, we were unable to make a similar comparison for the MH25 region, 
yet our results do not conflict with the published estimates taking into account  their large reported  uncertainty.
In general, our oxygen abundance estimates for Holmberg~I are consistent with the previously published values, yet we observe a significant spread of metallicity between different regions.

It is worth  noting that for some \HII regions (denoted by an asterisk in Table~\ref{tab:spec}) the obtained spectra cross only their outer
 part. The reported line fluxes for two  such regions ($\mathrm{6\_3}$ and $\mathrm{14\_17}$) correspond to the area of 
diffuse emission between the nearby \HII regions mentioned in their names. In all these cases, the observed emission
 lines come from the outer parts of \HII regions and, hence, could be highly contaminated by the DIG and biased towards a lower ionization state. Hence, these regions can show higher values of \SIIHa\, and \NIIHa\, as well as an incorrect value of $EW_\mathrm{H\beta}$. The measured metallicity of them can be unreliable. Additionally, shock waves might have high contribution into the excitation of the MH7 and d regions and, hence, the empirical metallicity calibration used in our analysis is not applicable to them.

Below we consider some interesting \HII regions in more details.

\subsubsection*{MH25 (Complex 2)}
The brightest and among the largest \HII regions in Holmberg~I  appears as a shell-like structure around six
 identified O stars. Our data show a single slightly broaden \Ha line profile, but the observed gradient of velocity dispersion
 towards the centre of the region points to the presence of an unresolved expanding superbubble there.

While we consider this region as a part of \HII complex \#2, it could not be related to it. It is located relatively far away from the majority of the nebulae in the complex and is clearly tied with \HI supershell \#9 instead of \#7 (see Figs.~\ref{fig:HoI_stars} and \ref{fig:HoI_pv}). Probably, star formation in MH25 was triggered by the influence of expanding \HI supershell \#9 with the ISM in the SGS.

This region was divided into MH25 and MH27 in \cite{MH94}, while recent images demonstrate that it is a single object. Differences in ionization conditions of the inner and outer parts of the region are clearly seen in our observed data. The gradient of \OIIIHb\, ratio (Fig.~\ref{fig:profs_all}, bottom right-hand panel) and the invert gradient of \SIIHa\, (Fig.~\ref{fig:profs_c1_4}, bottom right-hand panel) are typical of \HII regions with a central source of ionization. The central part of MH25 show the highest \OIIIHb\, in the galaxy. Because of that, in addition to the integrated spectrum of MH25, we also analysed separately its central (MH25c in Table~\ref{tab:spec} and Fig.~\ref{fig:bpt}), northwestern (MH25nw), and southeastern (MH25se) parts.

The central part of MH25 is the only place in Holmberg~I, where we have detected the He~\textsc{ii} 4686\AA\, line. It also appears in the integrated spectrum, yet it is not observed in the outer bright areas MH25nw and MH25se, and the He~\textsc{ii}/H$\beta$ ratio  is two times higher in the central part MH25c. Hence, we expect to find a source of ionization in the centre of MH25 having a sufficiently hard spectrum and high temperature to explain high ratios of \OIIIHb\, and He~\textsc{ii}/H$\beta$.

Given the emissivities and recombination coefficients for hydrogen and helium \citep{Osterbrock2006}, we obtain that $L(\mathrm{HeII} 4686)/L(\mathrm{H\beta}) \simeq 2.14 Q_2/Q_0$, where $Q_2$ and $Q_0$ are the amount of the emitted helium and hydrogen ionizing quanta per second. Hence, in order to explain the observed line ratio of  He~\textsc{ii}/H$\beta$ = 0.16 in the central part of MH25, the ionizing source should produce  $Q_2$ no less than 7 per cent of $Q_0$. According to the models by \cite{Smith2002}, 
all types of O stars provide less amount of $Q_2$ and are not available to produce the observed intensity of the He~\textsc{ii} 4686\AA\, line. The presence of the Wolf-Rayet (WR) star WN10-WN11 is necessary to get the observed flux ratio. Comparing the measured EW of the He~\textsc{ii} 4686\AA\, line ($\simeq 9$ \AA) and its $FWHM$ ($\simeq 3.5$ \AA\, corrected for instrumental broadening) with the empirical diagnostic diagram of \cite{Crowther1997} made for different types of WR stars (see their fig.~1b), we obtain that the MH25 central source falls into the region occupied by WN9-10 stars.

Summing up, we conclude that despite the fact that there are no  observed peculiarities in the ionized gas kinematics in the MH25 region, it represents a slowly expanding superbubble around a young star cluster with at least 6 O stars and one WN9-11 star.

\subsubsection*{MH20 and MH20a  (complex 4)}

We consider the region MH20 from the list of \cite{MH94} as two separate nebulae -- the compact \HII region MH20 and the connected faint shell MH20a because of quite a large separation between them (see Fig.~\ref{fig:hoi_loc}).

A shell-like structure of MH20a is clearly seen in our \Ha images, while no  O stars were found inside it. We obtained two spectra of different parts of this shell -- they point to photoionization as a mechanism of emission-line excitation according to the BPT diagrams (Fig.~\ref{fig:bpt}).
A single slightly asymmetric \Ha line profile is observed towards the central and northern part of MH20a; its blue
 wing  reveals a faint component at the level of about 10 per cent of the maximum and shifted by $58~\kms$ (profile \#22, Fig.~\ref{fig:profs_c1_4}).

The most interesting fact is that the only bright (1.4 mJy) radio continuum point-like source in the galaxy
 according to the VLA observations presented in \cite{Hindson18} (see fig.~20 there)  clearly coincides
 with the MH20a region. 
 The author has shown that this non-thermal radio
 emission dominates there, and proposed that its origin can be a SNR, yet it can also come from some background
 source. Our spectral analysis does not reveal any signs of shock waves there. Moreover, the broad lines corresponding to a distant quasar are clearly seen in our spectrum PA=259. Hence, MH20a is not an SNR, and a background source is responsible for the emission in the radio continuum there.

\subsubsection*{MH 10}

This bright isolated \HII region is located in the area of interaction of three \HI supershells -- \#2, 3, and 11.
Probably, star formation there has been triggered by their collision.

Two O stars detected there provide a sufficient amount of ionizing quanta to create the \HII regions, while the analysis
 of \SIIHa\,  and \OIIIHb\, line ratios point to the absence of the outer area of low-excitation ionized gas. Hence, this region seems to be density bound and should demonstrate an efficient leakage of ionizing quanta.  Indeed, a faint diffuse tail of ionized gas is observed elongated towards the centre of \HI supershell \#3 (see Fig.~\ref{fig:HoI_stars}).

\subsubsection*{Region f}

This nebula at the north of the SGS was detected for the first time in our \Ha images. It represents a faint bubble-like
 nebula with irregular density around the star identified as BHeB of an age from 30 to 55 Myr (see Fig.~\ref{fig:HoI_stars}).

The \Ha line profiles on the rim of the nebula represent a single narrow component, while some asymmetry is observed
 towards the centre of the nebula (see profiles \#6 and 7 in Fig.~\ref{fig:profs_all}). Note, however, that the redshifted component  detected there shows a low signal-to-noise ratio and, hence, could not be considered as a confident proof of the nebula expansion.

We observed this region with the long slit in order to reveal its nature. The obtained spectrum showed that the main source of emission-line excitation should be  photoionization. The estimated age of the region is about 7 Myr. Taking into account its large size (about 300 pc), the region f seems to be an ionized bubble around a poor OB association at the late stage of its evolution, yet we have no sufficient data to definitely state this.

\subsubsection*{Regions MH7 and d (Complex 3): SNR candidates}

The compact faint shell-like region d with a size of about 60~pc observed on the inner rim of the SGS in complex \#3 draws our particular attention. Its \Ha line profile is significantly broad and clearly multicomponent (see profiles \#39, 40 in Fig.~\ref{fig:profs_c3}). 
The estimated expansion velocity of the region is $74 \kms$ that corresponds to a kinematic age of 0.2~Myr. An analysis of the spectra crossing the region d show the enhanced flux ratio of \SIIHa = 0.43--0.49 indicating high contribution of shock excitation. As we noted above, despite the region falls into the area of pure photoionization in the BPT diagrams (Fig.~\ref{fig:bpt}), its position agrees well with the models of the shocks in the low-metallicity environment.

\begin{figure}
	\includegraphics[width=\linewidth]{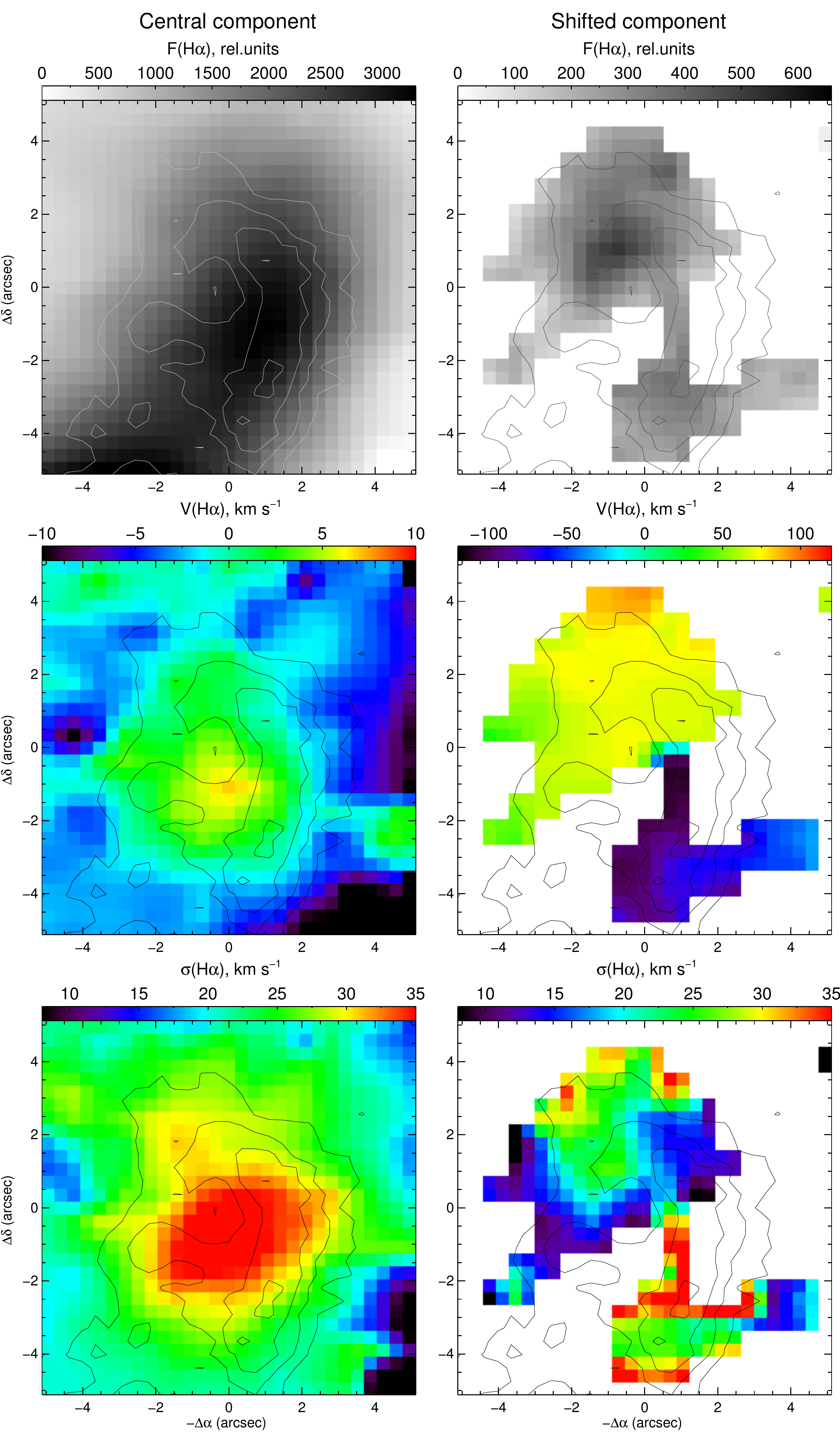}
	\caption{Region MH7: the map of the central and shifted components of the \Ha line profile (top panels), their line-of-sight velocities (middle panels), and velocity dispersions (bottom panel). The contours correspond to the \Ha intensities $(0.6, 1, 1.4, 1.8)\times10^{-16}\ \mathrm{erg\ s^{-1}\ cm^{-2}\ arcsec^{-2}}$. }\label{fig:reg7_decomp}
\end{figure}

Another region showing similar properties -- MH7 -- is located close to the region d in the same complex \#3.
 It also has a broad multicomponent \Ha line profile (\#38, 41--43 in Fig.~\ref{fig:profs_c3}) corresponding to a high
 expansion velocity ($85 \kms$) and a low kinematic age (0.3~Myr). The flux ratio of \SIIHa = 0.36--0.39 there
 is a bit lower than for the region d, yet the MH7 location in the BPT diagrams still agrees with the models of ionization by shocks.

Both regions MH7 and d stand out among all other \HII regions in the galaxy. They were classified according to
the  $I-\sigma$ diagram (Fig.~\ref{fig:isigma}) as the unique energetic objects. They appear to be the only regions in the galaxy with the clearly resolved multicomponent \Ha line profile corresponding to a high expansion velocity; they show the highest \SIIHa\, ratio in the galaxy. Because of that, we consider these regions as  good SNR candidates.

The emission in the region d might be wholly created by an old SNR, because there are no  ionizing sources revealed inside
 the region (see Fig.~\ref{fig:HoI_stars}). A more complicated situation is  in the region MH7 -- there are two O stars
 observed towards the centre of the region. The contribution of photoionization from them probably reduce the observed
 \SIIHa\ as compared with the region d.

Fig.~\ref{fig:reg7_decomp}  presents a spatially resolved kinematics of the region MH7 in details: maps of the
flux, line-of-sight velocity and, velocity dispersion of the central and shifted components are shown. It is clearly
seen that the highest velocity dispersion corresponds to a non-shifted component in the centre of this shell-like region. It is interesting that while the flux distribution of the central component sets the morphology of the region, the blue- and redshifted components are clearly separated and dominate in  different parts of MH7. We can explain this fact by  a high inhomogeneity of the \HI density in the region. 
Indeed, the local minimum of the \HI density is observed towards the centre of MH7, while it grows up at its edges.
 Hence, the shifted components may correspond to the expanding bubble, whose receding side has a higher density in its northern part, while the approaching side -- in the southern part.

Given the measured optical size and the expansion velocity of both SNR candidates d and MH7 (see Table~\ref{tab:shells}), we can derive their energy of explosion necessary to match the observed properties of the nebulae. A high expansion velocity suggests that the SNR still undergoes adiabatic expansion.	According to the self-similar solution of \citet{sedov}, the evolution of an SNR at this stage can be described by the following equation:
$$
	R_{s}=13.5(E_{51}/n_{o})^{0.2} (t/10^{4} \mathrm{yr})^{0.4} \ (\mathrm{pc}),
$$
	where $E_{51} = E_{o}/10^{51}\ \mathrm{erg}$ is the  explosion energy, $R_s$ is the  radius, $t$ is the  age of an SNR. The expansion velocity of an SNR evolves as $v_\mathrm{exp}=0.4 R_{s}/t$ during this stage.
	The unperturbed density  $n_{o} \simeq 0.35 \mathrm{cm^{-3}}$ is derived from the \HI data (helium contribution is taken into account). Using these relations, we obtained $E_{51} \simeq 0.07$ and 0.24 for the regions d and MH7, respectively, proposing the adiabatic stage of expansion.
	If a SNR currently undergo the post-adiabatic snow-plough stage of evolution, we can use the analysis performed by \citet{Chevalier1974}. According to this work,
$$E_{51} = 5.3\times 10^{-8} n_{o}^{1.12} v_{\mathrm{exp}}^{1.4}(\kms) R_s^{3.12}(\mathrm{pc}),
$$
that gives us $E_{51} \simeq 0.36$ and 0.87 for the regions d and MH7.
	
Such low values of $E_{51}$ derived from our analysis do not contradict  the hypothesis of the SNR nature of the objects d and MH7. Indeed, according to modern understanding, supernovae having even much lower explosion energy are observed \citep[see, e.g.,][]{Benetti2016, Chugai2016, Utrobin2017}. In addition, high radiation loses can also be responsible for the observed values of the SNR energy much lower, than the initial explosion energy \citep{Sharma2014, Vasiliev2015b}.

It is interesting that despite the fact that these two regions appear to be good SNR candidates, they do not exhibit any bright emission
 in the radio continuum. Only a faint source of the non-thermal radio continuum could be identified between the regions MH7 and d in the images presented by \cite{Hindson18}. This fact does not contradict 
 their identification as SNRs. As  follows from the analysis of detections of an extragalactic SNR performed by 
\cite{Bozzetto2017}, optical SNR candidates statistically tend not to be detected in the radio or X rays. Probably, deeper
 radio observations are needed to clarify the nature of the  MH7 and d regions.

\section{Discussion}\label{sec:discuss}

In this paper we present the deepest \Ha images of Holmberg~I  available to date, as well as a detailed study of the ionized gas kinematics. These data allow us to treat star formation in the galaxy as a process taking place in large (several hundreds of pc) unified complexes presenting the same age and gas kinematics
within them. All the individual \HII regions within a complex are tied together with faint filaments or diffuse clouds of ionized gas. 
We state that star formation in  Holmberg~I is a self-regulating process at the scales of such complexes, while the external factors like the SGS expansion influence  the whole star formation in the galaxy.

All regions of the ongoing star formation in Holmberg~I, similarly to other dIrr galaxies with SGSs, are located on the rim of the central \HI SGS in the areas of high \HI density. This is natural in the framework of the generally accepted theory of star formation, propagated within an expanding supershell. Complex \#2 represents another example of the same process but on a smaller scale. The bright nebulae of this complex are located in the walls of the local \HI supershell \#7.

Another well known mechanism of triggering of star formation is the collision of several expanding supershells. It was reproduced in numerical and analytical models by different authors \citep[e.g.,][]{Chernin1995, Ntormousi2011, Kawata2014, Vasiliev2017} and observed in a number of nearby galaxies \citep[e.g.][]{Lozinskaya2002, Egorov2017}. Several facts point to the conclusion that the supershell collision is a mechanism that triggered the ongoing star formation in the complex NW. Its related \HI supershell \#1  has the same age as the SGS (see Table~\ref{tab:shells}) which points to their previous independent evolution. Currently, the rims of both supershells are in contact, and the ionized gas is distributed between them. Moreover, the distribution of young stars of different ages (see Fig.~\ref{fig:HoI_stars}) inside the complex NW shows a gradient from the  centre of  \HI supershell \#1 towards its northeastern part that is the evidence for its independent evolution by the influence of stellar feedback from these stars. Such a gradient is absent in the southeastern part of the complex (towards the SGS) that allows one to propose  triggering of star formation there as a result of the supershell collision. \cite{Warren2012} have found several clumps of the cold low-dispersion \HI gas in Holmberg~I. The most extended of them are located on the rims of the NW shell (in particular, between  supershell \#1 and the SGS) 
that also points to the ongoing cooling and compression of the ISM that stimulates star formation.

\begin{figure}
	\includegraphics[width=0.93\linewidth]{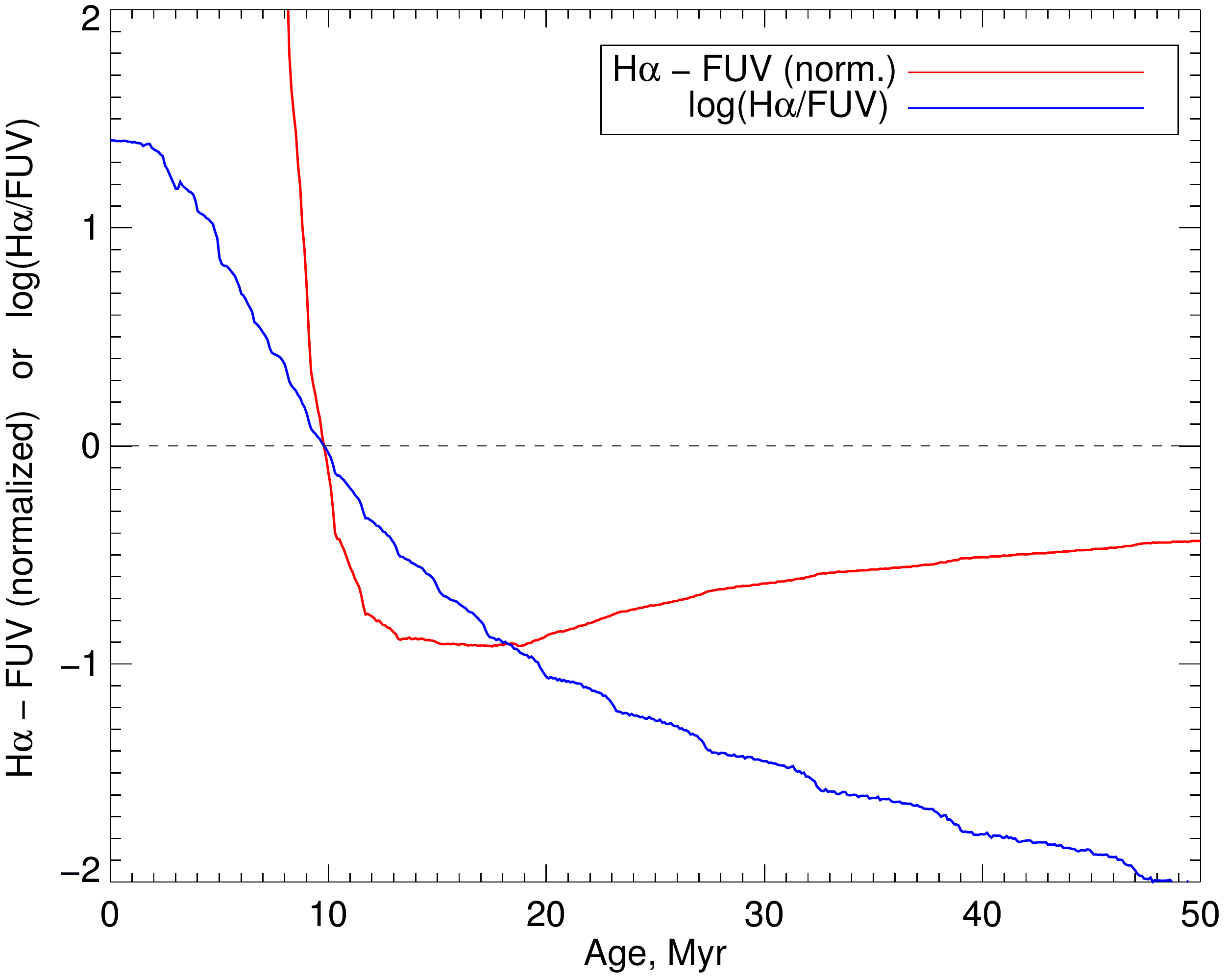}
	\caption{Evolution of the \Ha and FUV fluxes logarithm ratio (blue curve) and difference (red curve) with the age of the cluster according to STARBURST99 model. The flux difference is normalized to one-third of the maximum FUV flux over the whole timescale of the model in order to fit both curves to the same scale.}\label{fig:ha_fuv_model}
\end{figure}

Triggering of star formation by colliding shells can take place also for several other regions in the galaxy. Thus, the \HII regions MH10 and MH26 are located at the intersections of several \HI supershells. The first one probably formed in the collision of quite young \HI supersells \#2, 3 with an older one \#11. A collision of two old \HI supershells of almost an equal age -- SGS and \#8 -- can stimulate star formation in the MH26 region. Young OB stars are also observed in the intersection of the SGS and \HI supershell \#3 (see Fig.~\ref{fig:HoI_stars}).

Taking into account the old age of supershells \#9 and \#10 together with their location relative to the SGS, we may also propose that  star formation in \HII complexes \#3 and \#4 was also triggered by the collision of those shells with the SGS several Myr ago. However, we cannot exclude now an opposite scenario -- that these supershells on the rim of the SGS were produced by feedback of supernovae and stellar winds in the mentioned complexes supposing that star formation in them was taking place at least for several tens of Myr. The last scenario is supported by a large number of BHeB stars with ages of 50--200 Myr inside  supershell \#9, while only few of such stars are observed inside  supershell \#10.

\begin{figure}
	\includegraphics[width=\linewidth]{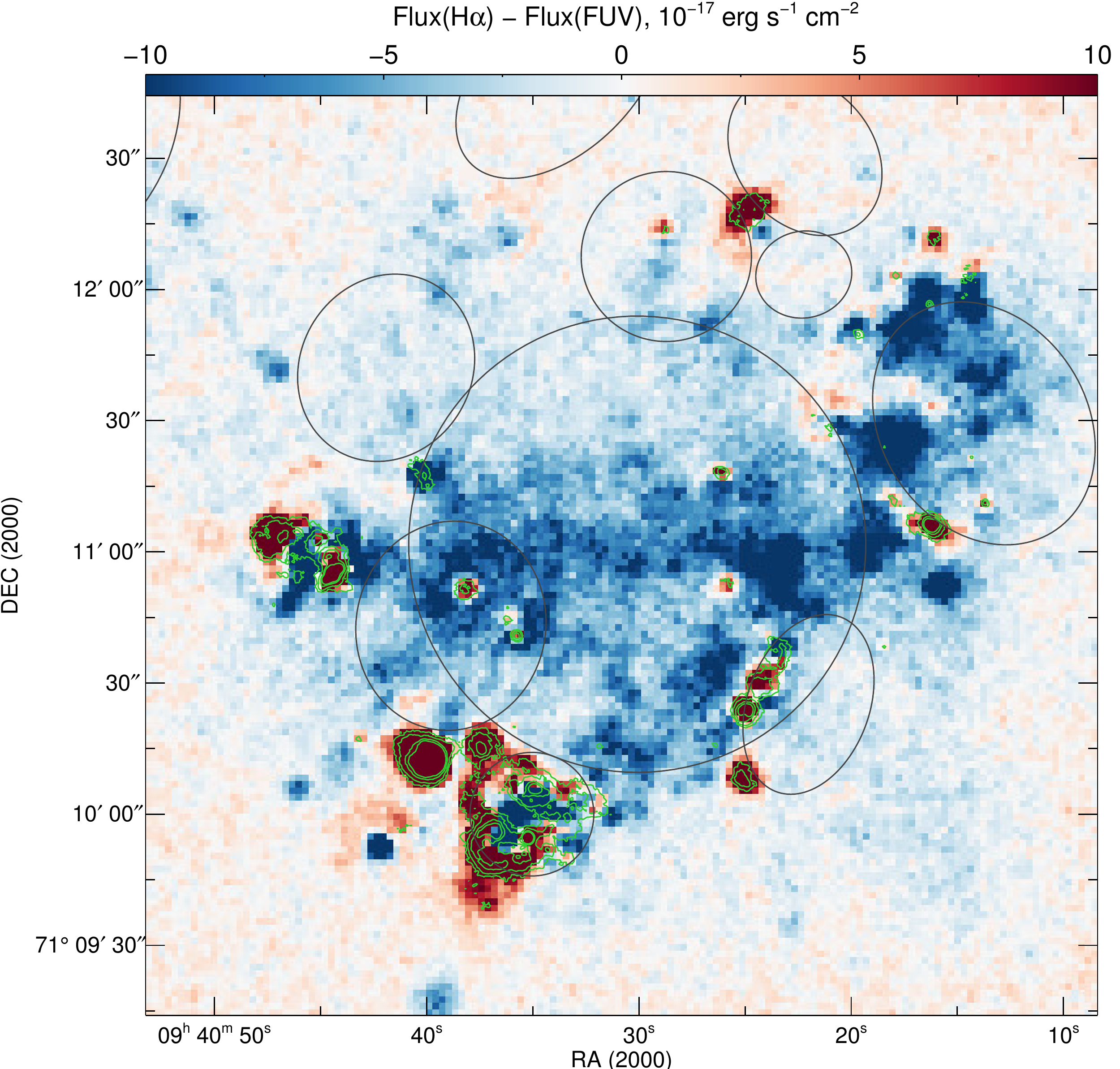}
	\caption{Map of the \Ha and FUV fluxes difference for Holmberg~I. The red colour denotes the regions with the dominated \Ha emission, while the blue colour traces the ones with enhanced contribution of the FUV emission. Green contours correspond to the \Ha intensities $(0.75, 1.5, 3, 4.5)\times10^{-16}\ \mathrm{erg\ s^{-1}\ cm^{-2}\ arcsec^{-2}}$. Gray ellipses denote the location of the detected \HI supershells. }\label{fig:ha_fuv_map}
\end{figure}

The signs of star-formation propagation and triggering in Holmberg~I are shown not only by the distribution of stars of different ages, but also by the comparison between \Ha and FUV images.
Both \Ha and FUV emission trace the young stellar population in galaxies but they dominate at  different ages. The \Ha emission is a good tracer of the current star formation (age < 10 Myr), while the FUV emission traces the recent star formation that took place on longer timescales (age < 100 Myr).
Fig.~\ref{fig:ha_fuv_model}  shows the evolution of the H$\alpha$/FUV flux ratio as well as their difference according to the \mbox{STARBURST99} simulations \citep{Leitherer1999} for the Holmberg~I metallicity $Z = 0.2Z\odot$. The red line traces the change of the normalized \Ha -- FUV flux difference with the age of the star cluster. This difference is positive at  ages less than about 9 Myr and negative for  older regions. The map of \Ha -- FUV (Fig.~\ref{fig:ha_fuv_map}) exhibits the regions with a negative value (older than 9 Myr) in blue  and with a positive value (younger than 9 Myr) in red. Hence, red colour in this image corresponds to the regions of the ongoing star formation, while blue regions -- to the areas where star formation has recently occurred. In many places, especially in complexes \#1, 2, and NW, the blue area is closely bordered by a younger red area, that points to star formation propagation within the complex.

The morphology of ionized and neutral gas together with the young-star distribution allow us to make some speculative assumptions on further evolution of the SGS.
Given a large number of \HII regions and young massive stars, we may expect a gradual fragmentation and destruction of the SGS in the area of its interaction with  supershell \#1 (NW complex), where the observed \HI density is lower than in other parts of the SGS. This process will form an open tunnel in the \HI distribution like it is already observed in the eastern part of the galaxy (towards supershell \#6).

As we have mentioned before, the origin of two expanding supershells \#9 and \#10 observed on the rim of the SGS is unclear now. But in any case, their presence demonstrates a current instability of the SGS there. If they were formed on the rim of the SGS during the late stages of its evolution, then these supershells could be considered as a consequence of the ongoing fragmentation of the SGS driven by stellar feedback. Or else, if they have evolved independently of the SGS, we now observe a process of dissipation of these two supershells as a result of their collision with the SGS -- this process should destabilize the SGS as well.


Summing up, we observe a propagation of star formation in Holmberg~I, and in some places it is triggered by the interaction of several \HI supershells. In its turn, the ongoing star formation on the rims of the SGS gradually leads to its destruction.

\section{Summary}\label{sec:summary}

We present a detailed multiwavelength analysis of the regions of star formation in the nearby dIrr galaxy Holmberg~I. 
Since all the \HII regions in the galaxy are located inside the galaxy-sized \HI supergiant shell,  we focused
 on  investigation of star-formation triggering on the rims of the SGS, as well as on the role of star
 formation in the evolution of the SGS.

 Our work was based on the data set obtained with the SCORPIO and SCORPIO-2 focal reducers at the 6-m SAO RAS telescope:
 deep H$\alpha$, [S~\textsc{ii}], and \OIII narrow-band images of the galaxy, four long-slit spectra crossing most of the \HII regions, 
and the high spectral resolution \Ha data cube obtained with the scanning Fabry--Perot interferometer. These data were
 complemented with the archival data in \HI 21-cm (VLA) and FUV (\textit{GALEX}) from the LITTLE THINGS archive, as well
 as by the photometric data of individual stars observed 
with the  \HST (ANGST archive). Using all these data, we performed an analysis of the ionized and neutral gas morphology and kinematics; described the  emission spectrum of \HII regions in terms of their excitation mechanism and metallicity; identified young massive stars and compared their energetics and distribution with  properties of the observed ionized gas; traced the spatial distribution of the ongoing and recent star formation in the galaxy.
Below we highlight the most important results obtained.

\begin{itemize}

\item Four expanding \HI supershells and 9 new faint \HII regions were found and analysed in addition to the previously known.

\item We show that star formation in the galaxy takes place in five large unified complexes rather than in the individual
 \HII regions. The \HII regions in these complexes are tied together by faint filaments of ionized gas or embedded into  common clouds of diffuse ionized gas. 
	
	\item Triggering of star formation on  scales of individual complexes, as well as of the whole galaxy, is observed. Most  \HII regions surround places of the recent star formation. In the northwestern complex, we also observed a gradient in the age distribution of individual stars. Also our observations point to the fact that the current star formation there was triggered by the collision of the SGS with other \HI supershells.

	\item The ionized gas distribution in the galaxy reveals 18 shell-like structures. Most of them appear to be expanding superbubbles (according to their enhanced velocity dispersion), yet only a half demonstrate weak high-velocity components in the \Ha line profile, and only two of them show a clearly multicomponent line profile.
	
	\item We identified two SNR candidates and one WR star in the galaxy. One object with the bright radio continuum emission previously proposed as a possible SNR candidate turned out to be a background quasar.
	
	 \item A large fraction of the escaping ionizing radiation from each \HII complex is observed. 
It points to the expected large amount of diffuse ionized gas  which appears to be only partially detected in our observations.
	
	 \item The presence of younger \HI supershells of a smaller size on the rims of the SGS 
together with a large fraction of escaping ionizing quanta allows us to propose that currently we observe a process
 of SGS fragmentation and destruction by the influence of  star formation  on its rim.
	
\end{itemize}

\section*{Acknowledgements}

We thank the anonymous referee for detailed comments that helped us to improved the clarity of this manuscript.

This work is based on the observations obtained with the 6-m telescope of the
Special Astrophysical Observatory of the Russian Academy of Sciences carried out with the financial
support of the Ministry of Education and Science of the Russian Federation
(agreement No. 14.619.21.0004, project ID RFMEFI61914X0004).
We thank A. Burenkov, D. Oparin, and R. Uklein for the assistance in 6-m telescope observations.

The study was supported by the Russian
Foundation for Basic Research (RFBR), project 18-02-00976. OE is also thankful to the European Astronomical Society for a travel grant for participation in the EWASS-2017 conference, where the first results of this work have been presented.

The study is partially based on the observations made with the NASA/ESA \textit{Hubble Space Telescope} obtained from the Data Archive at the Space Telescope Science Institute which is operated by the Association of Universities for Research in Astronomy, Inc., under the NASA contract NAS 5-26555. These observations are associated with program \# 10605.

\bibliographystyle{mnras}
\bibliography{HoI_aph}

\begin{thebibliography}{}
\makeatletter
\relax
\def\mn@urlcharsother{\let\do\@makeother \do\$\do\&\do\#\do\^\do\_\do\%\do\~}
\def\mn@doi{\begingroup\mn@urlcharsother \@ifnextchar [ {\mn@doi@}
  {\mn@doi@[]}}
\def\mn@doi@[#1]#2{\def\@tempa{#1}\ifx\@tempa\@empty \href
  {http://dx.doi.org/#2} {doi:#2}\else \href {http://dx.doi.org/#2} {#1}\fi
  \endgroup}
\def\mn@eprint#1#2{\mn@eprint@#1:#2::\@nil}
\def\mn@eprint@arXiv#1{\href {http://arxiv.org/abs/#1} {{\tt arXiv:#1}}}
\def\mn@eprint@dblp#1{\href {http://dblp.uni-trier.de/rec/bibtex/#1.xml}
  {dblp:#1}}
\def\mn@eprint@#1:#2:#3:#4\@nil{\def\@tempa {#1}\def\@tempb {#2}\def\@tempc
  {#3}\ifx \@tempc \@empty \let \@tempc \@tempb \let \@tempb \@tempa \fi \ifx
  \@tempb \@empty \def\@tempb {arXiv}\fi \@ifundefined
  {mn@eprint@\@tempb}{\@tempb:\@tempc}{\expandafter \expandafter \csname
  mn@eprint@\@tempb\endcsname \expandafter{\@tempc}}}

\bibitem[\protect\citeauthoryear{{Afanasiev} \& {Moiseev}}{{Afanasiev} \&
  {Moiseev}}{2005}]{scorpio}
{Afanasiev} V.~L.,  {Moiseev} A.~V.,  2005, \mn@doi [Astronomy Letters]
  {10.1134/1.1883351}, \href
  {http://adsabs.harvard.edu/abs/2005AstL...31..194A} {31, 194}

\bibitem[\protect\citeauthoryear{{Afanasiev} \& {Moiseev}}{{Afanasiev} \&
  {Moiseev}}{2011}]{scorpio2}
{Afanasiev} V.~L.,  {Moiseev} A.~V.,  2011, Baltic Astronomy, \href
  {http://adsabs.harvard.edu/abs/2011BaltA..20..363A} {20, 363}

\bibitem[\protect\citeauthoryear{Allen, Groves, Dopita, Sutherland  \&
  Kewley}{Allen et~al.}{2008}]{Allen2008}
Allen M.~G.,  Groves B.~A.,  Dopita M.~A.,  Sutherland R.~S.,   Kewley L.~J.,
  2008, \apjs, 178, 20

\bibitem[\protect\citeauthoryear{Bagetakos, Brinks, Walter, de Blok, Usero,
  Leroy, Rich  \& Kennicutt}{Bagetakos et~al.}{2011}]{Bagetakos2011}
Bagetakos I.,  Brinks E.,  Walter F.,  de Blok W. J.~G.,  Usero A.,  Leroy
  A.~K.,  Rich J.~W.,   Kennicutt R.~C.,  2011, \aj, 141, 23

\bibitem[\protect\citeauthoryear{{Baldwin}, {Phillips}  \&
  {Terlevich}}{{Baldwin} et~al.}{1981}]{BPT}
{Baldwin} J.~A.,  {Phillips} M.~M.,   {Terlevich} R.,  1981, \mn@doi [\pasp]
  {10.1086/130766}, \href {http://adsabs.harvard.edu/abs/1981PASP...93....5B}
  {93, 5}

\bibitem[\protect\citeauthoryear{Bastian et~al.,}{Bastian
  et~al.}{2011}]{Bastian2011}
Bastian N.,  et~al., 2011, \mnras, 412, 1539

\bibitem[\protect\citeauthoryear{{Benetti} et~al.,}{{Benetti}
  et~al.}{2016}]{Benetti2016}
{Benetti} S.,  et~al., 2016, \mn@doi [\mnras] {10.1093/mnras/stv2811}, \href
  {http://adsabs.harvard.edu/abs/2016MNRAS.456.3296B} {456, 3296}

\bibitem[\protect\citeauthoryear{{Berg} et~al.,}{{Berg}
  et~al.}{2012}]{Berg2012}
{Berg} D.~A.,  et~al., 2012, \mn@doi [\apj] {10.1088/0004-637X/754/2/98}, \href
  {http://adsabs.harvard.edu/abs/2012ApJ...754...98B} {754, 98}

\bibitem[\protect\citeauthoryear{{Bozzetto} et~al.,}{{Bozzetto}
  et~al.}{2017}]{Bozzetto2017}
{Bozzetto} L.~M.,  et~al., 2017, \mn@doi [\apjs] {10.3847/1538-4365/aa653c},
  \href {http://adsabs.harvard.edu/abs/2017ApJS..230....2B} {230, 2}

\bibitem[\protect\citeauthoryear{{Bressan}, {Marigo}, {Girardi}, {Salasnich},
  {Dal Cero}, {Rubele}  \& {Nanni}}{{Bressan} et~al.}{2012}]{bressan12}
{Bressan} A.,  {Marigo} P.,  {Girardi} L.,  {Salasnich} B.,  {Dal Cero} C.,
  {Rubele} S.,   {Nanni} A.,  2012, \mn@doi [\mnras]
  {10.1111/j.1365-2966.2012.21948.x}, \href
  {http://adsabs.harvard.edu/abs/2012MNRAS.427..127B} {427, 127}

\bibitem[\protect\citeauthoryear{Cannon et~al.,}{Cannon
  et~al.}{2011a}]{Cannon2011a}
Cannon J.~M.,  et~al., 2011a, \apj, 735, 35

\bibitem[\protect\citeauthoryear{Cannon et~al.,}{Cannon
  et~al.}{2011b}]{Cannon2011b}
Cannon J.~M.,  et~al., 2011b, \apj, 735, 36

\bibitem[\protect\citeauthoryear{Cardelli, Clayton  \& Mathis}{Cardelli
  et~al.}{1989}]{Cardelli1989}
Cardelli J.~A.,  Clayton G.~C.,   Mathis J.~S.,  1989, \apj, 345, 245

\bibitem[\protect\citeauthoryear{{Chen}, {Girardi}, {Bressan}, {Marigo},
  {Barbieri}  \& {Kong}}{{Chen} et~al.}{2014}]{chen14}
{Chen} Y.,  {Girardi} L.,  {Bressan} A.,  {Marigo} P.,  {Barbieri} M.,   {Kong}
  X.,  2014, \mn@doi [\mnras] {10.1093/mnras/stu1605}, \href
  {http://adsabs.harvard.edu/abs/2014MNRAS.444.2525C} {444, 2525}

\bibitem[\protect\citeauthoryear{{Chen}, {Bressan}, {Girardi}, {Marigo}, {Kong}
   \& {Lanza}}{{Chen} et~al.}{2015}]{chen15}
{Chen} Y.,  {Bressan} A.,  {Girardi} L.,  {Marigo} P.,  {Kong} X.,   {Lanza}
  A.,  2015, \mn@doi [\mnras] {10.1093/mnras/stv1281}, \href
  {http://adsabs.harvard.edu/abs/2015MNRAS.452.1068C} {452, 1068}

\bibitem[\protect\citeauthoryear{Chernin, Efremov  \& Voinovich}{Chernin
  et~al.}{1995}]{Chernin1995}
Chernin A.~D.,  Efremov Y.~N.,   Voinovich P.~A.,  1995, \mnras, 275, 313

\bibitem[\protect\citeauthoryear{{Chevalier}}{{Chevalier}}{1974}]{Chevalier1974}
{Chevalier} R.~A.,  1974, \mn@doi [\apj] {10.1086/152740}, \href
  {http://adsabs.harvard.edu/abs/1974ApJ...188..501C} {188, 501}

\bibitem[\protect\citeauthoryear{{Chugai}}{{Chugai}}{2016}]{Chugai2016}
{Chugai} N.~N.,  2016, \mn@doi [Astronomy Letters] {10.1134/S106377371602002X},
  \href {http://adsabs.harvard.edu/abs/2016AstL...42...82C} {42, 82}

\bibitem[\protect\citeauthoryear{{Copetti}, {Pastoriza}  \&
  {Dottori}}{{Copetti} et~al.}{1986}]{Copetti1986}
{Copetti} M.~V.~F.,  {Pastoriza} M.~G.,   {Dottori} H.~A.,  1986, \aap, \href
  {http://adsabs.harvard.edu/abs/1986A%26A...156..111C} {156, 111}

\bibitem[\protect\citeauthoryear{{Crowther} \& {Smith}}{{Crowther} \&
  {Smith}}{1997}]{Crowther1997}
{Crowther} P.~A.,  {Smith} L.~J.,  1997, \aap, \href
  {http://adsabs.harvard.edu/abs/1997A%26A...320..500C} {320, 500}

\bibitem[\protect\citeauthoryear{Croxall, van Zee, Lee, Skillman, Lee,
  C{\^o}t{\'e}, Kennicutt  \& Miller}{Croxall et~al.}{2009}]{Croxall2009}
Croxall K.~V.,  van Zee L.,  Lee H.,  Skillman E.~D.,  Lee J.~C.,  C{\^o}t{\'e}
  S.,  Kennicutt R.~C.,   Miller B.~W.,  2009, \apj, 705, 723

\bibitem[\protect\citeauthoryear{{Dalcanton} et~al.,}{{Dalcanton}
  et~al.}{2009}]{Dalcanton2009}
{Dalcanton} J.~J.,  et~al., 2009, \mn@doi [\apjs] {10.1088/0067-0049/183/1/67},
  \href {http://adsabs.harvard.edu/abs/2009ApJS..183...67D} {183, 67}

\bibitem[\protect\citeauthoryear{{Dohm-Palmer} et~al.,}{{Dohm-Palmer}
  et~al.}{1997}]{Dohm-Palmer1997}
{Dohm-Palmer} R.~C.,  et~al., 1997, \mn@doi [\aj] {10.1086/118665}, \href
  {http://adsabs.harvard.edu/abs/1997AJ....114.2527D} {114, 2527}

\bibitem[\protect\citeauthoryear{Egorov, Lozinskaya, Moiseev  \&
  Smirnov-Pinchukov}{Egorov et~al.}{2014}]{Egorov2014}
Egorov O.~V.,  Lozinskaya T.~A.,  Moiseev A.~V.,   Smirnov-Pinchukov G.~V.,
  2014, \mnras, 444, 376

\bibitem[\protect\citeauthoryear{Egorov, Lozinskaya, Moiseev  \&
  Shchekinov}{Egorov et~al.}{2017}]{Egorov2017}
Egorov O.~V.,  Lozinskaya T.~A.,  Moiseev A.~V.,   Shchekinov Y.~A.,  2017,
  \mnras, 464, 1833

\bibitem[\protect\citeauthoryear{Fitzpatrick}{Fitzpatrick}{1999}]{Fitzpatrick1999}
Fitzpatrick E.~L.,  1999, \pasp, 111, 63

\bibitem[\protect\citeauthoryear{Guerrero, Villaver  \& Manchado}{Guerrero
  et~al.}{1998}]{Guerrero1998}
Guerrero M.~A.,  Villaver E.,   Manchado A.,  1998, \apj, 507, 889

\bibitem[\protect\citeauthoryear{{Hindson} et~al.,}{{Hindson}
  et~al.}{2018}]{Hindson18}
{Hindson} L.,  et~al., 2018, \mn@doi [\apjs] {10.3847/1538-4365/aaa42c}, \href
  {http://adsabs.harvard.edu/abs/2018ApJS..234...29H} {234, 29}

\bibitem[\protect\citeauthoryear{{Hoessel} \& {Danielson}}{{Hoessel} \&
  {Danielson}}{1984}]{Hoessel1984}
{Hoessel} J.~G.,  {Danielson} G.~E.,  1984, \mn@doi [\apj] {10.1086/162584},
  \href {http://adsabs.harvard.edu/abs/1984ApJ...286..159H} {286, 159}

\bibitem[\protect\citeauthoryear{Hoopes \& Walterbos}{Hoopes \&
  Walterbos}{2003}]{Hoopes2003}
Hoopes C.~G.,  Walterbos R. A.~M.,  2003, /apj, 586, 902

\bibitem[\protect\citeauthoryear{{Hunter}, {Elmegreen}  \& {Ludka}}{{Hunter}
  et~al.}{2010}]{Hunter2010}
{Hunter} D.~A.,  {Elmegreen} B.~G.,   {Ludka} B.~C.,  2010, \mn@doi [\aj]
  {10.1088/0004-6256/139/2/447}, \href
  {http://adsabs.harvard.edu/abs/2010AJ....139..447H} {139, 447}

\bibitem[\protect\citeauthoryear{Hunter et~al.,}{Hunter
  et~al.}{2012}]{Hunter2012}
Hunter D.~A.,  et~al., 2012, The Astronomical Journal, 144, 134

\bibitem[\protect\citeauthoryear{{Karachentsev} \& {Kaisin}}{{Karachentsev} \&
  {Kaisin}}{2007}]{Karach2007}
{Karachentsev} I.~D.,  {Kaisin} S.~S.,  2007, \mn@doi [\aj] {10.1086/512127},
  \href {http://adsabs.harvard.edu/abs/2007AJ....133.1883K} {133, 1883}

\bibitem[\protect\citeauthoryear{{Karachentsev}, {Makarov}  \&
  {Kaisina}}{{Karachentsev} et~al.}{2013}]{Karach2013}
{Karachentsev} I.~D.,  {Makarov} D.~I.,   {Kaisina} E.~I.,  2013, \mn@doi [\aj]
  {10.1088/0004-6256/145/4/101}, \href
  {http://adsabs.harvard.edu/abs/2013AJ....145..101K} {145, 101}

\bibitem[\protect\citeauthoryear{{Kauffmann} et~al.,}{{Kauffmann}
  et~al.}{2003}]{Kauffmann2003}
{Kauffmann} G.,  et~al., 2003, \mn@doi [\mnras]
  {10.1111/j.1365-2966.2003.07154.x}, \href
  {http://adsabs.harvard.edu/abs/2003MNRAS.346.1055K} {346, 1055}

\bibitem[\protect\citeauthoryear{{Kawata}, {Gibson}, {Barnes}, {Grand}  \&
  {Rahimi}}{{Kawata} et~al.}{2014}]{Kawata2014}
{Kawata} D.,  {Gibson} B.~K.,  {Barnes} D.~J.,  {Grand} R.~J.~J.,   {Rahimi}
  A.,  2014, \mn@doi [\mnras] {10.1093/mnras/stt2267}, \href
  {http://adsabs.harvard.edu/abs/2014MNRAS.438.1208K} {438, 1208}

\bibitem[\protect\citeauthoryear{{Kewley} \& {Ellison}}{{Kewley} \&
  {Ellison}}{2008}]{Kewley2008}
{Kewley} L.~J.,  {Ellison} S.~L.,  2008, \mn@doi [\apj] {10.1086/587500}, \href
  {http://adsabs.harvard.edu/abs/2008ApJ...681.1183K} {681, 1183}

\bibitem[\protect\citeauthoryear{{Kewley}, {Dopita}, {Sutherland}, {Heisler}
  \& {Trevena}}{{Kewley} et~al.}{2001}]{Kewley2001}
{Kewley} L.~J.,  {Dopita} M.~A.,  {Sutherland} R.~S.,  {Heisler} C.~A.,
  {Trevena} J.,  2001, \mn@doi [\apj] {10.1086/321545}, \href
  {http://adsabs.harvard.edu/abs/2001ApJ...556..121K} {556, 121}

\bibitem[\protect\citeauthoryear{{Koleva}, {Prugniel}, {Bouchard}  \&
  {Wu}}{{Koleva} et~al.}{2009}]{Koleva2009}
{Koleva} M.,  {Prugniel} P.,  {Bouchard} A.,   {Wu} Y.,  2009, \mn@doi [\aap]
  {10.1051/0004-6361/200811467}, \href
  {http://adsabs.harvard.edu/abs/2009A%26A...501.1269K} {501, 1269}

\bibitem[\protect\citeauthoryear{{Leitherer} et~al.,}{{Leitherer}
  et~al.}{1999}]{Leitherer1999}
{Leitherer} C.,  et~al., 1999, \mn@doi [\apjs] {10.1086/313233}, \href
  {http://adsabs.harvard.edu/abs/1999ApJS..123....3L} {123, 3}

\bibitem[\protect\citeauthoryear{{Levesque} \& {Leitherer}}{{Levesque} \&
  {Leitherer}}{2013}]{Levesque2013}
{Levesque} E.~M.,  {Leitherer} C.,  2013, \mn@doi [\apj]
  {10.1088/0004-637X/779/2/170}, \href
  {http://adsabs.harvard.edu/abs/2013ApJ...779..170L} {779, 170}

\bibitem[\protect\citeauthoryear{{L{\'o}pez-S{\'a}nchez}, {Dopita}, {Kewley},
  {Zahid}, {Nicholls}  \& {Scharw{\"a}chter}}{{L{\'o}pez-S{\'a}nchez}
  et~al.}{2012}]{Lopez-Sanchez2012}
{L{\'o}pez-S{\'a}nchez} {\'A}.~R.,  {Dopita} M.~A.,  {Kewley} L.~J.,  {Zahid}
  H.~J.,  {Nicholls} D.~C.,   {Scharw{\"a}chter} J.,  2012, \mn@doi [\mnras]
  {10.1111/j.1365-2966.2012.21145.x}, \href
  {http://adsabs.harvard.edu/abs/2012MNRAS.426.2630L} {426, 2630}

\bibitem[\protect\citeauthoryear{{Lozinskaya}}{{Lozinskaya}}{2002}]{Lozinskaya2002}
{Lozinskaya} T.~A.,  2002, \mn@doi [Astronomical and Astrophysical
  Transactions] {10.1080/10556790215592}, \href
  {http://adsabs.harvard.edu/abs/2002A%26AT...21..223L} {21, 223}

\bibitem[\protect\citeauthoryear{{Markwardt}}{{Markwardt}}{2009}]{mpfit}
{Markwardt} C.~B.,  2009, in {Bohlender} D.~A.,  {Durand} D.,   {Dowler} P.,
  eds,  Astronomical Society of the Pacific Conference Series Vol. 411,
  Astronomical Data Analysis Software and Systems XVIII. p.~251 (\mn@eprint
  {arXiv} {0902.2850})

\bibitem[\protect\citeauthoryear{{Martins}, Schaerer  \& Hillier}{{Martins}
  et~al.}{2006}]{Martins2006}
{Martins} F.,  Schaerer D.,   Hillier D.~J.,  2006, \aap, 436, 1049

\bibitem[\protect\citeauthoryear{{McGaugh}}{{McGaugh}}{1991}]{McGaugh1991}
{McGaugh} S.~S.,  1991, \mn@doi [\apj] {10.1086/170569}, \href
  {http://adsabs.harvard.edu/abs/1991ApJ...380..140M} {380, 140}

\bibitem[\protect\citeauthoryear{McQuinn, Skillman, Cannon, Dalcanton, Dolphin,
  Stark  \& Weisz}{McQuinn et~al.}{2009}]{Mcquinn2009}
McQuinn K. B.~W.,  Skillman E.~D.,  Cannon J.~M.,  Dalcanton J.~J.,  Dolphin
  A.~E.,  Stark D.,   Weisz D.~R.,  2009, \apj, 695, 561

\bibitem[\protect\citeauthoryear{McQuinn et~al.,}{McQuinn
  et~al.}{2010a}]{Mcquinn2010a}
McQuinn K. B.~W.,  et~al., 2010a, \apj, 721, 297

\bibitem[\protect\citeauthoryear{McQuinn et~al.,}{McQuinn
  et~al.}{2010b}]{Mcquinn2010b}
McQuinn K. B.~W.,  et~al., 2010b, \apj, 724, 49

\bibitem[\protect\citeauthoryear{Miller \& Hodge}{Miller \& Hodge}{1994}]{MH94}
Miller B.~W.,  Hodge P.~W.,  1994, \apj, 427, 656

\bibitem[\protect\citeauthoryear{Miller \& Hodge}{Miller \& Hodge}{1996}]{MH96}
Miller B.~W.,  Hodge P.~W.,  1996, \apj, 458, 467

\bibitem[\protect\citeauthoryear{{Moiseev}}{{Moiseev}}{2002}]{Moiseev2002}
{Moiseev} A.~V.,  2002, Bulletin of the Special Astrophysics Observatory, \href
  {http://adsabs.harvard.edu/abs/2002BSAO...54...74M} {54, 74}

\bibitem[\protect\citeauthoryear{{Moiseev}}{{Moiseev}}{2015}]{Moiseev2015}
{Moiseev} A.~V.,  2015, \mn@doi [Astrophysical Bulletin]
  {10.1134/S1990341315040112}, \href
  {http://adsabs.harvard.edu/abs/2015AstBu..70..494M} {70, 494}

\bibitem[\protect\citeauthoryear{{Moiseev} \& {Egorov}}{{Moiseev} \&
  {Egorov}}{2008}]{Moiseev2008}
{Moiseev} A.~V.,  {Egorov} O.~V.,  2008, \mn@doi [Astrophysical Bulletin]
  {10.1134/S1990341308020089}, \href
  {http://adsabs.harvard.edu/abs/2008AstBu..63..181M} {63, 181}

\bibitem[\protect\citeauthoryear{Moiseev \& Lozinskaya}{Moiseev \&
  Lozinskaya}{2012}]{Moiseev2012}
Moiseev A.~V.,  Lozinskaya T.~A.,  2012, \mnras, 423, 1831

\bibitem[\protect\citeauthoryear{Moustakas, Kennicutt~Jr., Tremonti, Dale,
  Smith  \& Calzetti}{Moustakas et~al.}{2010}]{Moustakas2010}
Moustakas J.,  Kennicutt~Jr. R.~C.,  Tremonti C.~A.,  Dale D.~A.,  Smith
  J.-D.~T.,   Calzetti D.,  2010, \apjs, 190, 233

\bibitem[\protect\citeauthoryear{{Mu\~noz-Tu\~non}, {Tenorio-Tagle},
  {Casta\~neda}  \& {Terlevich}}{{Mu\~noz-Tu\~non}
  et~al.}{1996}]{MunozTunon1996}
{Mu\~noz-Tu\~non} C.,  {Tenorio-Tagle} G.,  {Casta\~neda} H.~O.,   {Terlevich}
  R.,  1996, \mn@doi [\aj] {10.1086/118129}, \href
  {http://adsabs.harvard.edu/abs/1996AJ....112.1636M} {112, 1636}

\bibitem[\protect\citeauthoryear{Ntormousi, Burkert, Fierlinger  \&
  Heitsch}{Ntormousi et~al.}{2011}]{Ntormousi2011}
Ntormousi E.,  Burkert A.,  Fierlinger K.,   Heitsch F.,  2011, \apj, 731, 13

\bibitem[\protect\citeauthoryear{{Oey} et~al.,}{{Oey} et~al.}{2007}]{Oey2007}
{Oey} M.~S.,  et~al., 2007, \mn@doi [\apj] {10.1086/517867}, \href
  {http://adsabs.harvard.edu/abs/2007ApJ...661..801O} {661, 801}

\bibitem[\protect\citeauthoryear{{Oh}, {de Blok}, {Brinks}, {Walter}  \&
  {Kennicutt}}{{Oh} et~al.}{2011}]{Oh2011}
{Oh} S.-H.,  {de Blok} W.~J.~G.,  {Brinks} E.,  {Walter} F.,   {Kennicutt} Jr.
  R.~C.,  2011, \mn@doi [\aj] {10.1088/0004-6256/141/6/193}, \href
  {http://adsabs.harvard.edu/abs/2011AJ....141..193O} {141, 193}

\bibitem[\protect\citeauthoryear{{Osterbrock} \& {Ferland}}{{Osterbrock} \&
  {Ferland}}{2006}]{Osterbrock2006}
{Osterbrock} D.~E.,  {Ferland} G.~J.,  2006, {Astrophysics of gaseous nebulae
  and active galactic nuclei}, 2nd edn.
University Science Books

\bibitem[\protect\citeauthoryear{Ott, Walter, Brinks, Van~Dyk, Dirsch  \&
  Klein}{Ott et~al.}{2001}]{Ott2001}
Ott J.,  Walter F.,  Brinks E.,  Van~Dyk S.~D.,  Dirsch B.,   Klein U.,  2001,
  \aj, 122, 3070

\bibitem[\protect\citeauthoryear{{Pilyugin} \& {Grebel}}{{Pilyugin} \&
  {Grebel}}{2016}]{Pilyugin2016}
{Pilyugin} L.~S.,  {Grebel} E.~K.,  2016, \mn@doi [\mnras]
  {10.1093/mnras/stw238}, \href
  {http://adsabs.harvard.edu/abs/2016MNRAS.457.3678P} {457, 3678}

\bibitem[\protect\citeauthoryear{Pilyugin \& Thuan}{Pilyugin \&
  Thuan}{2005}]{Pilyugin2005}
Pilyugin L.~S.,  Thuan T.~X.,  2005, \apj, 631, 231

\bibitem[\protect\citeauthoryear{{Puche} \& {Westpfahl}}{{Puche} \&
  {Westpfahl}}{1994}]{Puche1994}
{Puche} D.,  {Westpfahl} D.,  1994, in {Meylan} G.,  {Prugniel} P.,  eds,
  European Southern Observatory Conference and Workshop Proceedings Vol. 49,
  European Southern Observatory Conference and Workshop Proceedings. p.~273

\bibitem[\protect\citeauthoryear{{Rahner}, {Pellegrini}, {Glover}  \&
  {Klessen}}{{Rahner} et~al.}{2017}]{Rahner2017}
{Rahner} D.,  {Pellegrini} E.~W.,  {Glover} S.~C.~O.,   {Klessen} R.~S.,  2017,
  \mn@doi [\mnras] {10.1093/mnras/stx1532}, \href
  {http://adsabs.harvard.edu/abs/2017MNRAS.470.4453R} {470, 4453}

\bibitem[\protect\citeauthoryear{{Sargent}, {Sancisi}  \& {Lo}}{{Sargent}
  et~al.}{1983}]{Sargent1983}
{Sargent} W.~L.~W.,  {Sancisi} R.,   {Lo} K.~Y.,  1983, \mn@doi [\apj]
  {10.1086/160715}, \href {http://adsabs.harvard.edu/abs/1983ApJ...265..711S}
  {265, 711}

\bibitem[\protect\citeauthoryear{{Schaerer} \& {Vacca}}{{Schaerer} \&
  {Vacca}}{1998}]{Schaerer1998}
{Schaerer} D.,  {Vacca} W.~D.,  1998, \mn@doi [\apj] {10.1086/305487}, \href
  {http://adsabs.harvard.edu/abs/1998ApJ...497..618S} {497, 618}

\bibitem[\protect\citeauthoryear{{Sedov}}{{Sedov}}{1946}]{sedov}
{Sedov} L.~I.,  1946, Dokl. Akad. Nauk SSSR, 52, 17

\bibitem[\protect\citeauthoryear{Seon}{Seon}{2009}]{Seon2009}
Seon K.-I.,  2009, \apj, 703, 1159

\bibitem[\protect\citeauthoryear{{Sharma}, {Roy}, {Nath}  \&
  {Shchekinov}}{{Sharma} et~al.}{2014}]{Sharma2014}
{Sharma} P.,  {Roy} A.,  {Nath} B.~B.,   {Shchekinov} Y.,  2014, \mn@doi
  [\mnras] {10.1093/mnras/stu1307}, \href
  {http://adsabs.harvard.edu/abs/2014MNRAS.443.3463S} {443, 3463}

\bibitem[\protect\citeauthoryear{Simpson, Hunter  \& Knezek}{Simpson
  et~al.}{2005}]{Simpson2005}
Simpson C.~E.,  Hunter D.~A.,   Knezek P.~M.,  2005, The Astronomical Journal,
  129, 160

\bibitem[\protect\citeauthoryear{{Skillman}, {Terlevich}, {Teuben}  \& {van
  Woerden}}{{Skillman} et~al.}{1988}]{Skillman1988}
{Skillman} E.~D.,  {Terlevich} R.,  {Teuben} P.~J.,   {van Woerden} H.,  1988,
  \aap, \href {http://adsabs.harvard.edu/abs/1988A%26A...198...33S} {198, 33}

\bibitem[\protect\citeauthoryear{{Smith}, {Norris}  \& {Crowther}}{{Smith}
  et~al.}{2002}]{Smith2002}
{Smith} L.~J.,  {Norris} R.~P.~F.,   {Crowther} P.~A.,  2002, \mn@doi [\mnras]
  {10.1046/j.1365-8711.2002.06042.x}, \href
  {http://adsabs.harvard.edu/abs/2002MNRAS.337.1309S} {337, 1309}

\bibitem[\protect\citeauthoryear{{Stilp}, {Dalcanton}, {Warren}, {Skillman},
  {Ott}  \& {Koribalski}}{{Stilp} et~al.}{2013}]{Stilp2013}
{Stilp} A.~M.,  {Dalcanton} J.~J.,  {Warren} S.~R.,  {Skillman} E.,  {Ott} J.,
   {Koribalski} B.,  2013, \mn@doi [\apj] {10.1088/0004-637X/765/2/136}, \href
  {http://adsabs.harvard.edu/abs/2013ApJ...765..136S} {765, 136}

\bibitem[\protect\citeauthoryear{{Tang}, {Bressan}, {Rosenfield}, {Slemer},
  {Marigo}, {Girardi}  \& {Bianchi}}{{Tang} et~al.}{2014}]{tang14}
{Tang} J.,  {Bressan} A.,  {Rosenfield} P.,  {Slemer} A.,  {Marigo} P.,
  {Girardi} L.,   {Bianchi} L.,  2014, \mn@doi [\mnras]
  {10.1093/mnras/stu2029}, \href
  {http://adsabs.harvard.edu/abs/2014MNRAS.445.4287T} {445, 4287}

\bibitem[\protect\citeauthoryear{{Tully}, {Bottinelli}, {Fisher}, {Gougenheim},
  {Sancisi}  \& {van Woerden}}{{Tully} et~al.}{1978}]{Tully1978}
{Tully} R.~B.,  {Bottinelli} L.,  {Fisher} J.~R.,  {Gougenheim} L.,  {Sancisi}
  R.,   {van Woerden} H.,  1978, \aap, \href
  {http://adsabs.harvard.edu/abs/1978A%26A....63...37T} {63, 37}

\bibitem[\protect\citeauthoryear{{Utrobin} \& {Chugai}}{{Utrobin} \&
  {Chugai}}{2017}]{Utrobin2017}
{Utrobin} V.~P.,  {Chugai} N.~N.,  2017, \mn@doi [\mnras]
  {10.1093/mnras/stx2415}, \href
  {http://adsabs.harvard.edu/abs/2017MNRAS.472.5004U} {472, 5004}

\bibitem[\protect\citeauthoryear{{Vasiliev} \& {Shchekinov}}{{Vasiliev} \&
  {Shchekinov}}{2017}]{Vasiliev2017}
{Vasiliev} E.~O.,  {Shchekinov} Y.~A.,  2017, \mn@doi [Open Astronomy]
  {10.1515/astro- 2017- 0021}, 26, 233

\bibitem[\protect\citeauthoryear{{Vasiliev}, {Moiseev}  \&
  {Shchekinov}}{{Vasiliev} et~al.}{2015a}]{Vasiliev2015a}
{Vasiliev} E.~O.,  {Moiseev} A.~V.,   {Shchekinov} Y.~A.,  2015a, Baltic
  Astronomy, \href {http://adsabs.harvard.edu/abs/2015BaltA..24..213V} {24,
  213}

\bibitem[\protect\citeauthoryear{{Vasiliev}, {Nath}  \&
  {Shchekinov}}{{Vasiliev} et~al.}{2015b}]{Vasiliev2015b}
{Vasiliev} E.~O.,  {Nath} B.~B.,   {Shchekinov} Y.,  2015b, \mn@doi [\mnras]
  {10.1093/mnras/stu2133}, \href
  {http://adsabs.harvard.edu/abs/2015MNRAS.446.1703V} {446, 1703}

\bibitem[\protect\citeauthoryear{{Vorobyov} \& {Basu}}{{Vorobyov} \&
  {Basu}}{2005}]{Vorobyov2005}
{Vorobyov} E.~I.,  {Basu} S.,  2005, \mn@doi [\aap]
  {10.1051/0004-6361:20041324}, \href
  {http://adsabs.harvard.edu/abs/2005A%26A...431..451V} {431, 451}

\bibitem[\protect\citeauthoryear{Walter, Brinks, de Blok, Bigiel, Kennicutt,
  Thornley  \& Leroy}{Walter et~al.}{2008}]{things}
Walter F.,  Brinks E.,  de Blok W. J.~G.,  Bigiel F.,  Kennicutt R.~C.,
  Thornley M.~D.,   Leroy A.~K.,  2008, \aj, 136, 2563

\bibitem[\protect\citeauthoryear{Warren et~al.,}{Warren
  et~al.}{2011}]{Warren2011}
Warren S.~R.,  et~al., 2011, \apj, 738, 10

\bibitem[\protect\citeauthoryear{{Warren} et~al.,}{{Warren}
  et~al.}{2012}]{Warren2012}
{Warren} S.~R.,  et~al., 2012, \mn@doi [\apj] {10.1088/0004-637X/757/1/84},
  \href {http://adsabs.harvard.edu/abs/2012ApJ...757...84W} {757, 84}

\bibitem[\protect\citeauthoryear{Weaver, McCray, Castor, Shapiro  \&
  Moore}{Weaver et~al.}{1977}]{Weaver1977}
Weaver R.,  McCray R.,  Castor J.,  Shapiro P.,   Moore R.,  1977, \apj, 218,
  377

\bibitem[\protect\citeauthoryear{Weisz, Skillman, Cannon, Dolphin,
  Kennicutt~Jr., Lee  \& Walter}{Weisz et~al.}{2008}]{Weisz2008}
Weisz D.~R.,  Skillman E.~D.,  Cannon J.~M.,  Dolphin A.~E.,  Kennicutt~Jr.
  R.~C.,  Lee J.~C.,   Walter F.,  2008, \apj, 689, 160

\bibitem[\protect\citeauthoryear{Weisz, Skillman, Cannon, Walter, Brinks, Ott
  \& Dolphin}{Weisz et~al.}{2009a}]{Weisz2009b}
Weisz D.~R.,  Skillman E.~D.,  Cannon J.~M.,  Walter F.,  Brinks E.,  Ott J.,
  Dolphin A.~E.,  2009a, \apj, 691, L59

\bibitem[\protect\citeauthoryear{Weisz, Skillman, Cannon, Dolphin, Kennicutt,
  Lee  \& Walter}{Weisz et~al.}{2009b}]{Weisz2009a}
Weisz D.~R.,  Skillman E.~D.,  Cannon J.~M.,  Dolphin A.~E.,  Kennicutt R.~C.,
  Lee J.~C.,   Walter F.,  2009b, \apj, 704, 1538

\bibitem[\protect\citeauthoryear{{Westpfahl} \& {Puche}}{{Westpfahl} \&
  {Puche}}{1994}]{Westpfahl1994}
{Westpfahl} D.,  {Puche} D.,  1994, in {Meylan} G.,  {Prugniel} P.,  eds,
  European Southern Observatory Conference and Workshop Proceedings Vol. 49,
  European Southern Observatory Conference and Workshop Proceedings. p.~295

\bibitem[\protect\citeauthoryear{Wiebe, Khramtsova, Egorov  \&
  Lozinskaya}{Wiebe et~al.}{2014}]{Wiebe2014}
Wiebe D.~S.,  Khramtsova M.~S.,  Egorov O.~V.,   Lozinskaya T.~A.,  2014,
  Astronomy Letters, 40, 278

\bibitem[\protect\citeauthoryear{{Young} \& {Lo}}{{Young} \&
  {Lo}}{1997}]{Young1997}
{Young} L.~M.,  {Lo} K.~Y.,  1997, \mn@doi [\apj] {10.1086/304909}, \href
  {http://adsabs.harvard.edu/abs/1997ApJ...490..710Y} {490, 710}

\bibitem[\protect\citeauthoryear{Zhang et~al.,}{Zhang et~al.}{2017}]{Zhang2017}
Zhang K.,  et~al., 2017, \mnras, 466, 3217

\bibitem[\protect\citeauthoryear{{de Vaucouleurs}, {de Vaucouleurs}, {Corwin},
  {Buta}, {Paturel}  \& {Fouqu{\'e}}}{{de Vaucouleurs}
  et~al.}{1991}]{deVaucouleurs1991}
{de Vaucouleurs} G.,  {de Vaucouleurs} A.,  {Corwin} Jr. H.~G.,  {Buta} R.~J.,
  {Paturel} G.,   {Fouqu{\'e}} P.,  1991, {Third Reference Catalogue of Bright
  Galaxies. Volume I: Explanations and references. Volume II: Data for galaxies
  between 0$^{h}$ and 12$^{h}$. Volume III: Data for galaxies between 12$^{h}$
  and 24$^{h}$.}

\bibitem[\protect\citeauthoryear{{van Dyk}, {Puche}  \& {Wong}}{{van Dyk}
  et~al.}{1998}]{vanDyk1998}
{van Dyk} S.~D.,  {Puche} D.,   {Wong} T.,  1998, \mn@doi [\aj]
  {10.1086/300584}, \href {http://adsabs.harvard.edu/abs/1998AJ....116.2341V}
  {116, 2341}

\makeatother
\end{thebibliography}

\label{lastpage}

\end{document}